%% file: TRG-17-001_temp.tex
\pdfoutput=1

\documentclass[11pt,twoside,a4paper,cmspaper,final,collab]{cms-tdr}
\def\svnVersion{1de0389-D}\def\svnDate{2020/10/19}\def\cmsCernNoTag{CERN-EP-2020-065}\def\cmsCernDate{\today}\def\cmsMessage{"Published in the Journal of Instrumentation as \href{http://dx.doi.org/10.1088/1748-0221/15/10/P10017}{\doi{10.1088/1748-0221/15/10/P10017}.}"}

\begin{document}\cmsNoteHeader{TRG-17-001}

\hyphenation{had-ron-i-za-tion}
\hyphenation{cal-or-i-me-ter}
\hyphenation{de-vices}
\newlength\cmsFigWidth
\ifthenelse{\boolean{cms@external}}{\setlength\cmsFigWidth{0.49\textwidth}}{\setlength\cmsFigWidth{0.65\textwidth}}
\providecommand{\cmsLeft}{left\xspace}
\providecommand{\cmsRight}{right\xspace}

\newcommand{\utca}{\ensuremath{\mu}TCA\xspace} \newcommand{\uGT}{\ensuremath{\mu}GT\xspace}
\newcommand{\uGMT}{\ensuremath{\mu}GMT\xspace} \newcommand{\bmtfetalimit}{0.83}
\newcommand{\omtfetalimit}{1.24} \newcommand{\emtfetalimit}{2.5}
\newcommand{\ETT}{\ensuremath{ETT}}
\newcommand{\ETTpos}{\ensuremath{ETT_\text{pos}}}
\newcommand{\ETTneg}{\ensuremath{ETT_\text{neg}}}
\newcommand{\ETTtot}{\ensuremath{ETT_\text{tot}}}

\newcommand{\ptoffline}{\ensuremath{\pt^{\text{offline}}\xspace}}
\newcommand{\etaoffline}{\ensuremath{\eta^{\text{offline}}\xspace}}
\newcommand{\ptLone}{\ensuremath{\pt^{\text{L1}}\xspace}}
\newcommand{\egamma}{\ensuremath{\Pe/\PGg\xspace}}
\newcommand{\etiso}{\ensuremath{\ET^\text{iso}\xspace}}
\newcommand{\mmumu}{\ensuremath{m_{\PGm\PGm}\xspace}}
\newcommand{\mjj}{\ensuremath{m_\mathrm{jj}\xspace}}
\newcommand{\ntt}{\ensuremath{n_\mathrm{TT}\xspace}}
\newcommand{\MHT}{\ensuremath{H_{\mathrm{T}}^{\text{miss}}}\xspace}

\newlength\cmsTabSkip\setlength{\cmsTabSkip}{1ex}

\cmsNoteHeader{TRG-17-001}

\title{Performance of the CMS Level-1 trigger in proton-proton collisions at
$\sqrt{s} = 13\TeV$}

\date{\today}

\abstract{At the start of Run 2 in 2015, the LHC delivered proton-proton
collisions at a center-of-mass energy of 13\TeV.  During Run~2 (years
2015--2018) the LHC eventually reached a luminosity of $2.1\times
10^{34}\percms$, almost three times that reached during Run~1 (2009--2013) and a
factor of two larger than the LHC design value, leading to  events with up to a
mean of about 50 simultaneous inelastic proton-proton collisions per bunch
crossing (pileup).  The CMS Level-1 trigger was upgraded prior to 2016 to
improve the selection of physics events in the challenging conditions posed by
the second run of the LHC.  This paper describes the performance of the CMS
Level-1 trigger upgrade during the data taking period of 2016--2018.  The
upgraded trigger implements pattern recognition and boosted decision tree
regression techniques for muon reconstruction, includes pileup subtraction for
jets and energy sums, and incorporates pileup-dependent isolation requirements
for electrons and tau leptons. In addition, the new trigger calculates
high-level quantities such as the invariant mass of pairs of reconstructed
particles. The upgrade reduces the trigger rate from background processes and
improves the trigger efficiency for a wide variety of physics signals. }

\hypersetup{ pdfauthor={CMS Collaboration}, pdftitle={Performance of the CMS Level-1 Trigger in
proton-proton collisions at sqrt(s)=13 TeV}, pdfsubject={CMS}, pdfkeywords={CMS,
trigger, physics}}

\maketitle

\tableofcontents

\section{Introduction}

The CERN LHC collides bunches of particles in the CMS and ATLAS experiments at a
maximum rate of about 40\unit{MHz}, where the bunches are spaced 25\unit{ns}
apart. Of these only about 1000 per second can be recorded for further analysis.
The Level-1 trigger system uses custom hardware processors to select up to
100\unit{kHz} of the most interesting events with a latency of 4\unit{\mus}. The High
Level Trigger (HLT) then performs a more detailed reconstruction, including
particle tracking, on a commodity computing processor farm, reducing the rate by
another factor of 100 in a few hundred milliseconds.  Events passing the HLT
selection are sent to a separate computing farm for more accurate event
reconstruction and storage.

The LHC operation is organized into periods of physics production, where protons
or heavy ions are collided, and  periods of shutdown during which repairs and
upgrade work are performed. The original CMS trigger system performed
efficiently in the LHC Run~1 (between 2009 and 2013) and 2015. Its design is
described in Ref.~\cite{CMS:trigger-tdr-2000} and its performance in
Ref.~\cite{CMS:trigger-run1}. In 2015 the LHC increased the proton-proton
center-of-mass collision energy from 8 to 13\TeV. The instantaneous luminosity
steadily increased throughout Run~2, which ended in 2018. These changes were
designed to provide a larger data set for studies of rare interactions and
searches for new physics, but they also presented several challenges to the
trigger system. Improved trigger algorithms were needed to enhance the
separation of signal and background events and to provide more accurate energy
reconstruction in the presence of a larger number of simultaneous collisions per
LHC bunch crossing (pileup).

The CMS Collaboration undertook a major upgrade to the Level-1 trigger system
(Phase~1) between Run~1 and Run~2, and plans a second upgrade (Phase~2) after
Run~3 ends (expected in 2024).  The Phase~1 upgrade replaced all of the Level-1
trigger hardware, cables, electronics boards, firmware, and software, as
described in the Technical Design Report for the Level-1 trigger upgrade
\cite{CMS-L1T-TDR}.  Despite higher instantaneous luminosity, energy, and
pileup, the upgraded Level-1 trigger maintained or increased its efficiency to
separate the chosen signal events from background, because of finer detector
input granularity, enhanced object reconstruction (\eg, $\PGm$, $\egamma$, jet,
$\PGt$, and energy sums), and correlated multi-object triggers targeting
specific physics signatures.

This paper describes the trigger algorithms of the Phase~1 Level-1  trigger
upgrade and reports their performance, measured using Run~2 data.  A brief
overview of the CMS detector is given in Section \ref{sec:cms}.  Section
\ref{sec:lhc} describes the performance of the LHC and its impact on the CMS
trigger system in Run~2.  Section \ref{sec:menu} provides an overview of the
large collection of algorithms used to select events for physics measurements.
Section \ref{sec:design} describes the design of the upgraded Level-1 trigger,
including updates since Ref.~\cite{CMS-L1T-TDR}.  The reconstruction algorithms,
along with their performance, are described in detail for each subdetector: the
muon trigger in Section \ref{sec:muons}, and the calorimeter trigger in Section
\ref{sec:calo}; appendix \ref{sec:prefiring} provides details on a calorimeter
trigger issue that affected Run 2 data.  Section \ref{sec:global} provides two
examples of new multi-object global trigger algorithms, while Section
\ref{sec:certification} describes how the data quality of the Level-1 trigger is
monitored in real time.  Section \ref{sec:summary} summarizes and draws
conclusions regarding the achievements of the upgraded Level-1 trigger in Run~2.

\section{The CMS detector} \label{sec:cms}

The central feature of the CMS apparatus is a superconducting solenoid of
6\unit{m} internal diameter, providing a magnetic field of 3.8\unit{T}. Within
the solenoid volume are a silicon pixel and strip tracker, a lead tungstate
crystal electromagnetic calorimeter (ECAL), and a brass and scintillator hadron
calorimeter (HCAL), each composed of a barrel and two endcap sections. Forward
calorimeters extend the pseudorapidity ($\eta$) coverage provided by the barrel
and endcap detectors. Muons are detected in gas-ionization chambers embedded in
the steel flux-return yoke outside the solenoid.  Drift tubes (DTs) cover the
central region ($\abs{\eta} < 1.2$), cathode strip chambers (CSCs) are installed in
the endcaps ($0.9<\abs{\eta}<2.4$), and resistive plate chambers (RPCs) provide
overlap out to $\abs{\eta}<1.7$.  A more detailed description of the CMS detector,
together with a definition of the coordinate system used and the relevant
kinematic variables, can be found in Ref.~\cite{CMS-Detector}.

\section{The LHC in Run~2} \label{sec:lhc}

Trigger performance depends on the running conditions of the LHC, such as
instantaneous luminosity, number of colliding bunches, and even the structure of
the filling scheme.  The LHC was designed to collide protons with a
center-of-mass energy of 14\TeV and an instantaneous luminosity corresponding to $1.0 \times
10^{34} \percms$, but it initially operated at lower energies and intensities.
During Run~1 the center-of-mass energy was increased in steps up to 8\TeV with a
peak instantaneous luminosity near $8.0 \times 10^{33} \percms$.  At that time,
the LHC operated with a longer minimum bunch spacing of 50\unit{ns}, instead of
the originally foreseen 25\unit{ns}.

During the first long shutdown period of the LHC in 2013--2014, the accelerator
was modified to provide safe operation at 13\TeV with 25\unit{ns} bunch spacing,
and the CMS experiment underwent upgrades \cite{CMS:phase-1-upgrade} to prepare
for a dramatic increase in collision rate.  Run~2 of the LHC lasted from 2015
until the end of 2018 with peak instantaneous luminosities of about $2.1 \times
10^{34} \percms$.  A typical filling scheme for the LHC in Run~2 comprised 2556
proton bunches per beam out of 3564 possible bunch locations.  The bunches were
grouped in ``trains'' of 48 bunches with 25\unit{ns} spacing, with larger gaps
between trains.  Of these, 2544 bunches collided at the CMS interaction point.
In the second long shutdown of the LHC (2019--2020), upgrades to the accelerator
are planned, possibly increasing the center-of-mass energy for Run 3
(foreseen to start late 2021 or early 2022), and allowing the LHC to sustain a maximum
instantaneous luminosity of $2.0 \times 10^{34} \percms$ for longer periods of
time.

In 2017 the LHC suffered frequent beam dumps. These were caused when an electron
cloud generated by tightly packed bunches interacted with frozen gas in the beam
pipe. The gas had become trapped in one area of the LHC during the year-end
technical stop between 2016 and 2017 \cite{lhc-electron-cloud}. To mitigate this
effect, the LHC moved to a special ``8b4e'' filling scheme in September 2017. In
this scheme the standard 48 bunch trains are replaced by mini-trains of 8 filled
bunches followed by 4 empty slots, suppressing the formation of electron clouds.
Since the 8b4e filling scheme allows a maximum of 1916 filled bunches in the
LHC, the peak instantaneous luminosity was leveled to ${\approx}1.55 \times
10^{34} \percms$, so the average pileup would not exceed 60. The LHC delivered
41.0 and 49.8\fbinv of proton-proton collisions to CMS in 2016 and 2017,
respectively, during which 35.9 and 41.5\fbinv of good quality data were
recorded.

In 2018 the beam dump issues had largely been mitigated so that a return to the
preferred nominal scheme was possible. The advantage of this scheme is the use
of a larger number of colliding bunches, providing higher instantaneous
luminosity without increasing the pileup. The peak luminosity of about $2.0
\times 10^{34} \percms$ led to an average pileup of 55, similar to that at the
start of 2017. The LHC ran smoothly in 2018 and delivered an integrated
luminosity of 68.0\fbinv to CMS, which recorded 59.7\fbinv of good quality data.

The LHC periodically provides short special runs, such as the van der Meer
scans,  with nonstandard beam settings, which require dedicated triggers and
calibrations. The precise measurement of the integrated luminosity recorded by
CMS is a necessary ingredient for most of the CMS physics results, and the CMS
experiment has several detectors dedicated to this measurement. The van der Meer
scans provide data necessary to calibrate these measurements. 

During the van der Meer scans, the LHC beams are scanned across each other to
provide an accurate luminosity calibration. The trigger system is used to
measure the rate of the beam collisions, which is used to calculate the
luminosity, as described in Ref.~\cite{vdm-ichep18}.  For some periods of the
van der Meer scan, the Level-1 trigger system records, with high rate, only
events from selected bunch crossings in the LHC orbit bunch structure to improve
the precision of the luminosity calibration.

\section{The physics program and the trigger menu} \label{sec:menu}

The CMS physics program targets many areas of interest to the high-energy
physics community.  After the discovery of the Higgs boson
\cite{Aad:2012tfa,Chatrchyan:2012xdj,CMS:higgs-discovery-long-paper}, measuring
its properties, which are currently compatible with the standard model (SM)
predictions \cite{Aad:2015zhl}, became of central importance.  Searches for
supersymmetric and exotic particles, together with candidates for dark matter,
are also central to the CMS physics program and they require a high-performance
trigger.  Such a high-performance trigger also enables precision measurements of
SM properties in the electroweak, top quark, and quantum chromodynamics (QCD)
sectors, with special attention to the physics of bottom quarks, where
triggering objects often have low transverse momentum (\pt).  Heavy ion
collisions are included in the CMS physics program, expanding our knowledge of
quark-gluon plasma dynamics. 

The Level-1 trigger information from the muon and calorimeter detectors with
coarse granularity and precision is used to select collision events for
investigations in all of the previously mentioned physics areas.  The selection
is performed using a list of algorithms (known as ``seeds''), which  check
events against predetermined criteria, that are collectively called the
``menu''. Any event that satisfies the conditions of at least one seed in the
menu is accepted for further processing in the trigger chain. This initiates a
readout of the complete detector information from the data acquisition system,
and the data are sent to the HLT. The broad range of menu algorithms reflects
the wide variety of research interests of the CMS Collaboration. The Level-1
menu evolves with shifting CMS physics priorities and adapts to changes in beam
or detector performance.

The most straightforward trigger algorithms consist of criteria applied to one
or more objects of a single type, such as muons, hadronic jets, tau leptons,
photons or electrons, scalar sum of transverse energy (\HT), and the energy
corresponding to the vector sum of the transverse missing momentum  (\MET).
Typical criteria include thresholds on the transverse component of the object's
energy \ET (or momentum), and on its $\eta$.  Signal processes with massive
particles typically produce objects at high \pt and low $\abs{\eta}$ values
(central in the detector), whereas the vast majority of background objects are
low \pt and tend to have higher $\abs{\eta}$.  Single- and double-object seeds
form the majority of the menu and cover about 75\% of the available rate.  Muon
and electron thresholds are chosen to efficiently select leptonic {\PW} and \PZ
boson decays, and $\PGt\PGt$ thresholds are set to maximize the Higgs boson
acceptance in this decay channel.

\begin{figure} \centering {
		\includegraphics[width=0.9\textwidth]{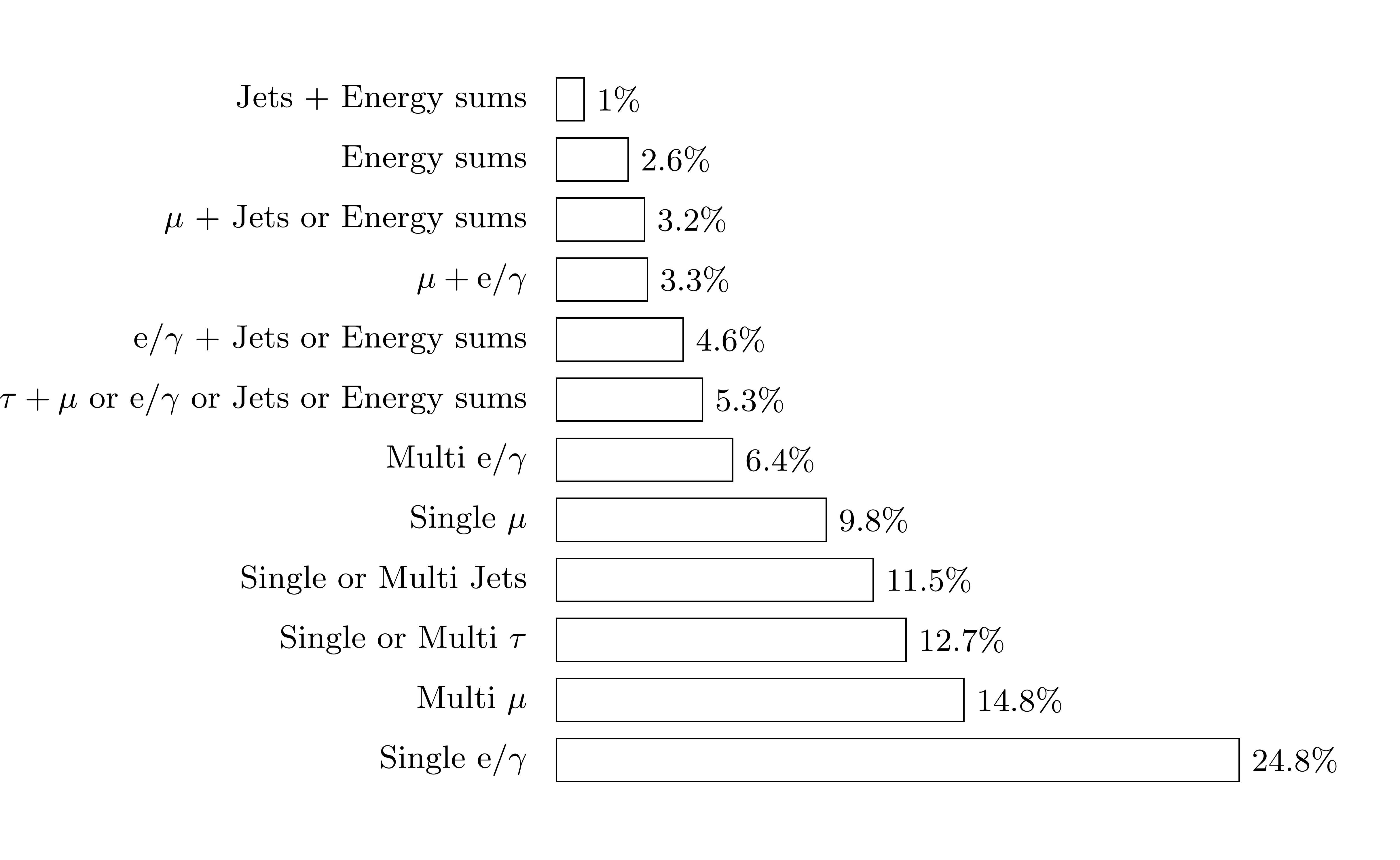} }
	\caption{Fractions of the 100\unit{kHz} rate allocation for single- and
	multi-object triggers and cross triggers in a typical CMS physics menu
	during Run~2.} \label{fig:menu-algorithm-share} \end{figure}

The ``cross'' seeds combine physics objects of different types, for example a
muon and a jet, allowing lower thresholds that target a diverse range of
signals.  More complex algorithms using correlations between multiple objects
select highly specific signal events, such as hadrons decaying to muons, or
Higgs bosons produced via vector boson fusion (VBF).  Finally, a small fraction
of events passing less restrictive algorithms are collected to calibrate the
detectors and measure trigger efficiencies.
Figure~\ref{fig:menu-algorithm-share} shows the ``proportional rate'', the
fraction of the maximum Level-1 trigger rate allocated to single-, multi- (same
type), and cross- (different type) object seeds. In the proportional rate
calculation, events triggered by $N$ different seeds are weighted by 1/$N$ to
ensure that the total sums to 100\%.

The menu algorithms are designed using a simulation of the Level-1 object
reconstruction using either Monte Carlo (MC) simulated collision events or,
where possible, previously collected data. The seed thresholds are adjusted to
achieve a total menu rate that is less than 100\unit{kHz}, estimated with data
collected with a trigger that  requires only a crossing of proton bunches,
referred to as a zero-bias trigger. The detection of the crossing of bunches
consists of the coincidence of two simultaneous signals from the two beam
pick-up monitors installed at the opposite ends of CMS along the beam line.

\begin{figure} \centering
	\includegraphics[width=0.45\textwidth]{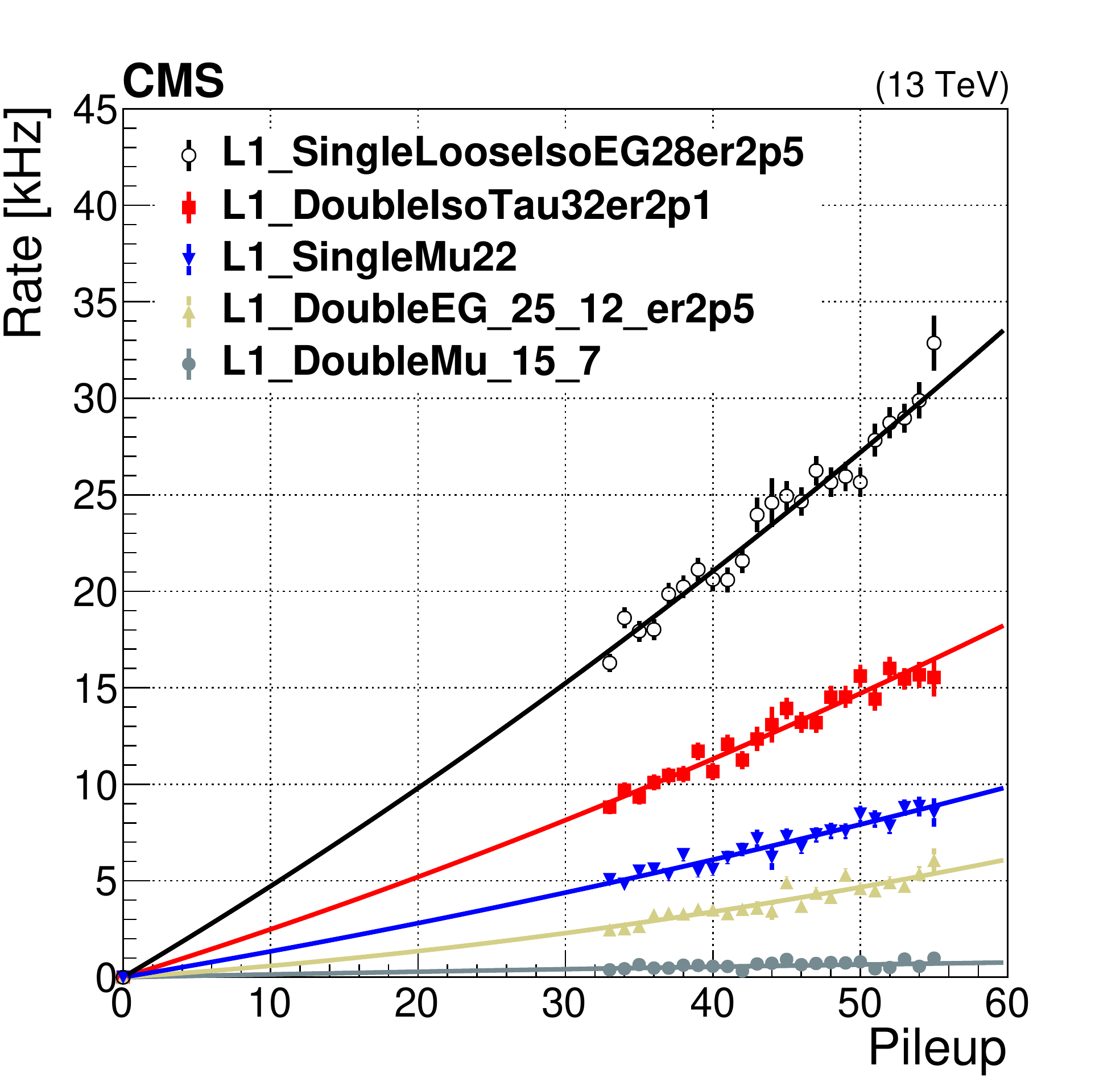}
	\includegraphics[width=0.45\textwidth]{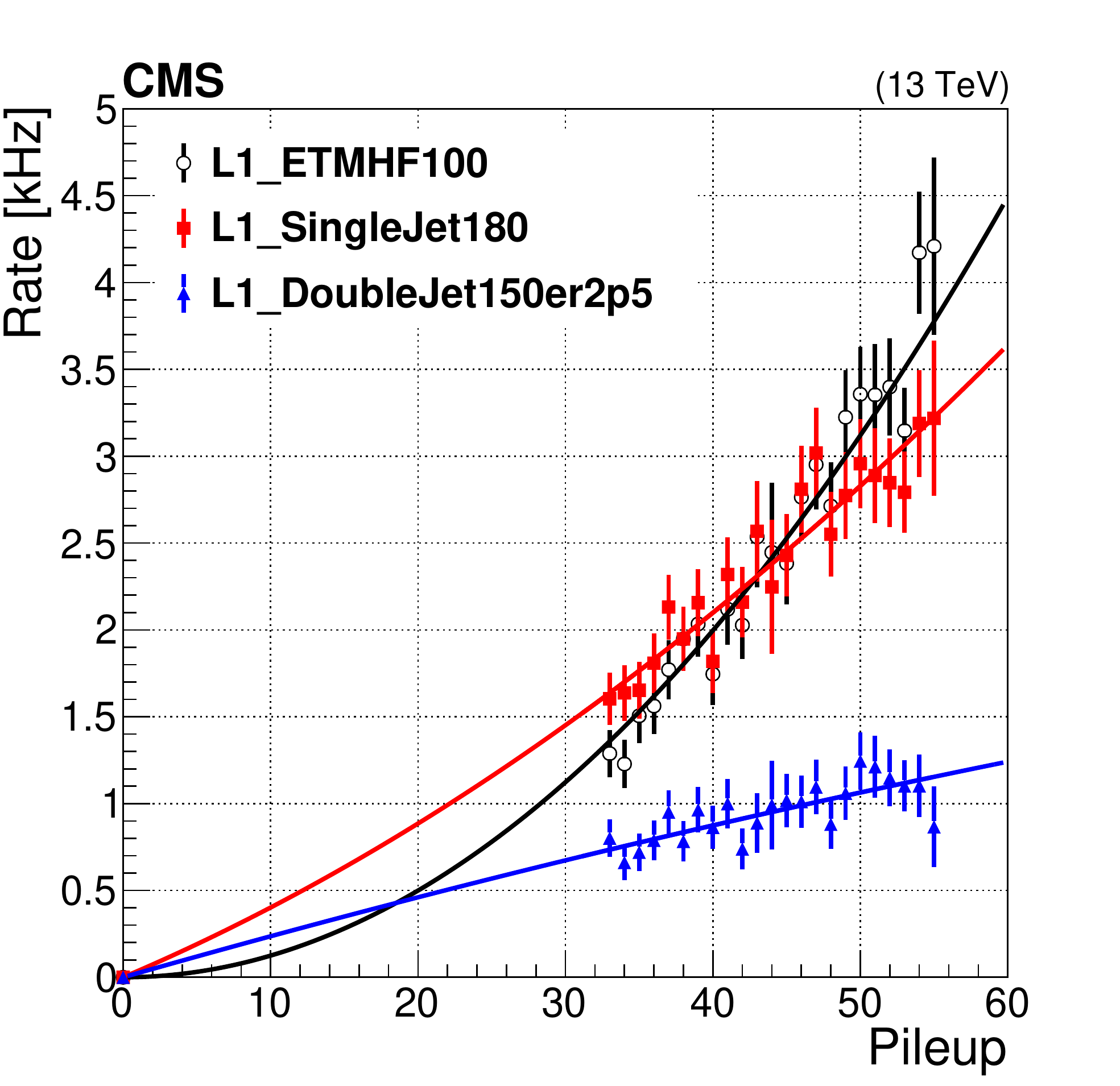}
	\caption{Level-1 trigger rates as a function of pileup for some
	benchmark seeds targeting leptons (\cmsLeft) and hadrons (\cmsRight).
	Rates are measured using data recorded during the 2018 LHC run.
	Definitions of the seed names are in Table~\ref{tab:algo-description}.
	The curves represent fits to the data points that are quadratic and
	constrained to pass through the origin.}
\label{fig:menu-algo-rate-vs-pu} \end{figure}

Trigger algorithm rates depend on the ability of the trigger reconstruction to
discriminate between signal objects, arising in hard collisions, from
backgrounds or misidentified objects. This becomes more difficult as pileup
increases. Figure~\ref{fig:menu-algo-rate-vs-pu} shows the rate of some
benchmark trigger seeds targeting leptons (\cmsLeft) and hadrons (\cmsRight) as
a function of pileup. Rate and pileup are measured in a time interval of a
``luminosity section'', corresponding to $2^{18}$ LHC orbits or 23.3 seconds of
data taking. In this and subsequent figures, error bands in the data points
represent their statistical uncertainty only. Single-object trigger rates
generally increase linearly with pileup, whereas double-object paths may have a
higher-order dependency. The largest dependence on pileup is shown by the seeds
based on the missing transverse energy.  The Level-1 trigger reconstruction
cannot distinguish between objects generated by different collisions within the
same bunch crossing. However, in offline reconstruction the objects are
associated with different reconstructed vertices that originate from different
collisions. This requires tracking information, which is not available in the
Level-1 trigger.

\begin{table}[] \topcaption{Detailed description of Level-1 trigger seed names
		used in Figure~\ref{fig:menu-algo-rate-vs-pu}}
		\begin{tabular}{l@{  }l} \hline Algorithm name &
			Description \\ \hline
			L1\_SingleLooseIsoEG28er2p5 \, & Single loosely isolated
			$\egamma$ with $\ET > 28\GeV$ and $\abs{\eta} < 2.5$\\
			L1\_DoubleIsoTau32er2p1 & Double isolated $\PGt$ with
			$\ET > 32\GeV$ and $\abs{\eta} < 2.1$ \\ L1\_SingleMu22 &
			Single muon with $\pt>22\GeV$ \\
			L1\_DoubleEG\_25\_12\_er2p5 & Double $\egamma$ with $\ET
			> 25 \GeV, 12\GeV$ and $\abs{\eta}<2.5$ \\
			L1\_DoubleMu\_15\_7 & Double muon with $\pt > 15\GeV,
			7\GeV$ \\ L1\_ETMHF100 & $\MET > 100\GeV$\\
			L1\_SingleJet180 & Single jet with $\ET > 180\GeV$ \\
			L1\_DoubleJet150er2p5 & Double jet with $\ET > 150\GeV$
			and $\abs{\eta}<2.5$ \\ \hline \end{tabular}
\label{tab:algo-description} \end{table}

\begin{figure} \centering {
		\includegraphics[width=0.6\textwidth]{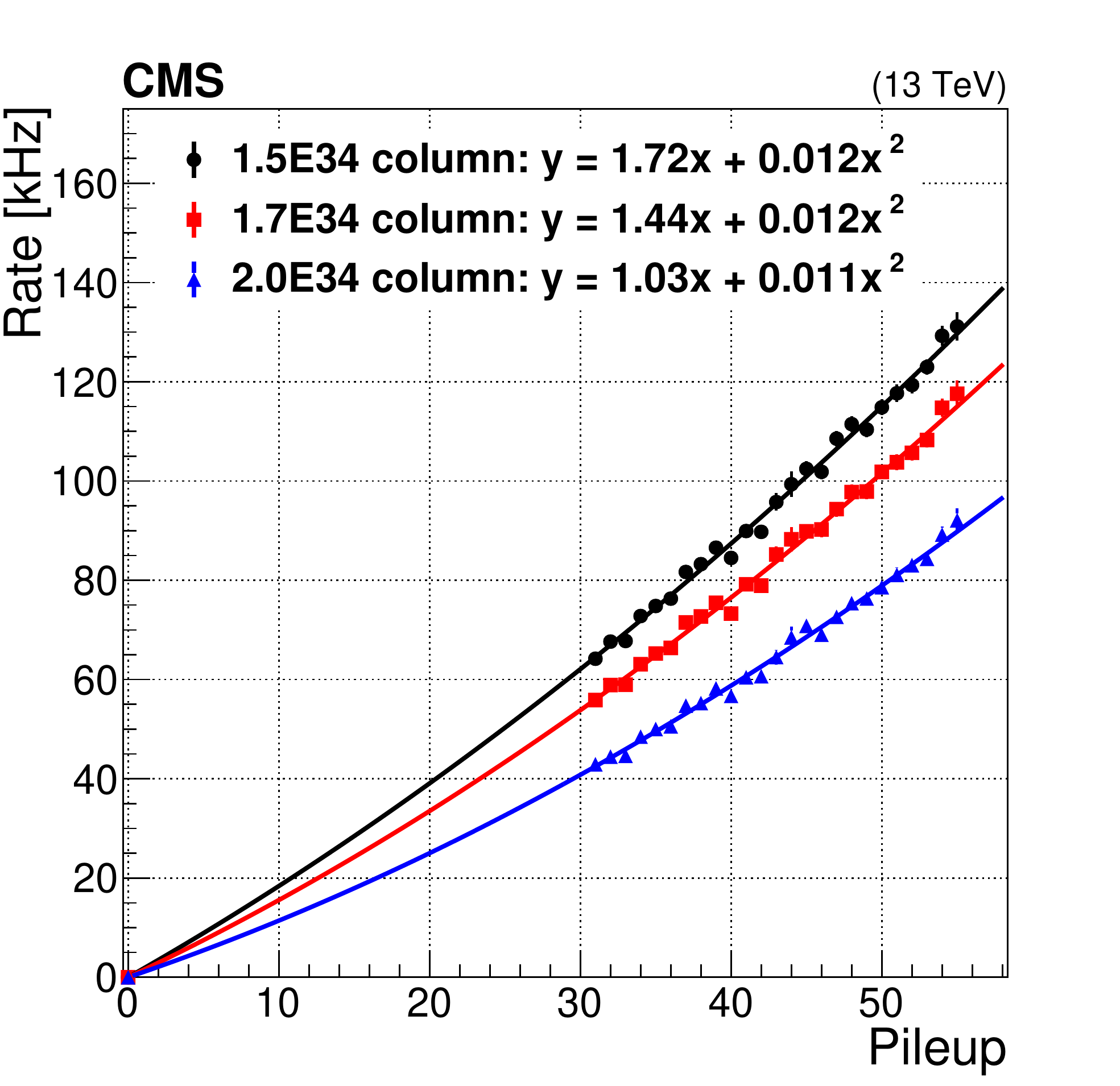} }
	\caption{Total Level-1 menu rates as a function of pileup for three sets
	of algorithms, or ``prescale columns'', defined in the text. The rates
	were recorded during a LHC fill with 2544 proton bunches. The
	instantaneous luminosities of $2.0 \times$, $1.7 \times$, and $1.5\times
	10^{34}\percms$  correspond to an average pileup of 55, 47, and 42
	respectively. The curves represent fits to the data points that are
	quadratic and constrained to pass through the origin. }
	\label{fig:menu-rate-vs-pu} \end{figure}

The rate of an algorithm can be reduced by applying a ``prescale'' that
determines what fraction of events selected by the seed will pass the trigger. A
prescale of $N$ means that only one in every $N$ events satisfying the condition
is accepted. Prescale values can only be positive integer numbers.

With a prescale of two, for example, only half of the events selected by the
seed will propagate to the HLT.  A ``prescale column'' is a set of prescale
values applied to each of the seeds in a particular menu.  During an LHC fill
the beam intensity decreases with time, so multiple prescale columns with
decreasing prescale values are used, to maximize signal efficiency while keeping
the rate under 100\unit{kHz}.

Trigger algorithms used for most physics analyses have a prescale value of one
in all columns, whereas high rate calibration triggers generally have prescale
values that are greater than one. Figure~\ref{fig:menu-rate-vs-pu} shows the
trigger rate as a function of pileup, defined as for
Fig.~\ref{fig:menu-algo-rate-vs-pu}, for a few benchmark prescale columns of the
trigger menu. These were tuned to reach a total Level-1 trigger rate of
100\unit{kHz} for three different target instantaneous luminosity values. The
prescale columns for luminosities of $1.5\times 10^{34}\percms$ and $1.7\times
10^{34}\percms$, represented by the black dots and red squares, respectively,
were not used to collect data at the highest pileup, but were activated only
when their corresponding Level-1 trigger rate was lower than 100\unit{kHz}.
Although quadratic functions fit the data points well, the very small quadratic
coefficient of these fits indicates a mostly linear dependence of the rate on
pileup suggesting a negligible contamination from pileup events.

The total number of algorithms in the CMS Level-1 menu used in proton-proton
collisions is between 350 and 400; the system architecture is limited to 512,
which is a factor 4 larger than the Run 1 system.  There were about $150$
unprescaled seeds in the menu at the end of 2018, of which approximately $100$
were ``contingency'' seeds with stricter selection requirements.  The remaining
50 were responsible for collecting data for all physics analyses that used the
full integrated luminosity delivered by the LHC. The other 250 algorithms were
prescaled and used for calibrations, monitoring, trigger efficiency
measurements, and other ancillary measurements.  Tables~\ref{tab:menu-slim} and
\ref{tab:menu-slim-2} show the unprescaled algorithms and their corresponding
thresholds. Algorithms with no $\pt$ threshold for muons have an effective
minimum $\pt$ that varies as a function of $\abs{\eta}$, because very low-$\pt$
muons do not reach the muon chambers. 

\begin{table}[] \centering \topcaption{List of the most used unprescaled Level-1
	trigger algorithms (seeds) during Run~2 and their requirements.}
	\label{tab:menu-slim} \begin{tabular}{@{}ll@{}} \hline
		Algorithm & Requirements (\pt, \ET, $\mmumu$,
		and $\mjj$ in \GeV{}) \\ \hline
         
	 \textit{Muons} & \\ Single $\PGm$       &  $\pt > 22$ \& Tight quality
		\\
         
	 Double $\PGm$       &  $\pt > 15, 7$ \& Medium quality \\

	 Double $\PGm$       &  $\pt > 15, 5$ \& Tight quality \\

	 Double $\PGm$       &  $\pt > 8, 8$ \& Tight quality \\

	 Double $\PGm$ + mass &  $\pt>4.5$ \& $\abs{\eta}<2.0$ \& Tight quality \&
		OS \& $\mmumu> 7$ \\

	 Double $\PGm$ + $\Delta R$ &  $\pt>4$ \& Tight quality \& OS \& $\Delta
		R<1.2$ \\

	 Double $\PGm$ + $\Delta R$ &  $\pt>0$ \& $\abs{\eta}<1.5$ \& Tight quality
		\& OS \& $\Delta R <1.4$ \\
         
	 Double $\PGm$ + BX  &  $\pt>0$ \& $\abs{\eta}<1.4$ \& Medium quality \&
		Non-colliding BX \\
          
	 Triple $\PGm$       &  $\pt > 5, 3, 3$ \& Medium quality \\

	 Triple $\PGm$       &  $\pt > 3, 3, 3$ \& Tight quality \\
         
	 Triple $\PGm$ + mass &  $\pt > 5, 3.5, 2.5$ \& Med.\ qual.; two $\PGm$
		OS \& $\pt > 5, 2.5$ \& $5 < \mmumu < 17$ \\
         
	 Triple $\PGm$ + mass & Three $\PGm$ any qual.; two $\PGm$  \& $\pt > 5,
		3$ \& Tight qual.\ \& OS \& $\mmumu < 9$ \\ [\cmsTabSkip]
		\textit{Electrons / photons} & ($\egamma$) \\

	 Single $\egamma$     &  $\pt > 60$ \\

	 Single $\egamma$     &  $\pt > 36$ \& $\abs{\eta}<2.5$ \\

	 Single $\egamma$     &  $\pt > 28$ \& $\abs{\eta}<2.5$ \& Loose isolation
		\\
          
	 Double $\egamma$     &  $\pt > 25, 12$ \& $\abs{\eta}<2.5$ \\

	 Double $\egamma$     &  $\pt > 22, 12$ \& $\abs{\eta}<2.5$ \& Loose
		isolation \\
          
	 Triple $\egamma$     &  $\pt > 18, 17, 8$ \& $\abs{\eta}<2.5$ \\
          
	 Triple $\egamma$     &  $\pt > 16, 16, 16$ \& $\abs{\eta}<2.5$ \\
		[\cmsTabSkip] \textit{Tau leptons} ($\PGt$) & \\ Single $\PGt$
		&  $\pt > 120$ \& $\abs{\eta}<2.1$ \\
         
	 Double $\PGt$     &  $\pt>32$ \& $\abs{\eta}<2.1$ \& Isolation \\
		[\cmsTabSkip] \textit{Jets} & \\ Single jet        &  $\pt>180$
		\\
          
	 Single jet + BX   &  $\pt>43$ \& $\abs{\eta}<2.5$ \& Non-colliding BX \\
          
	 Double jet        &  $\pt>150$ \& $\abs{\eta}<2.5$ \\
          
	 Double jet + $\Delta\eta$ &  $\pt>112$ \& $\abs{\eta}<2.3$ \&
	 $\Delta\eta<1.6$ \\
          
	 Double jet + mass &  $\pt> 110, 35$; two jets $\pt>35$ \& $\mjj>620$ \\
         
	 Double jet + mass &  $\pt> 30$ \& $\abs{\eta}<2.5$ \& $\Delta\eta<1.5$ \&
	 $\mjj>300$ \\
          
	 Triple jet    &  $\pt> 95, 75, 65$; two jets $\pt> 75, 65$ \&
	 $\abs{\eta}<2.5$ \\ [\cmsTabSkip] \textit{Energy sums} & \\ \MET  &  $\MET
	 > 100$ (Vector sum of \pt of calorimeter deposits with $\abs{\eta}<5.0$) \\
         
	 \HT   &  $\HT > 360$ (Scalar sum of \pt of all jets with $\pt>30$ and
	 $\abs{\eta}<2.5$) \\
         
	 \ET &  $\ET > 2000$ (Scalar sum of \pt of calorimeter deposits with
 $\abs{\eta}<5.0$) \\ \end{tabular} \begin{tabular}{@{}p{0.968\linewidth}@{}} \hline
         
	 Terms used \\ Tight quality: muons with hits in at
	 least 3 different muon stations. \\ Medium quality: muons with
	 hits in at least 2 different muon stations. \\ The
	 ``non-colliding BX" requirement selects beam-empty events. \\
	 $\Delta R \equiv ((\Delta \phi)^{2} + (\Delta
	 \eta)^{2})^{1/2}$, and phi is the azimuthal angle in radians.\\
	 OS: Opposite Sign (of electric charge).\\
	 $\ET{}$: Scalar sum of \pt of calorimeter deposits.\\
	 $\HT{}$: Scalar sum of \pt of jets. \\ Isolation
	 and loose isolation: The isolation requires an upper limit on the
	 transverse calorimeter energy surrounding the candidate. The limit
	 depends on the pileup, the Level-1 candidate \ET and $\abs{\eta}$.  Details
	 are given in Sections \ref{sec:egamma} and \ref{sec:tau}.
	 \end{tabular} \end{table}

    \begin{table}[] \centering \topcaption{List of
	the most used cross object unprescaled Level-1 trigger algorithms
	(seeds) during Run~2 and their corresponding requirements.}
	\setlength\extrarowheight{1.5pt}
	\begin{tabular}{@{}ll@{}} \hline Algorithm &
		Requirements \\ & (\pt, \ET, $\mmumu$, and $\mjj$ in
		GeV) \\ \hline \textit{Two objects} & \\
         
	 Single $\PGm$ + Single $\egamma$ & $\pt(\PGm) > 20$ \& Tight
		quality$(\PGm)$ \& $\pt(\egamma) > 10$ \& $\abs{\eta(\egamma)} <
		2.5$ \\
         
	 Single $\PGm$ + Single $\egamma$ & $\pt(\PGm) > 7$ \& Tight
		quality$(\PGm)$ \& $\pt(\egamma) > 20$ \& $\abs{\eta(\egamma)} <
		2.5$ \\
         
		Single $\PGm$ + & $\pt(\PGm) > 18$ \&  $\abs{\eta(\PGm)} < 2.1$ \& Tight
		quality$(\PGm)$ \& \\ \, Single $\PGt$ & \, $\pt(\PGt) > 24$ \&
		$\abs{\eta(\PGt)} < 2.1$ \\
         
	 Single $\PGm$ + \HT & $\pt(\PGm) > 6$ \& Tight quality$(\PGm)$ \& $\HT
		> 240$ \\
         
		Single $\egamma$ + & $\pt(\egamma) > 22$ \& $\abs{\eta(\egamma)} < 2.1$ \&
		Loose isolation($\egamma$) \& \\ \, Single $\PGt$ & \, $\pt(\PGt)
		> 26$ \& $\abs{\eta(\PGt)} < 2.1$ \& Isolation($\PGt$) \& $\Delta R >
		0.3$ \\

		Single $\egamma$ + & $\pt(\egamma) > 28$ \& $\abs{\eta(\egamma)} < 2.1$ \&
		Loose isolation($\egamma$) \& \\ \, Single jet & \,
		$\pt(\text{jet}) > 34$ \& $\abs{\eta(\text{jet})} < 2.5$ \& $\Delta
		R > 0.3$ \\
         
	 Single $\egamma$ + \HT & $\pt(\egamma) > 26$ \& $\abs{\eta(\egamma)} < 2.1$
		\& Loose isolation(\egamma) \& $\HT > 100$ \\
         
	 Single $\PGt$ + \MET & $\pt(\PGt) > 40$ \& $\abs{\eta(\PGt)} < 2.1$ \&
		$\MET > 90$\\ 
         
	 Single jet + \MET & $\pt(\text{jet}) > 140$ \& $\abs{\eta(\text{jet})} <
		2.5$ \& $\MET > 80$ \\ [\cmsTabSkip] \textit{Three objects} & \\

	 Single $\PGm$ & $\pt(\PGm) > 12$ \& $\abs{\eta(\PGm)} < 2.3$ \& Tight
		quality($\PGm$) \& \\ \, Double jet + $\Delta R$ & \,
		$\pt(\text{jet}) > 40$ \&  $\Delta \eta(\text{jet,jet}) < 1.6$
		\& $\abs{\eta(\text{jet})} < 2.3$ \& $\Delta R(\PGm,\text{jet}) <
		0.4$  \\ 
         
	 Single $\PGm$ + & $\pt(\PGm) > 3$ \& $\abs{\eta(\PGm)} < 1.5$ \& Tight
		quality ($\PGm$) \& \\ \, Single jet + \MET & \,
		$\pt(\text{jet}) > 100$ \& $\abs{\eta(\text{jet})} < 2.5$ \& $\MET >
		40$\\

	 Double $\PGm$ + $\HT$ &$\pt(\PGm) > 3$ \&  Tight quality$(\PGm)$ \&
		$\HT > 220$ \\
         
	 Double $\PGm$ + & $\pt(\PGm) > 0$ \& Medium quality($\PGm$) \& $\Delta
		R(\PGm,\PGm) < 1.6$ \& \\ \, Single jet + $\Delta R$ & \,
		$\pt(\text{jet}) > 90$ \& $\abs{\eta(\text{jet})} < 2.5$ \& $\Delta
		R(\PGm,\text{jet}) < 0.8$ \\

	 Double $\PGm$ + Single $\egamma$ & $\pt(\PGm) > 5$ \& Tight
		quality$(\PGm)$ \& $\pt(\egamma) > 9$ \& $\abs{\eta(\egamma)} < 2.5$
		\\
         
	 Double $\egamma$ + Single $\PGm$ & $\pt(\egamma) > 12$ \&
	 $\abs{\eta(\egamma)}< 2.5$ \& $\pt(\PGm) > 6$ \& Tight quality($\PGm$) \\
         
	 Double $\egamma$ + $\HT$ & $\pt(\egamma) > 8$ \& $\abs{\eta(\egamma)} <
	 2.5$ \& $\HT > 300$ \\ [\cmsTabSkip] \textit{Four objects} & \\
         
	 Double $\PGm$ + Double $\egamma$ & $\pt(\PGm) > 3$ \& Medium
	 quality($\PGm$) \& OS($\PGm$) \& $\pt(\egamma) > 7.5$ \\
         
	 Double $\PGm$ + Double $\egamma$ & $\pt(\PGm) > 5$ \& Medium
	 quality($\PGm$) \& OS($\PGm$) \& $\pt(\egamma) > 3$ \\ [\cmsTabSkip]
	 \textit{Five objects} & \\

	 Double $\PGm$ + \MET  + & $\pt(\PGm) > 3$ \& Tight quality($\PGm$) \&
	 $\MET > 50$ \& \\ \, Single jet OR & \, ($\pt(\text{jet}) > 60$ \&
	 $\abs{\eta(\text{jet})} < 2.5$) OR \\ \, Double jet  & \, ($\pt(\text{jet})
	 > 40$ \& $\abs{\eta(\text{jet})} < 2.5$) \\
         
	 $\HT$ + Quad jet & $\HT > 320$ \& $\pt(\text{jet}) > 70, 55, 40, 40$ \&
	 $\abs{\eta(\text{jet})} < 2.4$ \\

	 \hline \end{tabular} \label{tab:menu-slim-2} \end{table}

\section{The Level-1 trigger architecture} \label{sec:design}

\begin{figure} \centering {
		\includegraphics[width=0.8\textwidth]{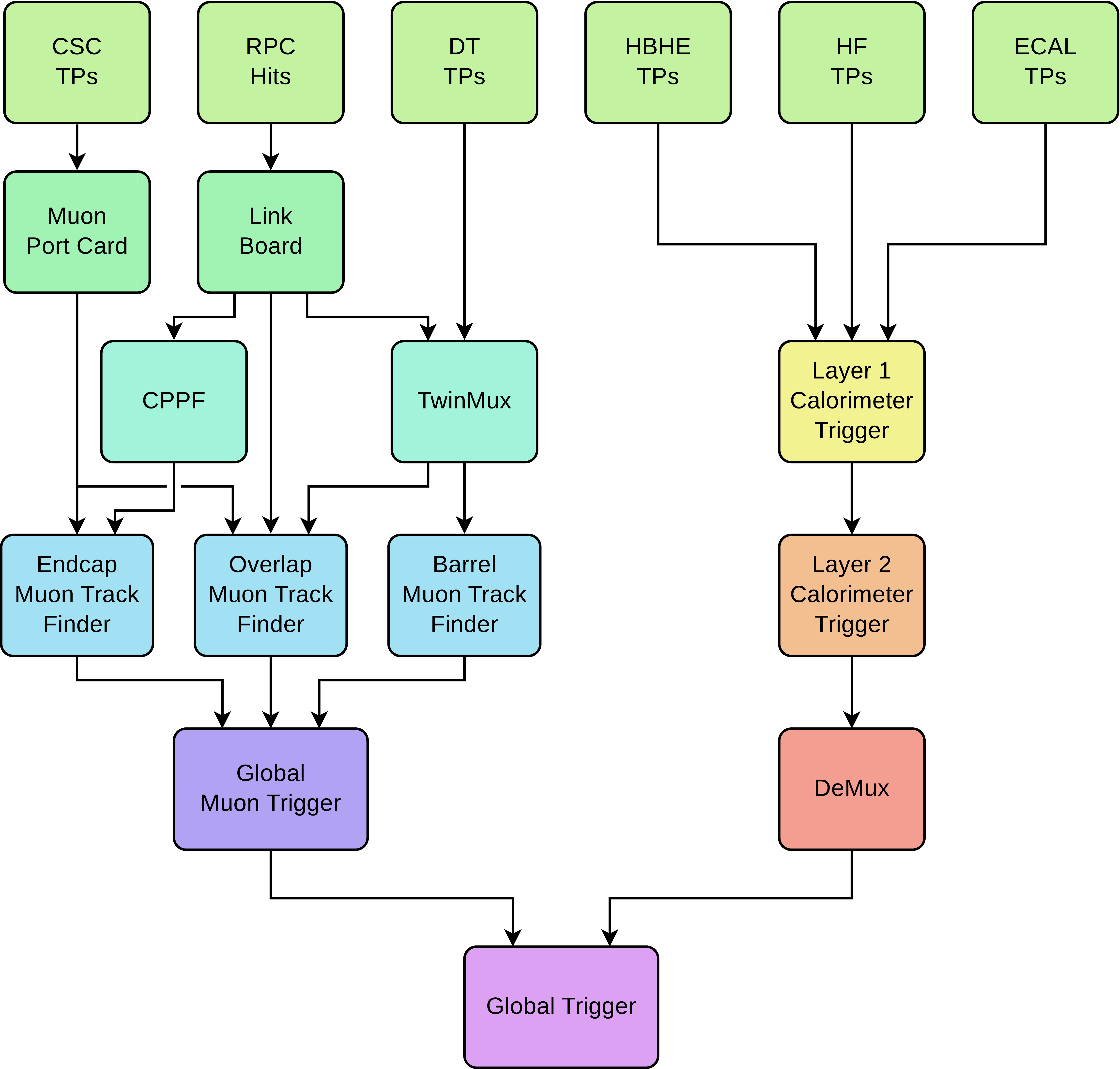} }
	\caption{Diagram of the upgraded CMS Level-1 trigger system during
	Run~2. More details about the muon and calorimeter trigger systems in
	Section~\ref{sec:muons} and \ref{sec:calo} respectively. Labels in the
	diagram correspond to trigger primitives (TPs), cathode strip chambers
	(CSC), drift tubes (DT), resistive plate chambers (RPC), concentration
	preprocessing and fan-out (CPPF), hadron calorimeter barrel (HB) and
	endcap (HE), hadron calorimeter forward (HF), electromagnetic
        calorimeter (ECAL), demultiplexing card (DeMux). }
\label{fig:design-diagram} \end{figure}

During the first LHC long shutdown and extending into 2015, the new CMS Level-1
trigger was installed to run in parallel with the Run~1 (legacy) Level-1
trigger, and eventually replaced it.  The upgraded Level-1 trigger is described
in detail in Ref.~\cite{CMS-L1T-TDR}, with the exception of two new muon
systems: the concentrator and preprocessor fanout (CPPF) and the TwinMux, which
are described in Section~\ref{sec:muons}.  Section~\ref{sec:design} summarizes
the overall design of the upgraded trigger, shown in
Fig.~\ref{fig:design-diagram}. 

In contrast to the Run~1 system that used the Versa Module Eurocard (VME)
standard and many parallel electrical cables for the interconnects, the upgraded
trigger uses  Advanced Mezzanine Cards (AMC) based on MicroTCA
technology~\cite{micro-tca} and multi-Gb/s serial optical links for
data transfer between modules.  The MicroTCA crate provides a high-bandwidth
backplane, system monitoring capabilities, and redundant power modules.  The
number of distinct electronics board types is greatly reduced because many
components are based on common hardware designs.

\begin{figure} \centering \includegraphics[width=0.9\textwidth]{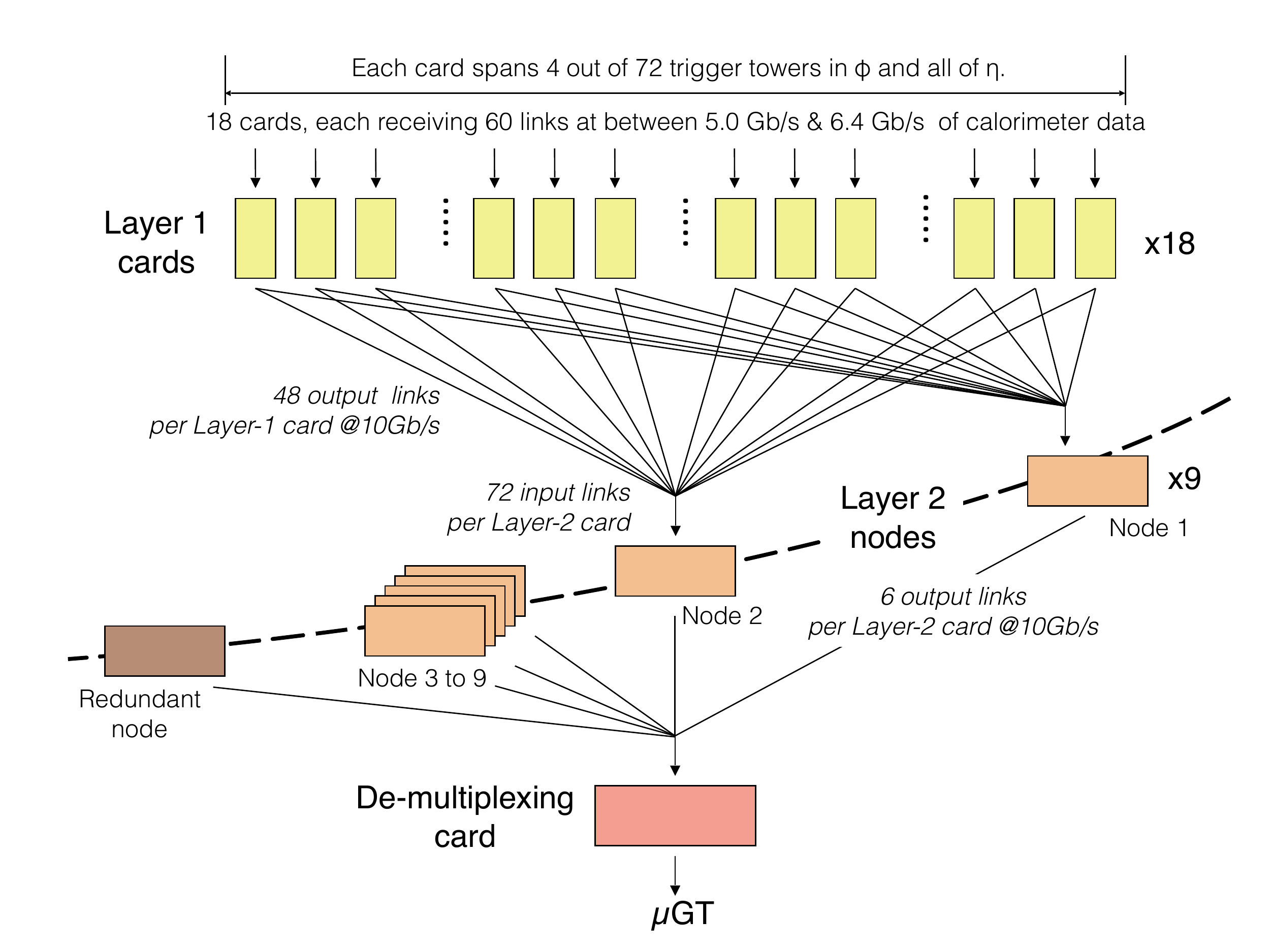}
\caption{The time-multiplexed trigger architecture of the upgraded CMS
calorimeter trigger.} \label{fig:tmt} \end{figure}

The calorimeter trigger consists of two layers: Layer-1 receives, calibrates,
and sorts the local energy deposits (``trigger primitives'') which are sent to
the trigger by the ECAL and HCAL; Layer-2 uses these calibrated trigger
primitives to reconstruct and calibrate the physics objects such as electrons,
tau leptons, jets, and energy sums.  The calorimeter trigger follows a
time-multiplexed trigger design~\cite{tmt-architecture} illustrated in
Fig.~\ref{fig:tmt}. Each main processing node has access to a whole event with a
granularity of $\Delta\eta{\times}\Delta\phi$ of $0.087{\times}0.087$ radians
(where phi is azimuthal angle) in most of the calorimeter acceptance (a
slightly coarser granularity is used at high $\abs{\eta}$). A demultiplexer
(DeMux) board then reorders, reserializes, and formats the events for the
global trigger (\uGT{}, which is pronounced micro-GT to emphasize the
connection to the MicroTCA technology used in this upgrade) processing.
Because the volume of incoming data and the algorithm latency are fixed, the
position of all data within the system is fully deterministic and no complex
scheduling mechanism is required. The benefits of time multiplexing include
removal of regional boundaries for the object reconstruction and full
granularity when computing energy sums. The multiplicity of processing nodes
provides the flexibility to add nodes as required by complex trigger algorithms.
These algorithms are fully pipelined and start processing as soon as the
minimum amount of data is received. 

The muon trigger system includes three muon track finders (MTF) which reconstruct muons
in the barrel (BMTF), overlap (OMTF), and endcap (EMTF) regions of the
detector, and the global muon trigger (\uGMT, pronounced
micro-GMT) for final muon selection.  The \uGT finally collects muons and
calorimeter objects and executes every algorithm in the menu in parallel for
the final trigger decision.

In the upgraded trigger, the BMTF, \uGMT, \uGT, and Layer-2 use the same
type of processor card.  The OMTF and EMTF electronic boards similarly share a
common design, whereas Layer-1, TwinMux, and CPPF each use a different design.
All processor cards, however, use a Xilinx Virtex-7 Field Programmable Gate
Array (FPGA). Thus many firmware and control software components, \eg, data
readout and link monitoring, can be reused by several systems, reducing the
workload for development and maintenance.

An advanced mezzanine card called the AMC13~\cite{amc13} provides fast control
signals from the trigger control and distribution system to the trigger AMCs
over the MicroTCA backplane.  If an event is selected, the trigger AMCs send
their data over the backplane to the AMC13, which also connects to the central
CMS data acquisition system via 10\unit{Gb/s} optical links.  More details on
the hardware can be found in Ref.~\cite{CMS-L1T-TDR}.

\section{The Level-1 muon trigger and its performance} \label{sec:muons} 

\begin{figure} \centering
		\includegraphics[width=0.9\textwidth]{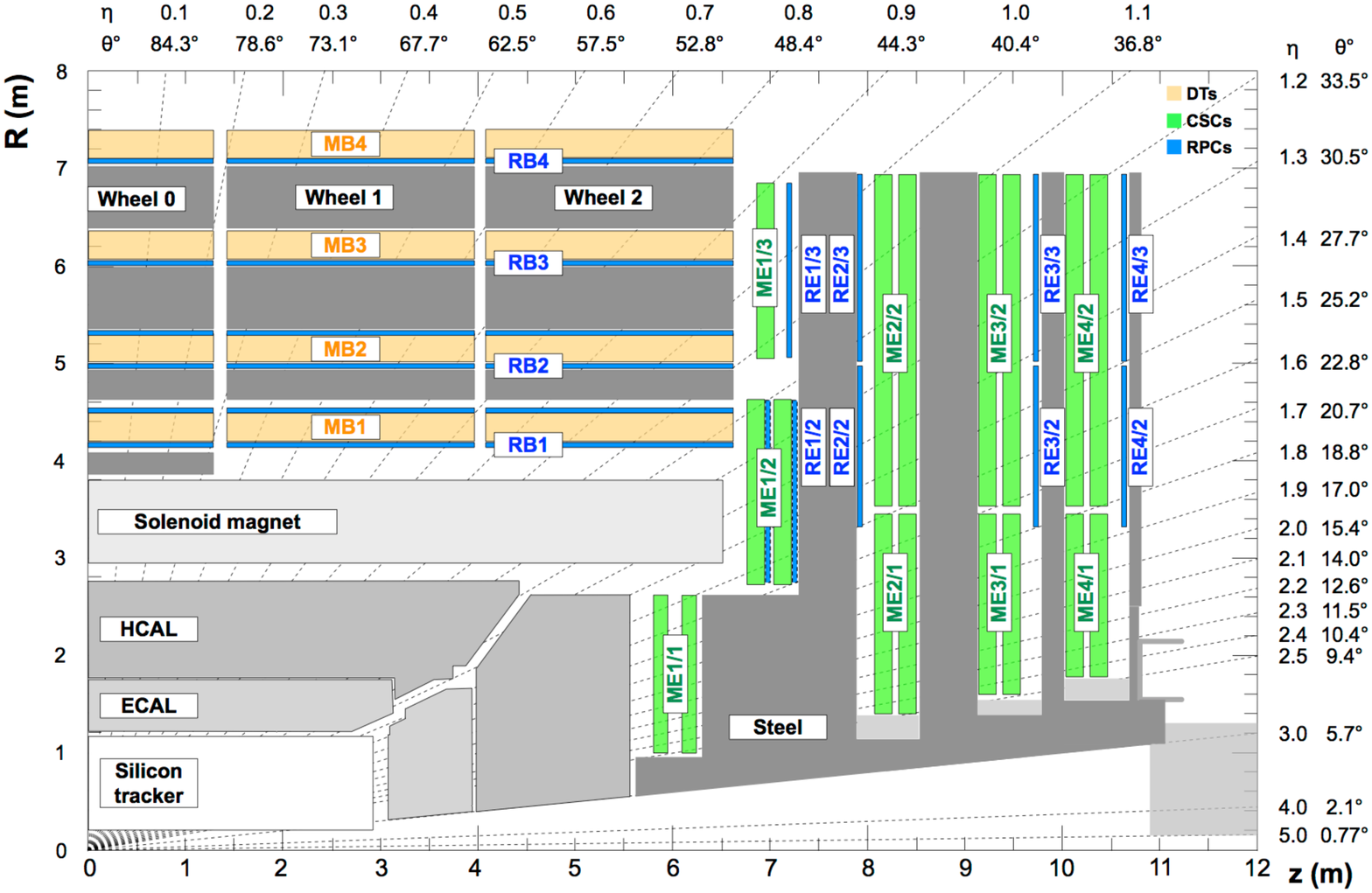}
	\caption{An $R$-$z$ slice of a quadrant of the CMS
	detector~\cite{cms:muon-performance-paper}. The origin of the axes
	represents the interaction point. The proton beams travel along the
	$z$-axis and cross at the interaction point. The three CMS muon
	subdetectors are shown: four stations of DTs in yellow, labelled MB;
	four stations of CSCs in green, labelled ME; and four stations of RPCs
	in blue, labelled RB or RE.} \label{fig:muon-detectors} \end{figure}

The CMS muon detector is composed of three partially overlapping subdetectors
(CSCs, DTs, and RPCs, as described in Section~\ref{sec:cms}), whose signals are
combined together into ``trigger primitives'' (TPs) to reconstruct muons and
measure their \pt{}. Trigger primitives provide coordinates, timing, and
quality information from detector hits. Figure~\ref{fig:muon-detectors} shows
the geometrical arrangement of the three muon subdetectors in a quadrant of the
CMS detector.

In the legacy trigger, data from each of the three subdetectors were used
separately to build independent muon tracks, which were combined by a global
muon trigger.  The upgraded Level-1 trigger combines information from all
available subdetectors  to reconstruct tracks in three distinct pseudorapidity
regions, improving the muon reconstruction efficiency and resolution while
reducing the misidentification rate.

The BMTF takes inputs from DT and RPC chambers in the barrel; all three muon
subsystems contribute to the OMTF tracks in the overlap between barrel and
endcap; and the EMTF uses CSC and RPC information to reconstruct endcap muons.
Detector symmetry allows each track finder to run the same algorithm in parallel
for different regions in $\phi$.  The BMTF is segmented in twelve sectors of
$30^\circ$ each, and both the OMTF and EMTF are segmented into 12 sectors of
$60^\circ$, six on each end of the experiment. A single board builds tracks in
one sector, plus 20--$30^\circ$ of overlap to account for muon bending in
$\phi$. 

The track finders use muon detector TPs to build muon track candidates, assign
a quality to each, and measure the charge and the \pt of each candidate
from the bending in the fringe field of the magnet yoke. Each track finder uses
muon finding and \pt assignment logic optimized for its region, and assigns the
track quality corresponding to the estimated \pt resolution.

Each track finder transmits up to 36 muons to the \uGMT, which resolves
duplicates from different boards, and sends the data for a maximum of eight
muons of highest rank (a linear combination of \pt and a quality value) to the
\uGT{}, where they are used in the final Level-1 trigger decision.

\subsection{Barrel muon trigger primitives } 

The DT and RPC barrel systems consist of four cylindrical stations wrapped
around the solenoid, each split into 12 wedges in $\phi$ and 5 wheels along the
beam direction.  In the upgraded Level-1 trigger, a new layer called the TwinMux
merges DT trigger primitives and RPC hits from the same station (\ie, detector
layer) into ``superprimitives''. Superprimitives combine the better spatial
resolution of the DT and the more precise timing from the RPC. Each
superprimitive is assigned a quality, which depends on the location of its
inputs, $\eta$ and $\phi$ coordinates, and an internal bending angle $\phi_{\mathrm{b}}$.
The TwinMux then sends superprimitives to the BMTF. The TwinMux also transmits
unmerged DT TPs and RPC hits to the OMTF.  In both cases the TwinMux increases
the bandwidth of the data links used to transmit TPs, thus reducing the number
of data links.  Merging DT and RPC hits also improves the TP efficiency and
timing in each station, which results in improved BMTF performance.  The TwinMux
is described in detail in Ref.~\cite{twinmux}.

\subsection{Endcap RPC trigger primitives}

The CPPF consists of eight MicroTCA boards with FPGA processors, designed to
concentrate endcap RPC TPs for transmission onto higher-bandwidth optical links.
The CPPF clusters RPC hits in adjacent strips into a single TP, and computes
their $\theta$ and $\phi$ coordinates before transmitting up to two clusters per
$10^\circ$ chamber to the EMTF.  The CPPF was commissioned in 2017. A detailed
description is given in Ref.~\cite{CPPF}.

\subsection{Barrel muon track finder}

The BMTF reconstructs muons in the barrel region ($\abs{\eta}<\bmtfetalimit$). The
BMTF track finding and \pt assignment algorithms are similar to their
predecessors running on the DTTF~\cite{CMS:trigger-run1,dttf}. Look-up tables
(LUTs) use the bending angle and the quality of the superprimitives of an inner
station to form an acceptance window for the outer station through an
extrapolation unit.  Each extrapolation unit receives superprimitives from one
thirty-degree sector/wheel and its five neighbors, i.e. the two adjacent sectors
in the same wheel and the corresponding three in the neighboring wheel.  The
track assembler unit receives the paired superprimitives for all stations and
combines them. Tracks with more stations, especially inner stations where the
magnetic field is stronger, are assigned higher quality.  

The assignment unit uses LUTs to assign \pt, $\phi$, and $\eta$ of a track. The
\pt value is assigned based on the difference of the $\phi$ coordinates of TPs
in neighboring stations, $\Delta \phi$, for the majority of tracks. However,
$\Delta \phi$ by itself cannot distinguish high- and low-\pt tracks because of
the inversion of their curvature due to the inversion of the magnetic field
direction in the yoke with respect to the inner solenoid region.  For this
reason two LUTs encode the \pt value for either the high- or low-\pt case, and
the internal bending angle of the superprimitive, $\phi_{\mathrm{b}}$, is used
to select the appropriate result. A LUT based purely on the bending angle
$\phi_{\mathrm{b}}$ augments the \pt assignment for tracks reconstructed from
only two superprimitives, where at least one of the TPs is assigned good
quality by the TwinMux. The \pt assigned by this LUT is compared to the one
obtained using the TP $\Delta \phi$ and the smaller value is selected.

\subsection{Overlap muon track finder}
 
The OMTF receives data from three DT and five RPC stations in the barrel, plus
four CSC and three RPC stations in the endcap, giving 18 total "layers" that
are used to build tracks (since each DT station has two layers).  Track
reconstruction occurs independently in each sector in $\phi$.  The OMTF uses
detector hits directly from the RPC system and trigger primitives from the DT and
CSC systems. In the following section the word ``hits" is used to indicate either.
Each track is constructed starting from a single reference hit in one layer, so
the first step is to select up to four reference hits, favoring hits from inner
layers and those with good $\phi$ resolution.  Up to two reference hits may
come from the same layer, enabling efficient reconstruction of nearby muons.

The algorithm uses patterns generated from simulated events to associate hits in
other layers with the reference hit. For each muon charge there are twenty-six
patterns corresponding to different \pt ranges, from 2 to 140\GeV. Each pattern
encapsulates information about the average muon track propagation between layers
and the probability density function of hit spread in $\phi$ in each layer, with
respect to the reference hit. The patterns differ depending on the reference
layers used. When multiple patterns match a given hit, a statistical estimator
based on the $\phi$ distribution of the hits resolves the ambiguity, preferring
patterns with a larger number of matched layers.  The OMTF reconstruction
algorithm can be regarded as a naive Bayes classifier.

Properties of the best matched patterns, together with the reference hit
$\phi$, are passed to the internal muon sorter, which removes possible
duplicates from a single muon producing multiple reference hits. The three best
muons per board are transmitted to the \uGMT, giving a maximum of 36 muons. A
more detailed description of the algorithm is found in Ref.~\cite{omtf}.

\subsection{Endcap muon track finder} 

The EMTF builds muon tracks from CSC and RPC TPs in the endcap. Both detectors
are composed of four stations separated in $z$ and covering $360^\circ$ in
$\phi$.  The CSCs have complete four-station coverage in the pseudorapidity
range $1.2 < \abs{\eta} < 2.4$ in two or three concentric rings of detectors per
station, whereas the endcap RPCs cover approximately $1.2 < \abs{\eta} < 1.7$ in two
rings of detectors per station.  The CSCs deliver up to two local charged tracks
per BX from each $10^\circ$ or $20^\circ$ chamber in each station and ring, with
${\approx}1/16^\circ$ precision in $\phi$ and ${\approx}1/4^\circ$ precision in
$\theta$.  The RPCs send hits from chambers with similar geometry, which are
clustered by the CPPF into TPs with ${\approx}1/4^\circ$ precision in $\phi$ and
${\approx}1^\circ$ precision in $\theta$.

The EMTF builds tracks using at most one TP (CSC or RPC) per station. The
algorithm first looks for CSC TPs correlated in $\phi$ in multiple stations
consistent with the presence of a muon track, matching at least one of the five
predefined patterns.  The pattern recognition runs in parallel in four zones in
$\theta$. After the patterns are found, the CSC or RPC TP in each station
closest to the pattern is taken for further processing. Resulting tracks are
ranked according to their straightness and the number of stations with hits.
Stations 1 and 2 are prioritized because the magnetic field is much stronger
between stations 1 and 2 than beyond station 2.  A muon track with TPs in these
two stations therefore has a more precise \pt assignment. The three hit patterns
with highest quality from each sector are kept for the \pt assignment, and the
others are discarded.

The bending angles in $\phi$ and $\theta$ of the muon track are used to
calculate the track \pt.  However, this relationship is complicated by several
factors. At low \pt, muons can experience significant multiple scattering and
energy loss and at high \pt, they can initiate electromagnetic showers. In
addition, the CMS magnetic field strength and direction varies with $\eta$
outside the solenoid, so muons of similar momenta can have different behavior in
the more central region ($\abs{\eta} < 1.55$) than in the more forward region
($\abs{\eta} > 2.1$). The complicated dependencies make this an ideal case for
machine learning. A boosted decision tree (BDT) regression technique is used to
provide an estimate of the track \pt, taking these dependencies into account.
The BDT input variables are compressed into 30~bits, and training parameters are
optimized using MC simulation of single-muon events. The BDT output values are
pre-evaluated and stored in a LUT loaded in a ${\approx}1\unit{GB}$ memory
module of the EMTF for fast determination.  Additional details about the design,
training, and implementation of the BDT can be found in Ref.~\cite{emtf}.

\subsection{Global muon trigger} \label{sec:uGMT}

The \uGMT receives up to 108 muon candidates (3 per sector) sent from the
three muon track finders.  The \uGMT sorts the muons and identifies and
removes duplicates, sending up to eight muons to the \uGT. Such duplicate muons
would significantly increase the trigger rate for multimuon trigger algorithms
and must be removed while keeping a high efficiency for events with two genuine
muons. In parallel to the duplicate removal and sorting stage, the \uGMT also
corrects the spatial coordinates of each muon by extrapolating the track from
the muon stations back to the interaction region.

The \uGMT uses the \pt and the quality of input muons to define an initial
ranking, separately sorting muons from the positive and negative $\eta$ sides of
the OMTF and EMTF, as well as from the BMTF.  It keeps the four highest ranked
muons coming from each endcap of the OMTF and EMTF, along with the highest
ranked eight BMTF muons. The second sorting stage compares the ranks of muons
coming from the first stage and selects the eight with the highest rank.
 
Because of the overlap between adjacent wedges or sectors of the track finders
(TFs), a muon traversing the detector in these overlap regions can be found by
the TF processors of both sides on the overlap.  In addition to this overlap in
$\phi$, the different regional TFs also have an overlap in $\eta$ where a muon
can be found by both the BMTF and OMTF, or by the OMTF and EMTF. Two different
methods are used for the identification of duplicates. The first method makes
use of the ``track address'' of the muon, which encodes the TPs used to build
the muon track, to find duplicates between BMTF wedges. The second method uses
the muon track coordinates, which are applied to find duplicates between
adjacent sectors in the OMTF and the EMTF, and between different regional TFs.
For the second method, simulated events are used to determine the optimal size
and shape of the regions in which tracks should be marked as duplicates.

Because the TF systems measure the muon coordinates within the muon systems, the
\uGMT extrapolates all input muon track parameters back to the collision
point. The extrapolation corrections are derived from MC simulation as a
function of \pt, $\phi$, $\eta$, and charge of the muon, and are stored in a
LUT. The corrections have a coarse granularity since they are limited to 4 bits:
they have steps of 0.05 radians in $\Delta\phi$ and 0.01 in $\Delta \eta$ and
are applied to muons with $\pt < 64\GeV$. These corrected coordinates are then
propagated to the \uGT to improve the performance of trigger algorithms
relying on the invariant mass or difference in spatial coordinates between
multiple muons.

The \uGMT also transmits the track quality to the \uGT as a selection option
for specific trigger paths.  Quality is also used for cancellation in case
duplicates are found.  Muons passing the ``tight'' quality criteria have good
\pt resolution, and are used in single-muon seeds.  All BMTF tracks pass the
tight criteria, thanks to the strong magnetic bending effect in the barrel
region, whereas OMTF and EMTF tracks must have TPs in at least three layers,
and in EMTF one of those TPs must be in the innermost layer.  The ``medium''
and ``loose'' criteria are used in OMTF and EMTF to increase the trigger
efficiency for events with multiple muon tracks by including tracks with fewer
TPs, or without a TP in the first layer.

\subsection{Performance} 
\begin{figure} \centering
	\includegraphics[width=0.48\textwidth]{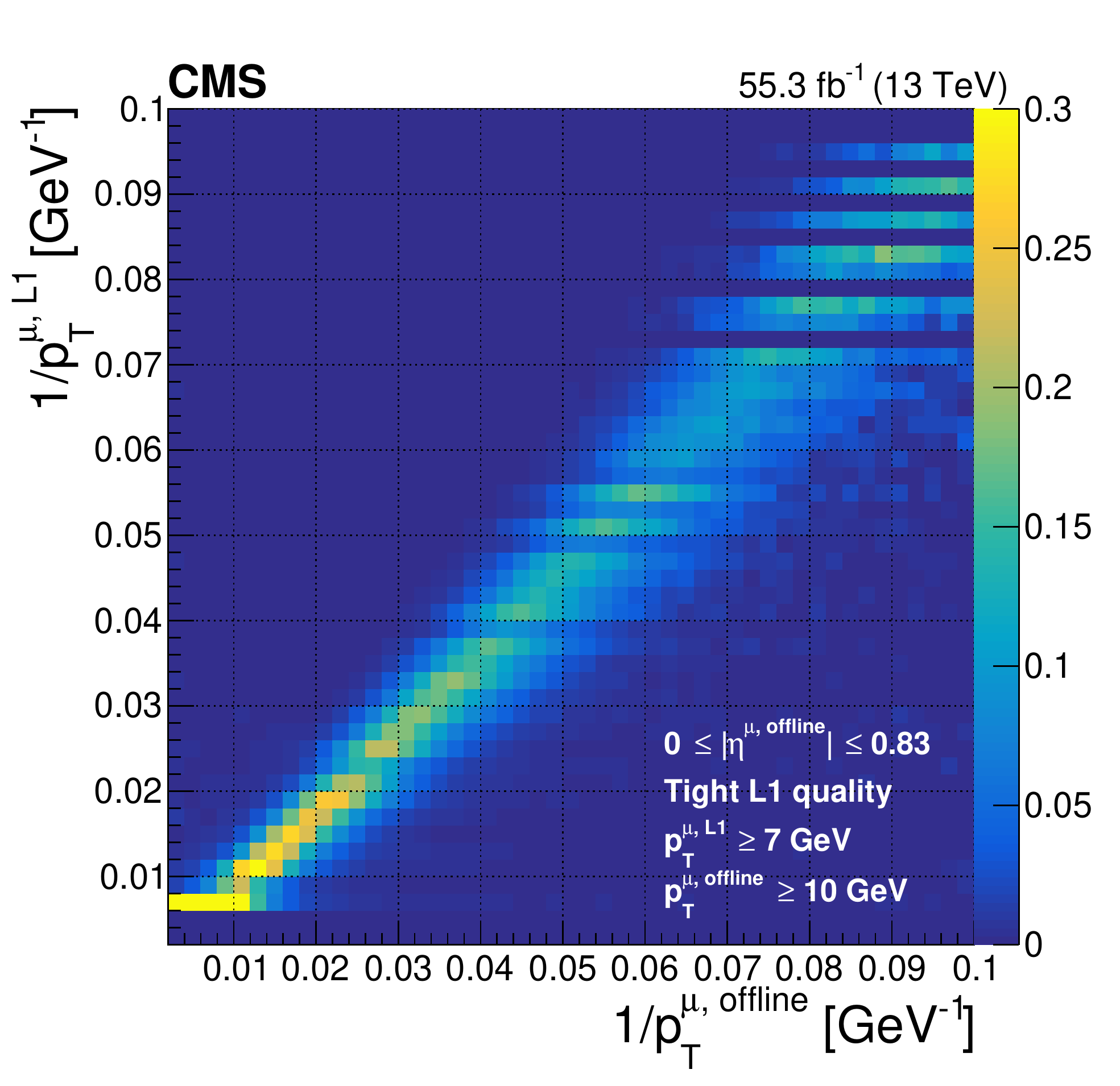}
	\includegraphics[width=0.48\textwidth]{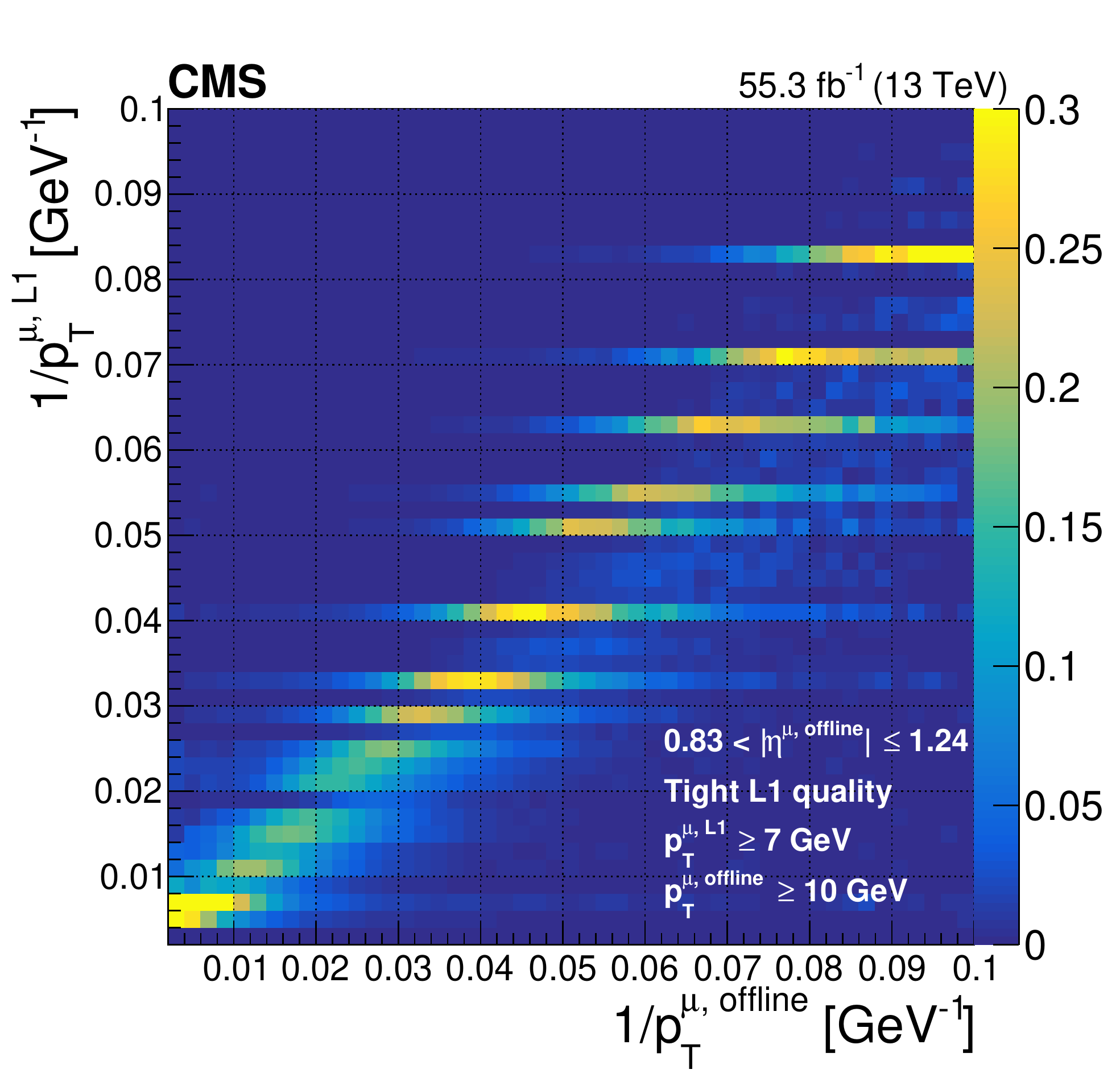}
	\includegraphics[width=0.48\textwidth]{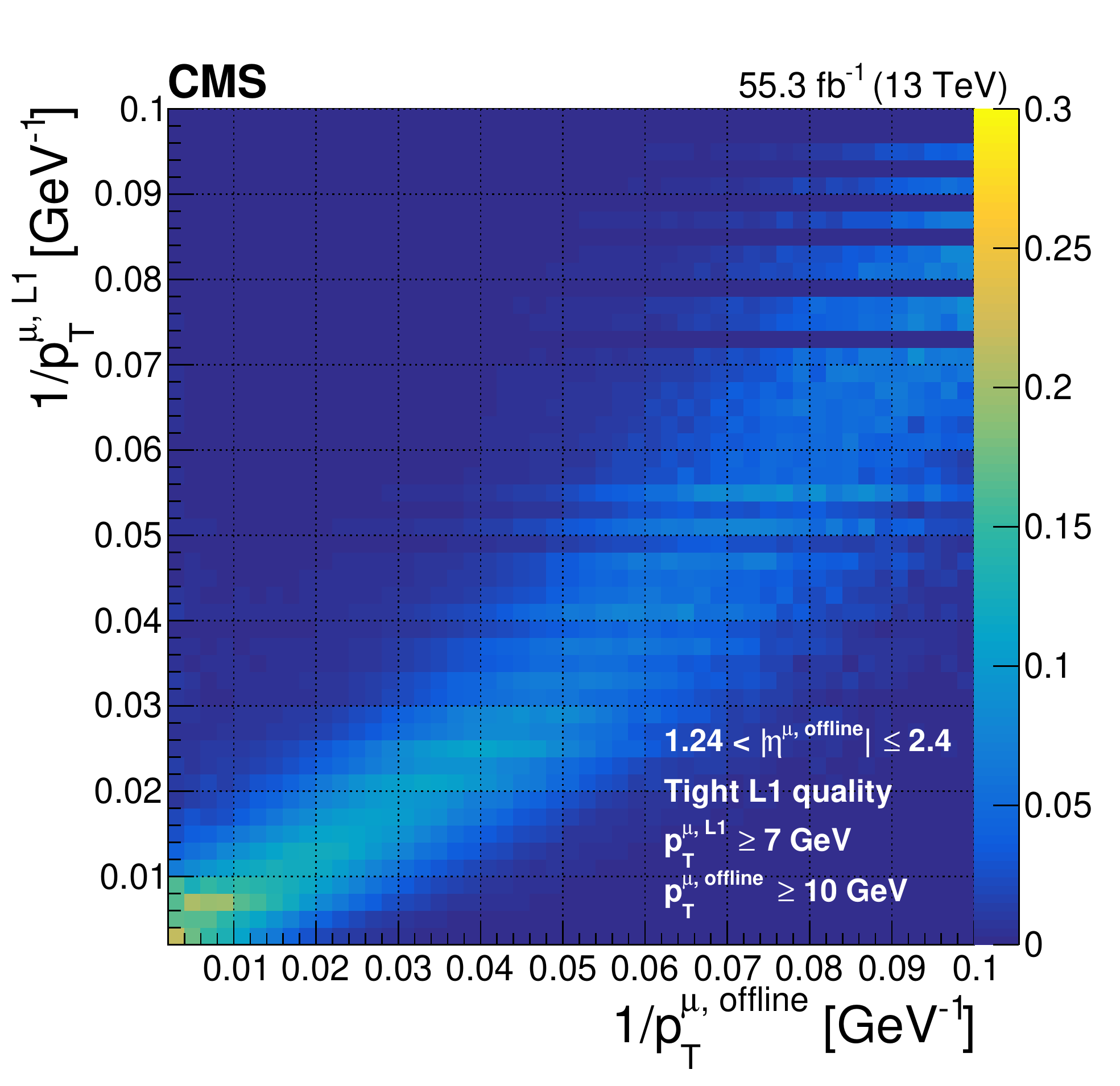}
	\caption{Correlation between $1/\pt$ of the muon (proportional to
	curvature) as assigned at Level-1 vs.\ offline for three $\abs{\eta}$
	regions: barrel (top \cmsLeft), overlap (top \cmsRight), and endcap
	(bottom). The measurements come from a data set enriched with events
	with a Z boson.  Distinct bands in the overlap region come from more
	discrete \pt assignment with the OMTF patterns. }
\label{fig:muon-pt-correlation} \end{figure}

The data recorded since the start of Run 2 are used to
study the performance of the upgraded muon trigger. The performance studies
presented in this chapter use data collected during 2018. Data collected during
2016 and 2017 give similar results. Figure~\ref{fig:muon-pt-correlation} shows
the correlation between the inverse of the muon \pt assigned at Level-1,
proportional to the track curvature, and the inverse of the offline
reconstructed muon \pt for the three $\eta$ regions of interest. The
correlation is linear but slightly off-diagonal, because Level-1 muon \pt
values are scaled up to provide 90\% efficiency for any given trigger \pt
threshold. The resolution in the barrel shows better resolution because the
orientation of the magnetic field with respect to the muon track causes less
bending in the forward regions.  The figure uses a data set triggered by a
single isolated muon, with two oppositely charged muons consistent with a \PZ
boson decay.

The efficiency measurements use a tag-and-probe \cite{CMS:tag-and-probe}
technique with offline reconstructed muons from preselected Drell--Yan events.
The tag muon is reconstructed with the CMS particle-flow algorithm
\cite{CMS:particle-flow}, and it is required to have $\pt > 26$\GeV and be
isolated such that nearby calorimeter energy deposits must sum to less than 15\%
of the muon \pt.  The tag muon must match within a cone of $\Delta R =
\sqrt{\smash[b]{(\Delta\eta)^2+(\Delta\phi)^2}} < 0.1$ to a muon reconstructed by the
single isolated muon HLT algorithm with $\pt > 24\GeV$.  The HLT muon must be
seeded by the single-muon Level-1 trigger with a \pt threshold of 22\GeV. 
 
\begin{figure} \centering 
	\includegraphics[width=0.48\textwidth]{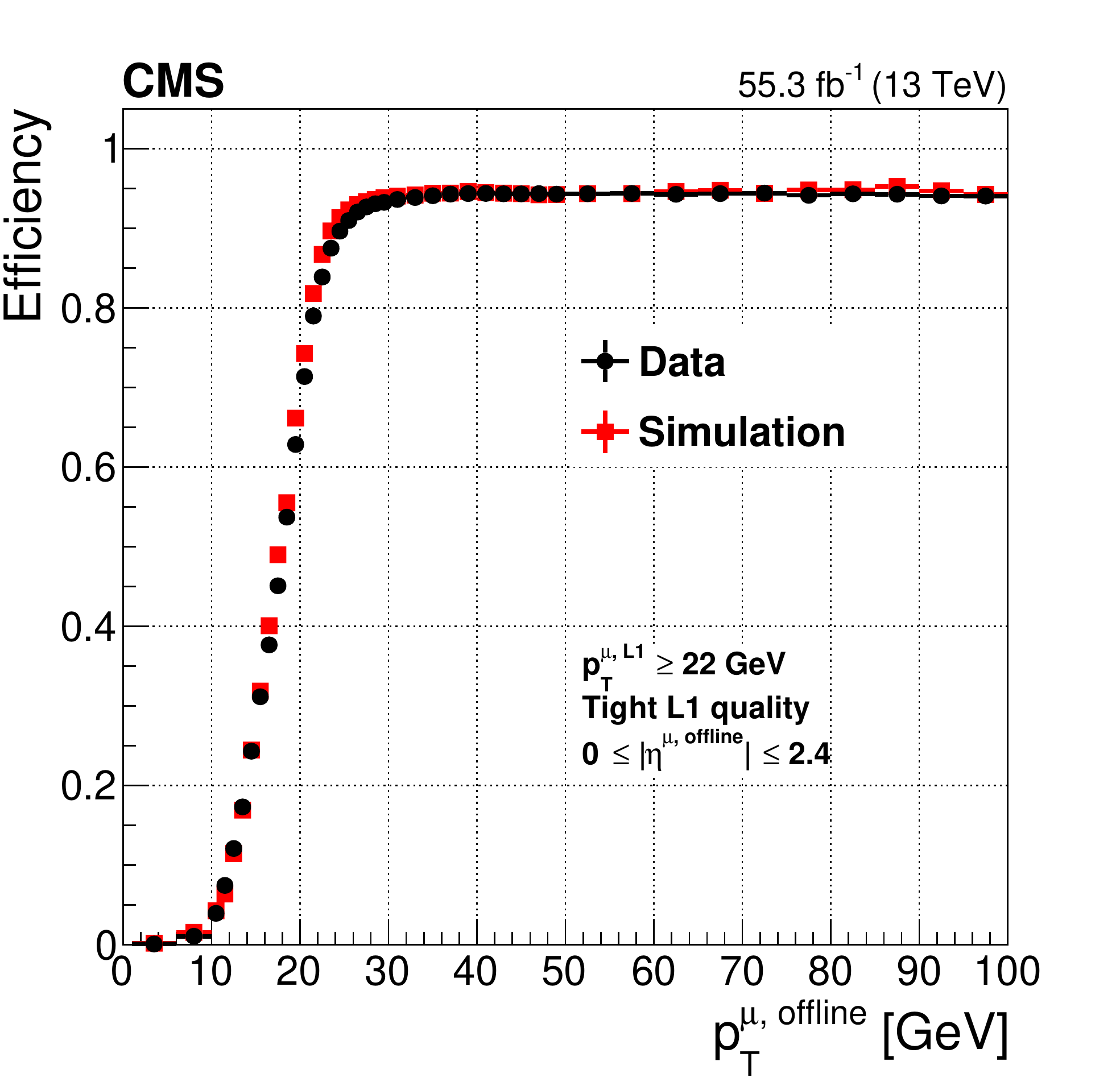} 
	\includegraphics[width=0.48\textwidth]{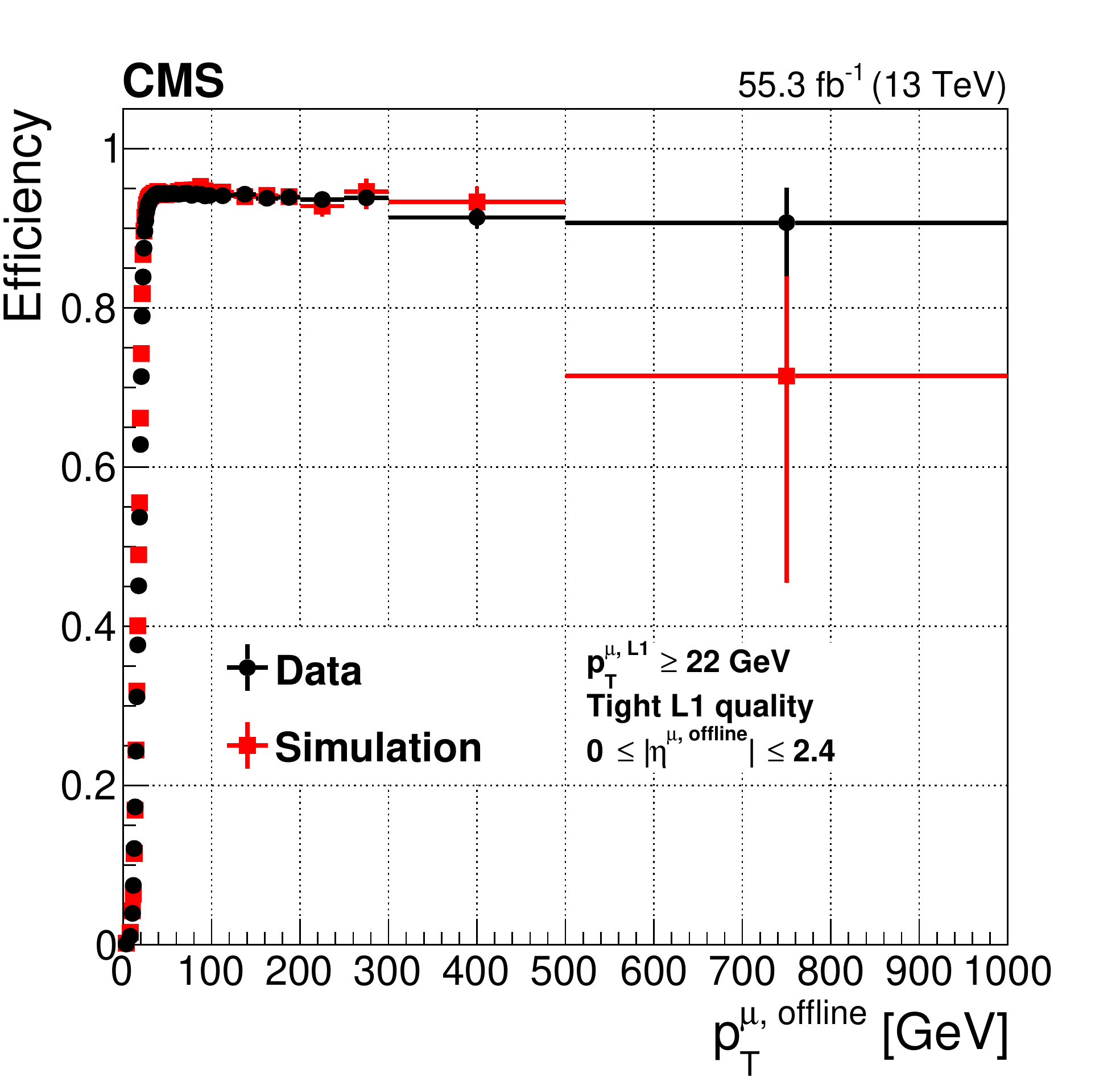} 
	\caption{Level-1 trigger efficiency, for data and simulation as a
	function of \ptoffline, for all reconstructed muons in the CMS
	acceptance $(\abs{\etaoffline} < 2.4)$ for the most commonly used
	single-muon trigger  during Run 2 $(\ptLone > 22\GeV)$, measured with
	the tag-and-probe method described in the text with the full 2018 data
	set. The \cmsLeft plot focuses on the steep increase part of the curve
	close to the trigger threshold. The \cmsRight plot shows the full
	momentum range up to 1\TeV. The simulation reproduces the data within a
	few percent accuracy. The Level-1 trigger efficiency plateau is stable
	as a function of the muon transverse momentum, retaining a high triggering 
	efficiency for muon $\ptoffline \leq 1\TeV$.}
\label{fig:gmt-muon-eff} \end{figure}

\begin{figure} \centering 
	\includegraphics[width=0.48\textwidth]{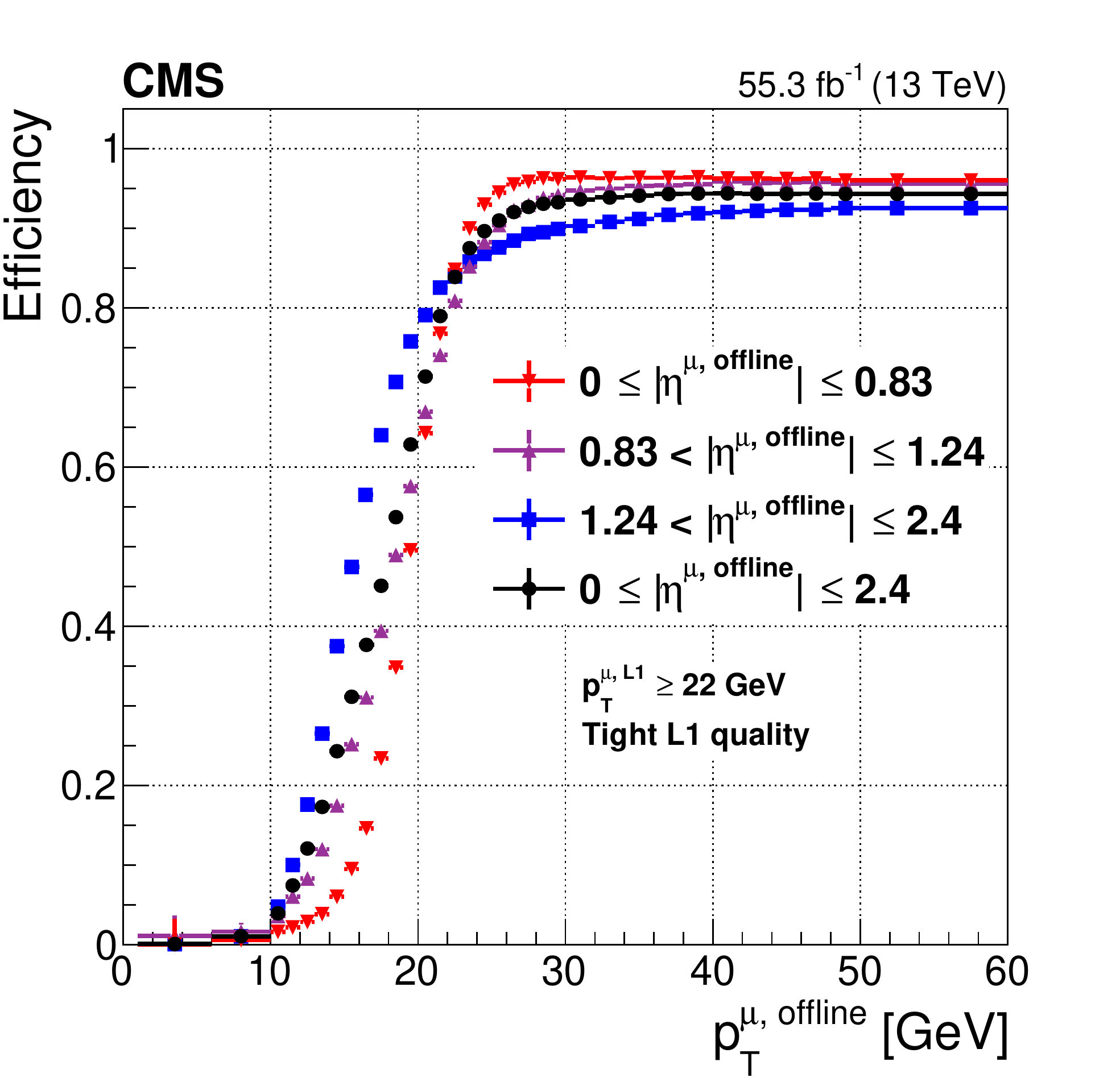} 
	\includegraphics[width=0.48\textwidth]{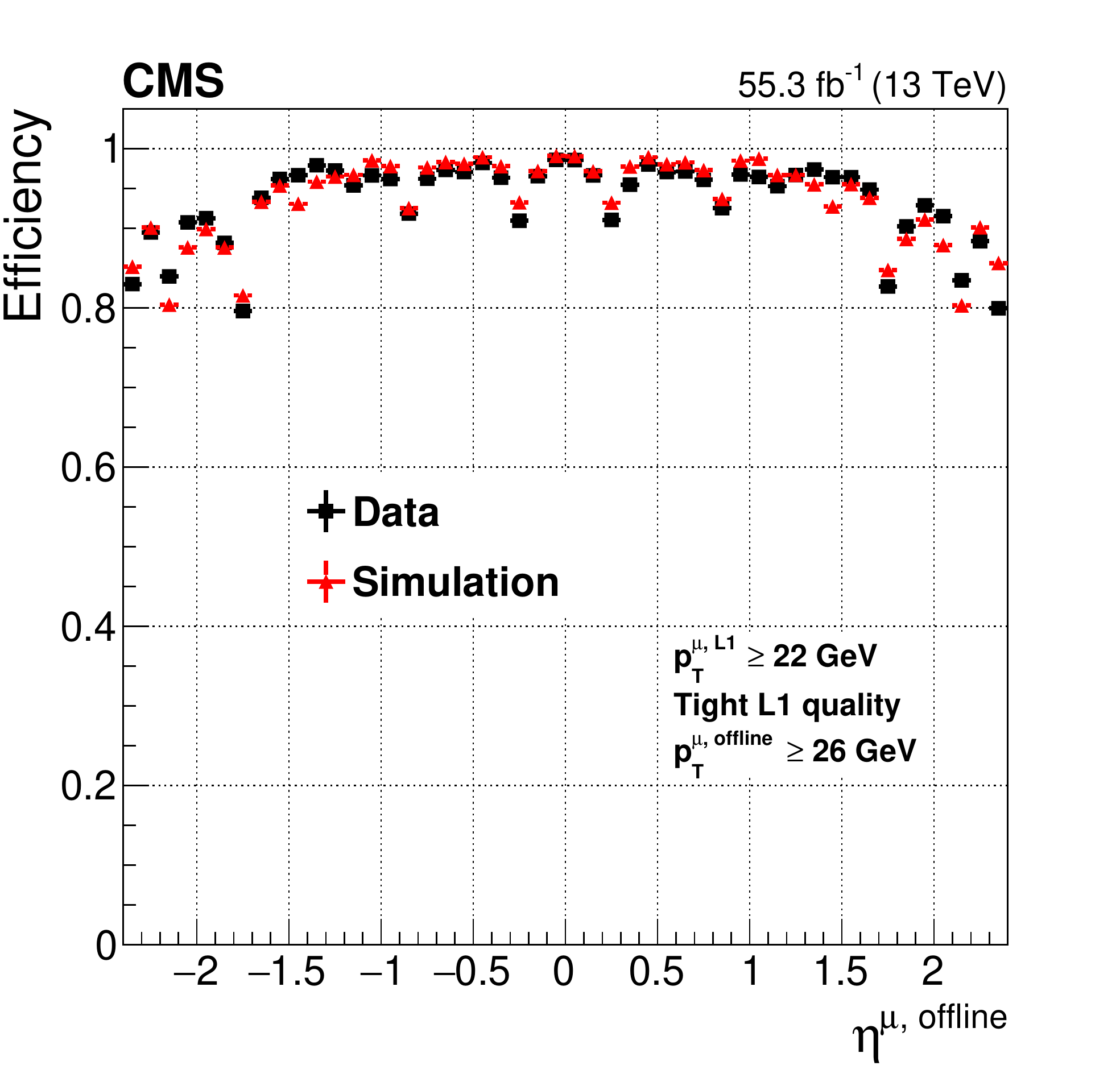} 
	\caption{ The \cmsLeft plot shows the Level-1 muon trigger efficiency
	for data as a function of the offline reconstructed muon \ptoffline for
	each $\eta$ region : barrel region in red, overlap region in purple,
	endcap region in blue, and the total in black. Turn-on curves for more
	central muons rise faster primarily because of improved momentum
	resolution from increased bending in the magnetic field of the yoke. The
	\cmsRight plot shows the Level-1 muon efficiency for data and simulation
	as a function of the offline reconstructed muon $\eta$. The modulation
	of the efficiency in $\eta$ is because of the acceptance of the muon
	systems. The efficiency is measured with the tag-and-probe method
	described in the text with the full 2018 data set.}
\label{fig:tf-muon-eff} \end{figure}

\begin{figure} \centering
	\includegraphics[width=0.48\textwidth]{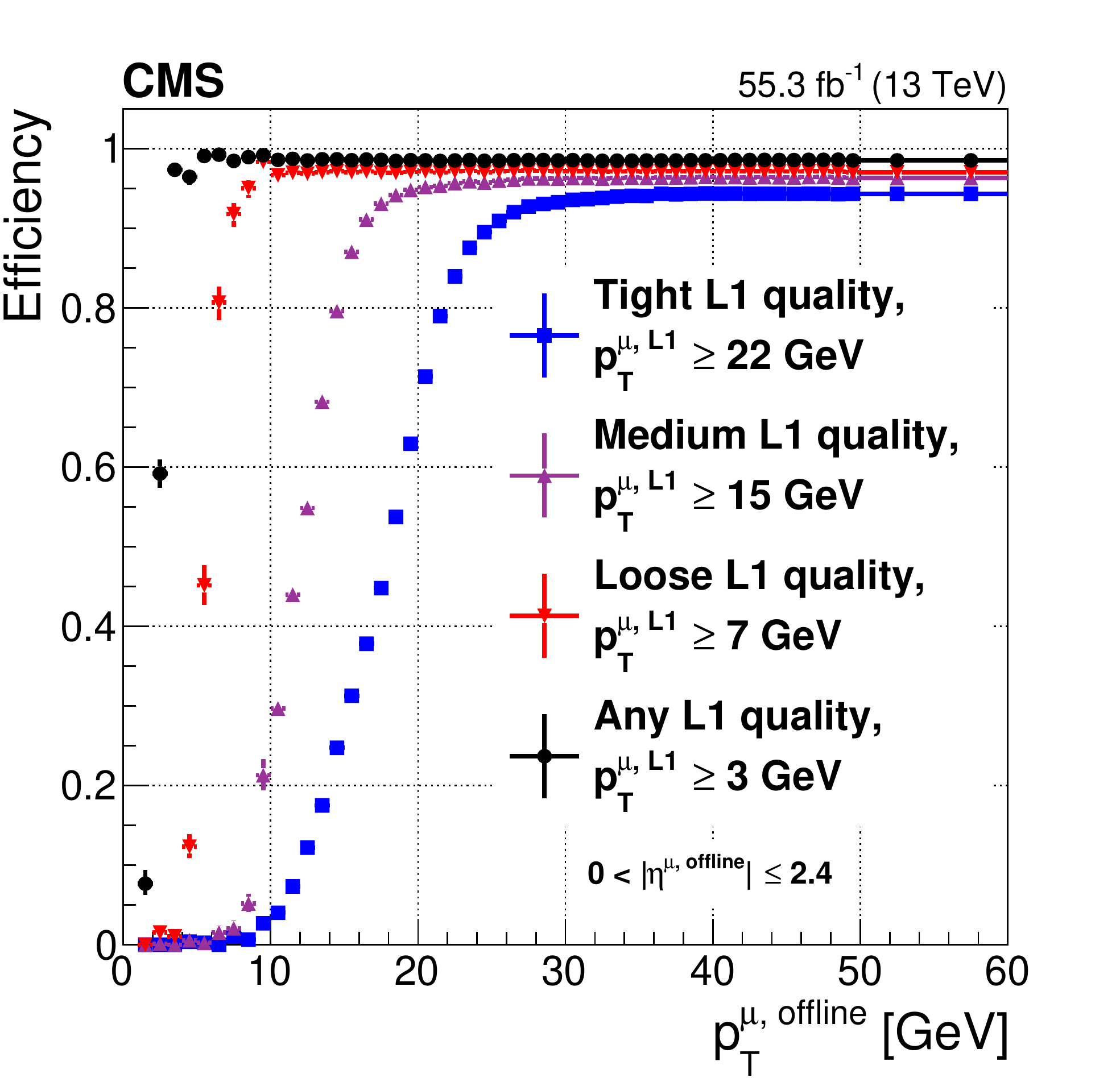}
	\includegraphics[width=0.48\textwidth]{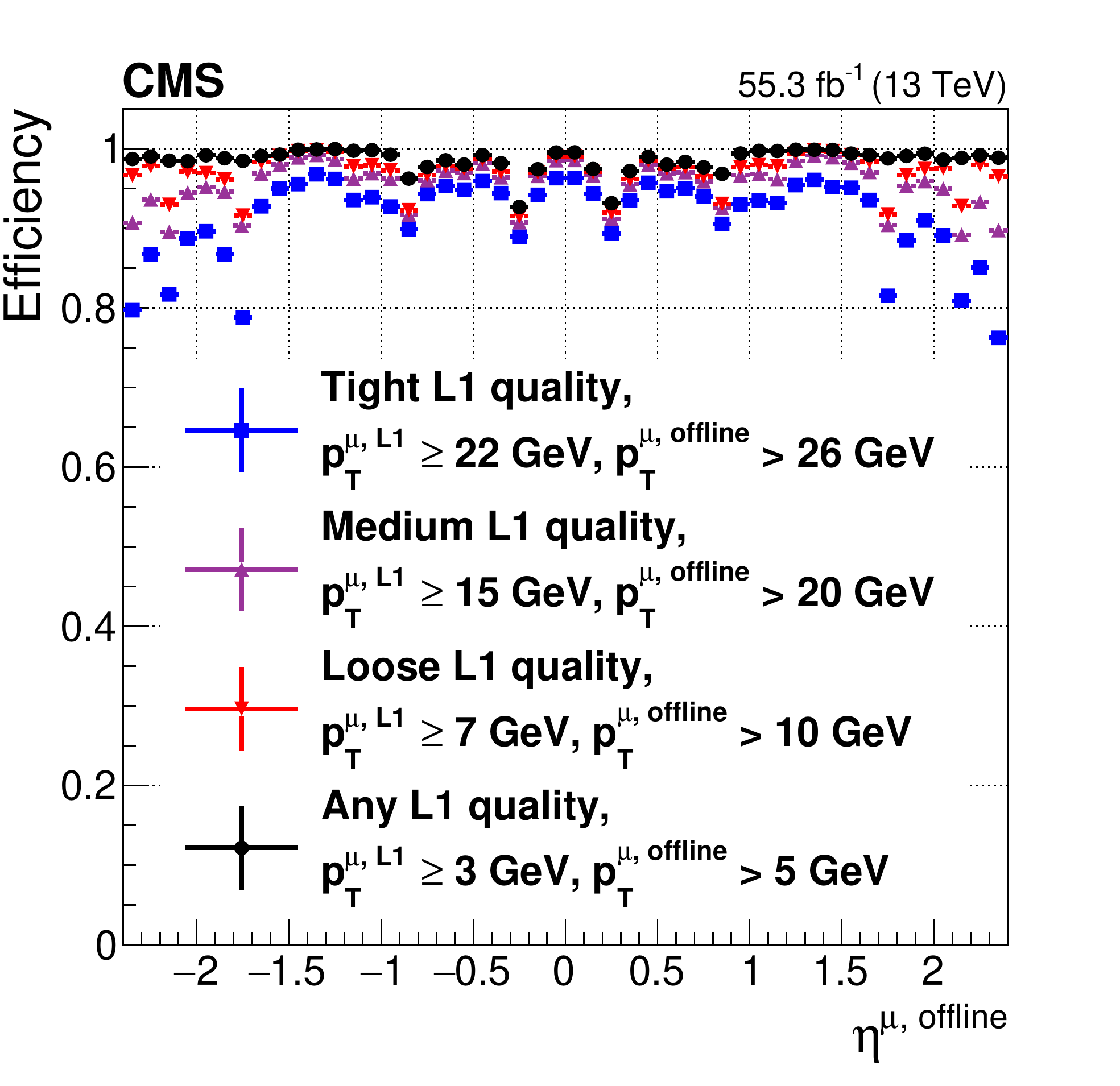}
	\caption{Level-1 muon trigger efficiency for all possible Level-1 muon
	qualities as a function of \ptoffline (\cmsLeft) and $\etaoffline$
	(\cmsRight), for all reconstructed muons in the CMS acceptance
	$(\abs{\etaoffline} < 2.4)$, measured with the tag-and-probe method
	described in the text with the full 2018 data set. The \ptLone 
	thresholds and muon qualities shown are the most commonly used during
	Run 2. The efficiency in the \cmsRight plot is for muons with \ptoffline
	in the plateau region, well above the \ptLone threshold.}
\label{fig:ugmt_efficiency_allQ} \end{figure}

\begin{figure} \centering
	\includegraphics[width=0.48\textwidth]{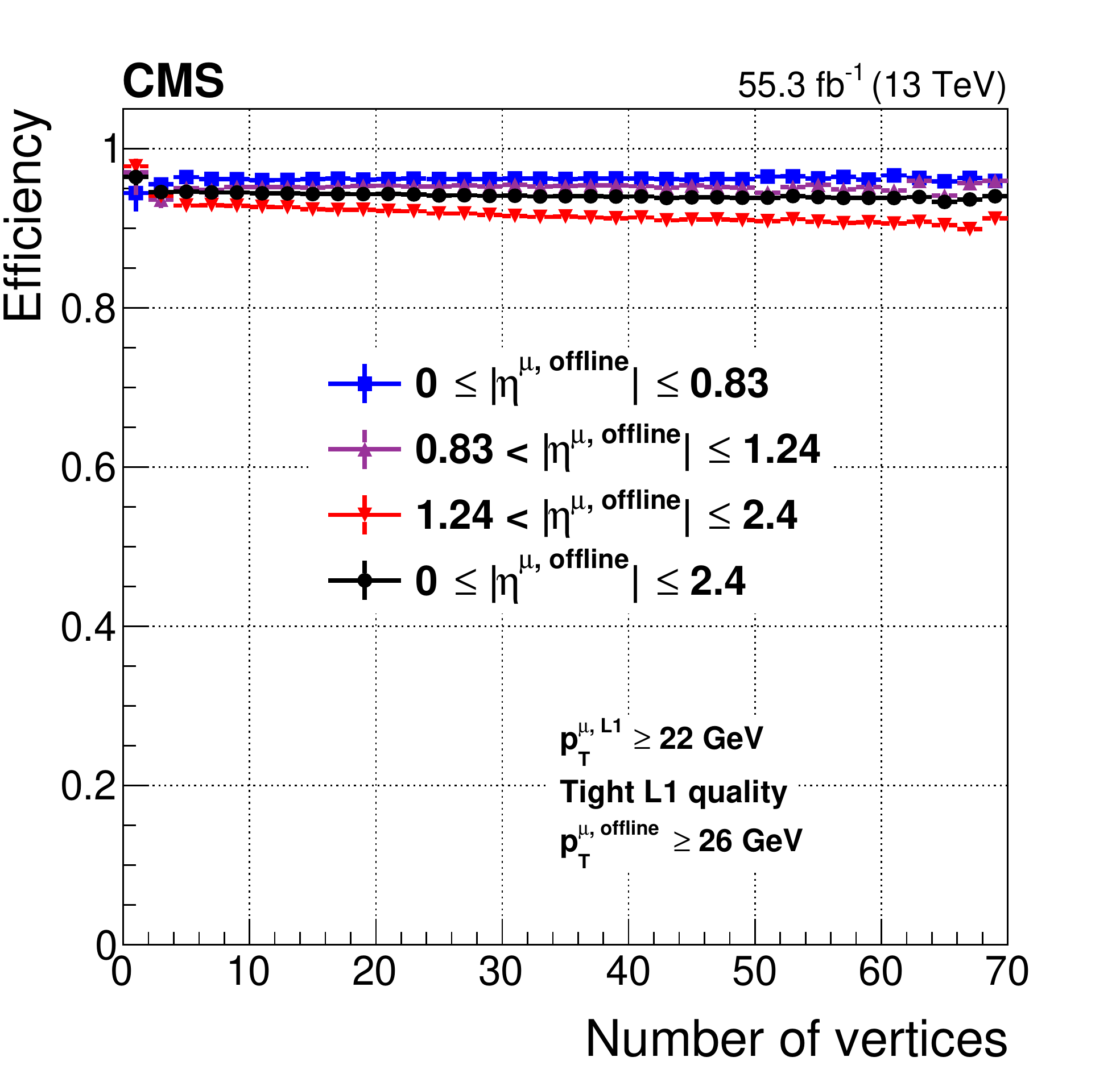}
	\includegraphics[width=0.48\textwidth]{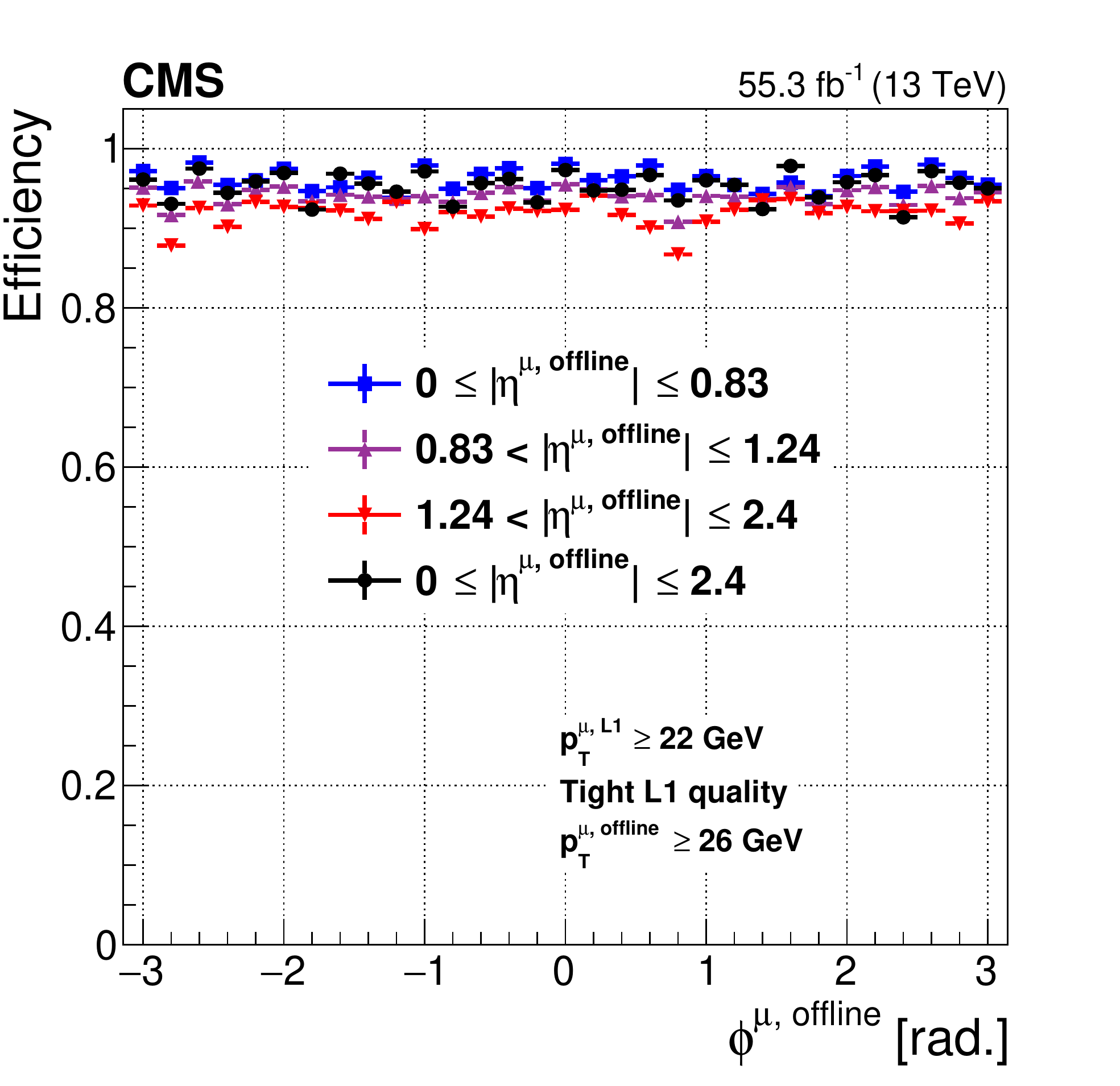}
	\caption{Level-1 trigger efficiency of the muon track finders as a
	function of the number of offline reconstructed vertices  (\cmsLeft) and
	muon $\phi$ (\cmsRight), measured with the tag-and-probe method
	described in the text with the full 2018 data set. These measurements
	are shown for the most commonly used single-muon trigger threshold in
	2018 $(\ptLone > 22\GeV)$. The efficiency has no dependence on the
	number of vertices for central muons, and a very mild dependence for
	endcap muons. The efficiency modulation in $\phi$ follows the
	geometrical acceptance of the muon detector: the efficiency is higher in
	the regions where the detector layers overlap. The efficiency drops at
	$\phi=-2.8$ and $0.8$ are caused by detector inefficiencies.}
\label{fig:muon-eff-vs-pu} \end{figure}

The numerator of the efficiency measurement includes events where a Level-1
muon from the triggering bunch crossing matches a probe muon, reconstructed
using the particle-flow information, within $\Delta R < 0.2$. The denominator
includes all events with a tag muon. The tag and the probe muons must be
separated by $\Delta R > 0.4$. This guarantees that the tag and the probe are
two different muons.  Figure~\ref{fig:gmt-muon-eff} shows trigger efficiencies
measured for a single-muon trigger with a \pt threshold of 22\GeV as a function
of the offline reconstructed muon \pt.  At the threshold value the efficiency
reaches about 86\% of the plateau, which is measured to be ${\approx}93\%$. A
more detailed description of the trigger performance at high muon \pt, where
radiative showering complicates the reconstruction, is given in
Ref.~\cite{cms-high-pt-muon-17-001}.  Figure~\ref{fig:tf-muon-eff} shows the
efficiency as a function of the reconstructed \pt of the probe muon, \ptoffline, for
the three track finder regions (\cmsLeft), and as a function of $\eta$
(\cmsRight).  The three track finders reach an efficiency plateau over 90\% for
the same $\pt^{\text{reco}}$ value, with the barrel track finder exhibiting the
sharpest turn-on curve.  Figure~\ref{fig:ugmt_efficiency_allQ} includes
efficiency measurements for different quality thresholds versus muon \pt and
$\eta$. The detector geometry is responsible for the reduction of trigger
efficiency in certain $\eta$ regions.  Figure~\ref{fig:muon-eff-vs-pu} shows
the efficiency in different $\abs{\eta}$ regions as a function of the number of
pileup vertices and muon $\phi$. In events with high pileup, extra tracks can
confuse the endcap muon reconstruction, causing the trigger efficiency to drop
by a few \% in the far forward region.

\begin{figure} \centering
		\includegraphics[width=0.48\textwidth]{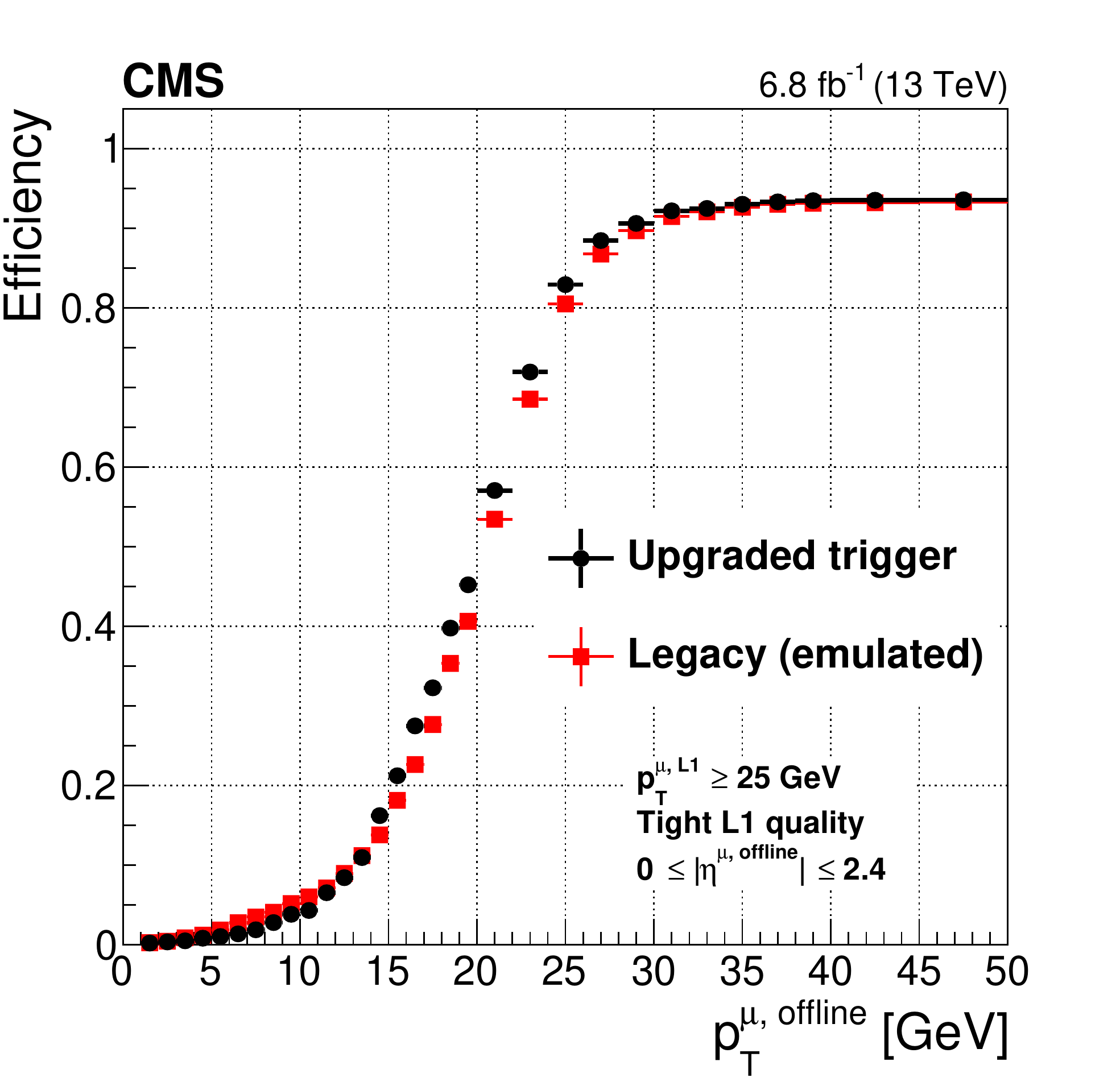}
		\includegraphics[width=0.48\textwidth]{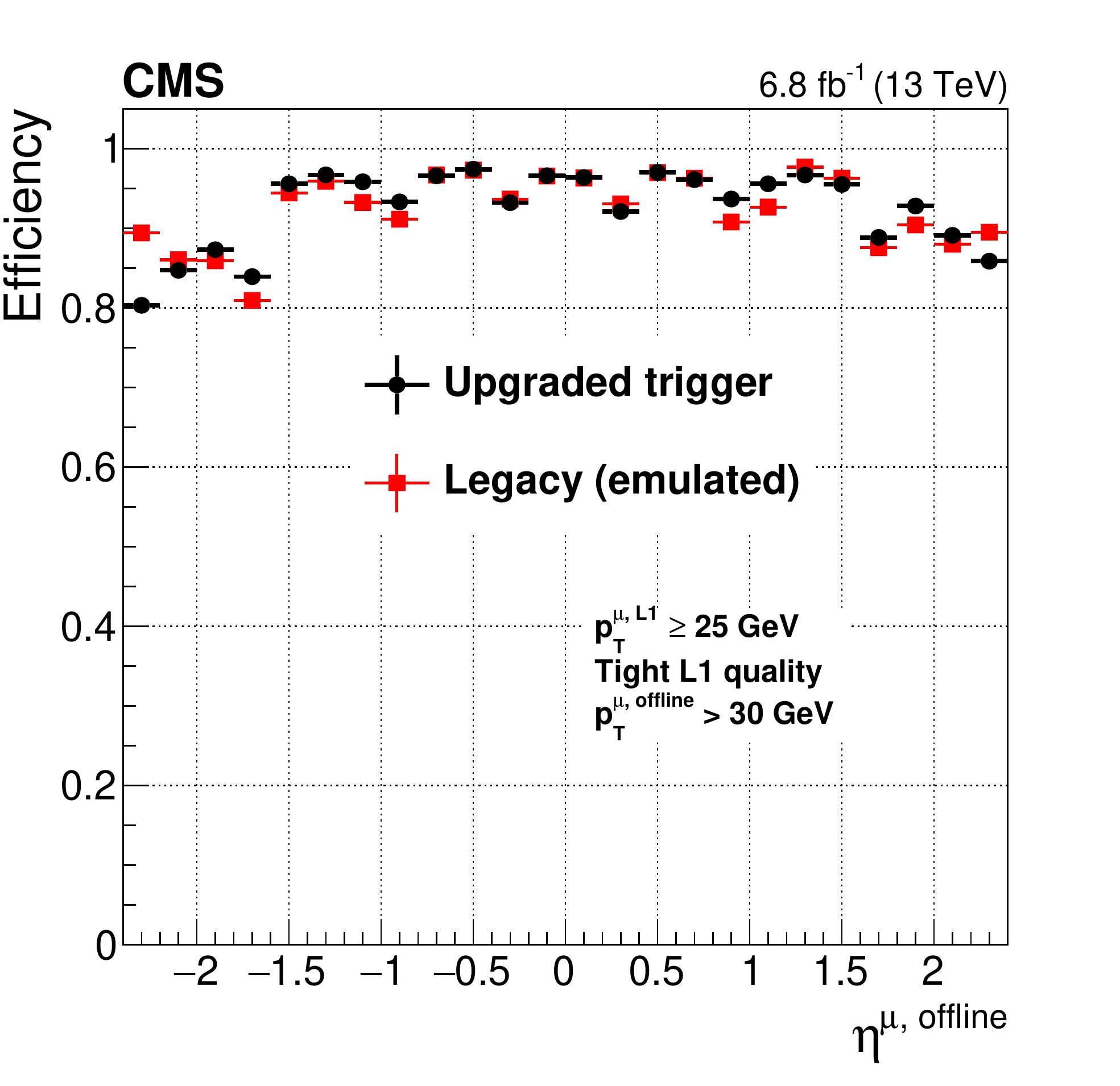}
			\caption{Efficiency of the re-emulated legacy Run~1
			algorithms compared with the upgraded Run~2 algorithms,
			measured using a tag-and-probe technique described in
			the text, plotted as a function of the offline
			reconstructed muon \pt (\cmsLeft) and $\eta$
			(\cmsRight). The \cmsLeft figure shows a sharper turn-on
			efficiency for the upgraded system for muons with \pt
			between 5 and 25\GeV.}
\label{fig:muon-legacy-comparison-eff} \end{figure}

\begin{figure} \centering
	\includegraphics[width=0.48\textwidth]{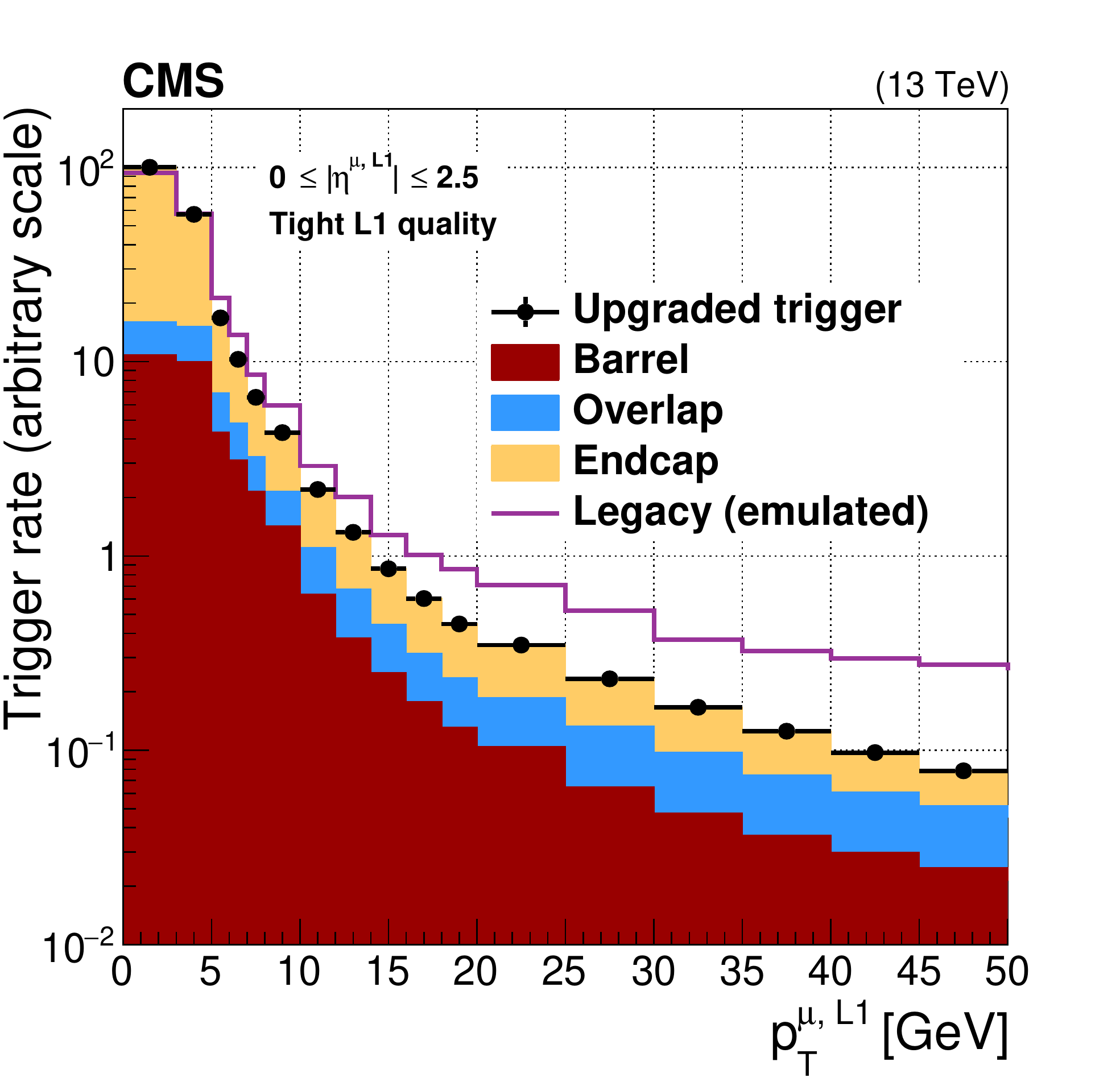}
	\includegraphics[width=0.48\textwidth]{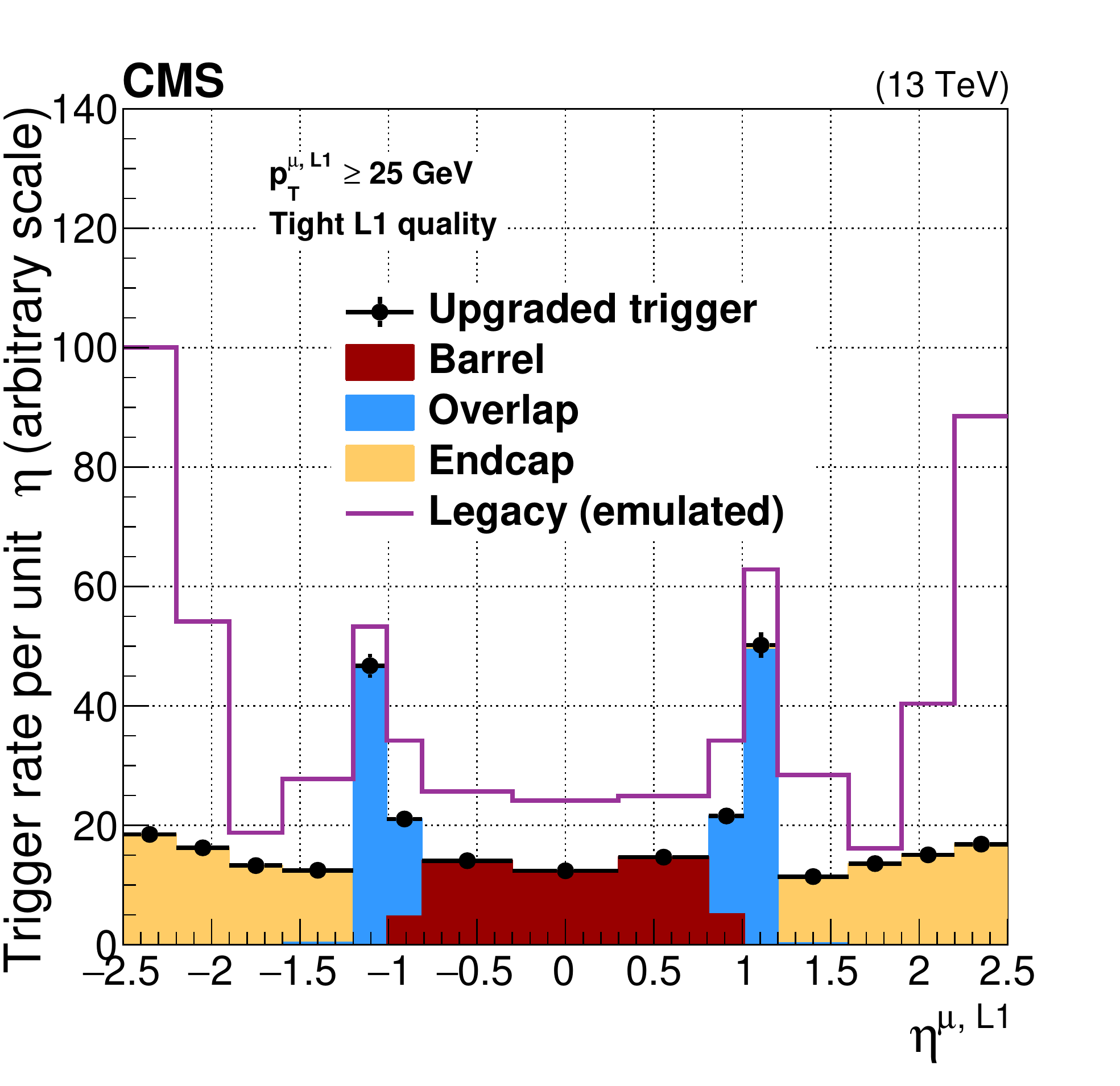} \caption{Rates
	of the re-emulated legacy Run~1 algorithms compared to the upgraded
	Run~2 algorithms, as a function of the Level-1 muon trigger \pt 
threshold (\cmsLeft) and $\eta$ (\cmsRight).  The most common Level-1
single-muon trigger threshold used in 2017 was $p_{\mathrm{T}}^{\mu\text{, L1}}
{\geq} 25\GeV$.} \label{fig:muon-legacy-comparison-rates} \end{figure}

In comparison to the legacy trigger system, the efficiency from the upgraded
muon trigger is similar or higher, depending on the $\eta$ region, as seen in
Fig.~\ref{fig:muon-legacy-comparison-eff}.
Figure~\ref{fig:muon-legacy-comparison-rates} overlays the re-emulated Run~1
(legacy) single-muon algorithm rates and Run~2 (upgrade) rates as a function of
Level-1 muon \pt (\cmsLeft) and $\eta$ (\cmsRight).  The muon trigger rate was
studied with an unbiased Run~2 data sample taken with a prescaled trigger that
only required colliding bunches for triggering. For the single-muon trigger with
a 22\GeV threshold, the rate is approximately a factor of 2 lower than for the
legacy trigger system, estimated from studies with simulated events. The rate
reduction improves at higher trigger thresholds, giving flexibility for tuning
in higher instantaneous luminosity conditions. The use of more sophisticated
\pt assignment algorithms, also exploiting multivariate analysis tools allowed
by the more powerful trigger firmware and hardware, result in a significant
rate reduction compared to the legacy system.

\section{The Level-1 calorimeter trigger and its performance} \label{sec:calo}

The calorimeter trigger was partially upgraded before data taking in the spring
of 2015, and was completed in March 2016.

It is organized in two layers: Layer-1 collects and calibrates the trigger
primitives coming from the calorimeters. Layer-2 receives the output from
Layer-1 and reconstructs and calibrates further physics objects like electrons,
photons, tau leptons, jets, and energy sums.  The following sections describe
the algorithms developed to reconstruct and identify electrons and photons, tau
leptons, and hadron jets, and to assign accurate energies and positions to each.

\subsection{Input calorimeter trigger primitive processing}

Calorimeter trigger towers (TTs) group $5{\times}5$ crystals in the ECAL barrel
(EB) along with the HCAL barrel (HB) tower directly behind them, with a
$\Delta\eta{\times}\Delta\phi$ size of $0.087{\times}0.087$. In the endcaps (EE
crystals, HE, and HF), the grouping logic is more complicated because of the
layout of the crystals, which results in TTs with $\Delta\eta{\times}\Delta\phi$
sizes of up to $0.17{\times}0.17$.  Look-up tables are implemented in Layer-1 to
calibrate electromagnetic energy deposits in the ECAL, as well as hadronic
energy deposits in both ECAL and HCAL towers. This calibration is performed in
addition to calibrations already applied by the ECAL and HCAL electronics, and
accounts for the changing calorimeter response over time, in particular, from
radiation damage. An unforeseen timing effect of the  changing crystal response
is discussed in Appendix~\ref{sec:prefiring}. The Layer-1 calibrations
compensate for various effects including, but not limited to, the average
particle energy loss in the tracker material in front of the calorimeters.  The
calibration factors for ECAL (HCAL) are binned in $\eta$ and \ET, and are
derived from single-photon (single-pions) simulations.

\begin{figure} \centering
	\includegraphics[width=0.495\textwidth]{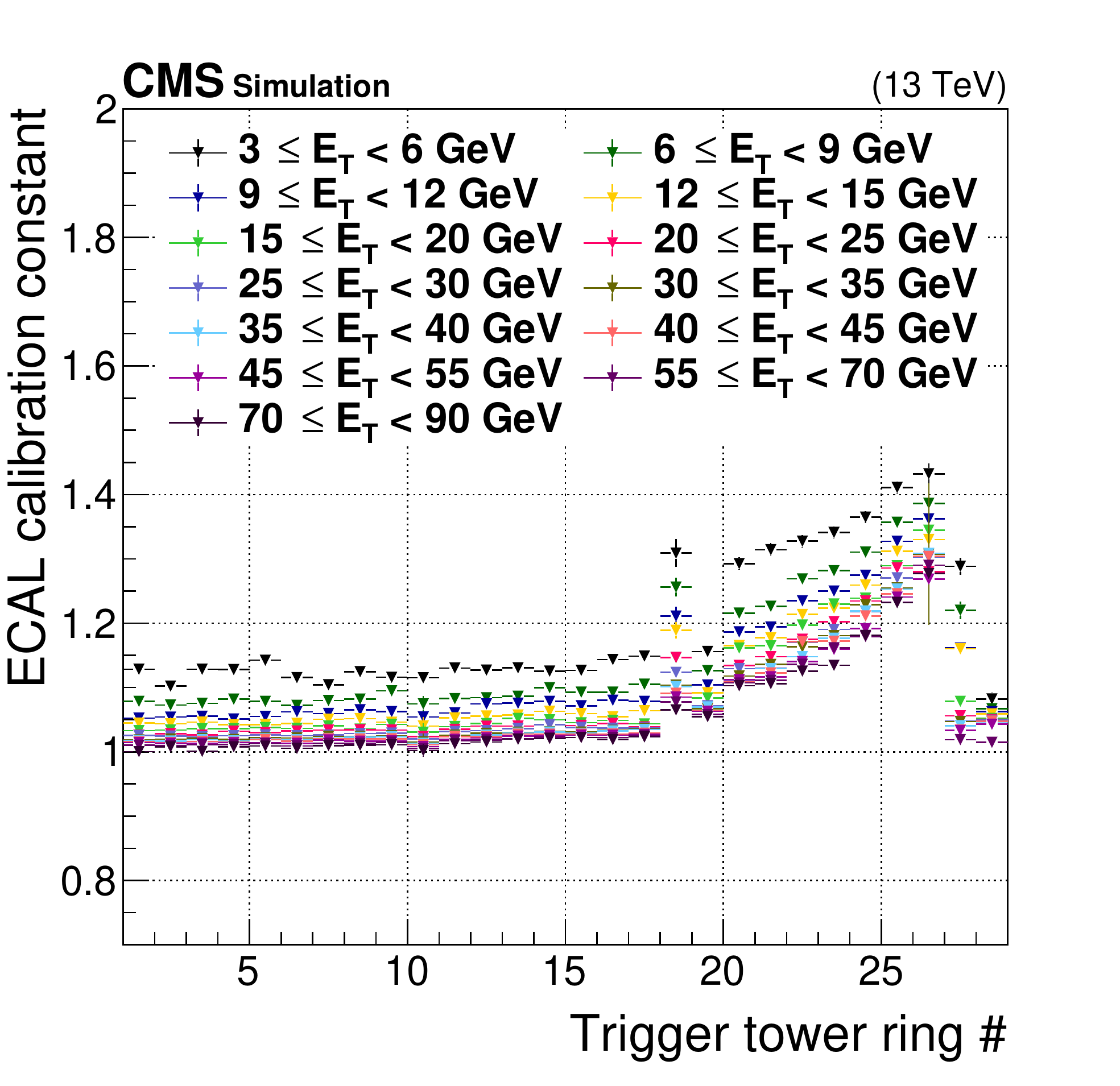}
	\includegraphics[width=0.495\textwidth]{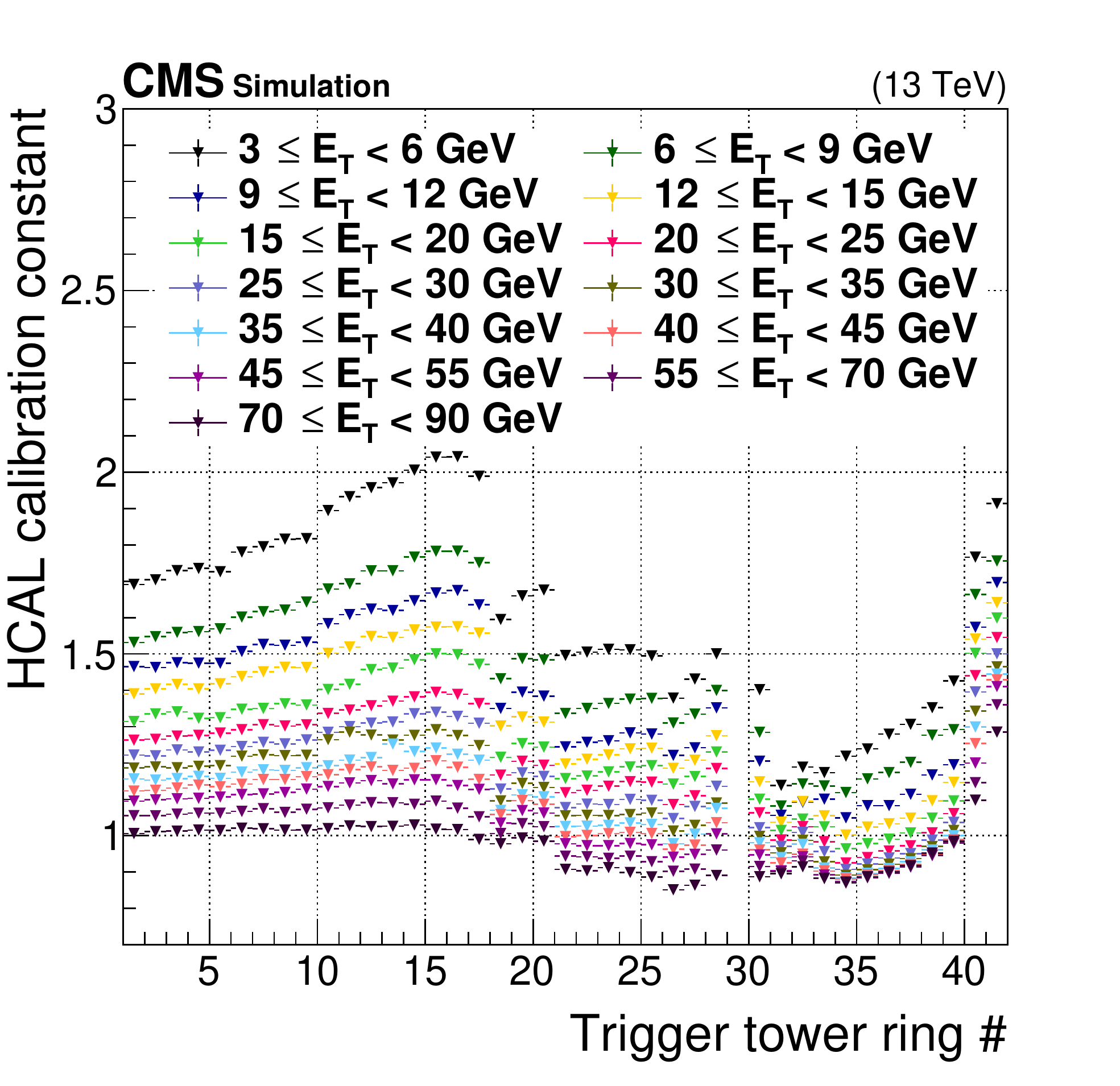}
	\caption{Layer-1 energy scale factors for ECAL (\cmsLeft) and HCAL
	(\cmsRight), shown for each constant-$\abs{\eta}$ ring of trigger
	towers. As specified in the legend, the color of each point corresponds
	to a range of uncalibrated trigger primitive transverse energy values
	received by the Layer-1 calorimeter trigger. Because of the HCAL geometry,
	the signals from trigger tower ring 29 are divided between rings 28 and
	30, and no scale factors are applied.  }
	\label{fig:layer1-scalefactors}
\end{figure}

Figure~\ref{fig:layer1-scalefactors} shows the scale factors derived for both
ECAL and HCAL trigger tower inputs, as a function of $\eta$, for various bins in
\ET. The increase of the calibration factors with $\eta$ reflects the profile
of the detector material in front of the calorimeters. The choice of the
binning of the scale factors respects the hardware limitation and takes into
account the dependency of the resolution in \ET.

The ECAL and HCAL TT information sent to the Layer-2 contains the combined ECAL
plus HCAL energy sum, the ECAL/HCAL energy ratio, and additional flags,
such as the fine-grain veto bit described in Section~\ref{sec:egamma}, and a
minimum-bias collision bit based on the HF detector used for some special runs.
The TT information, which constitutes the calorimeter trigger primitives, is
streamed with a 9-fold time multiplexing, and sent via asynchronous
10\unit{Gb/s} optical links to the Layer-2 trigger. 

\subsection{The electron and photon trigger algorithm} \label{sec:egamma}

\begin{figure} \centering \includegraphics[width=0.79\textwidth]{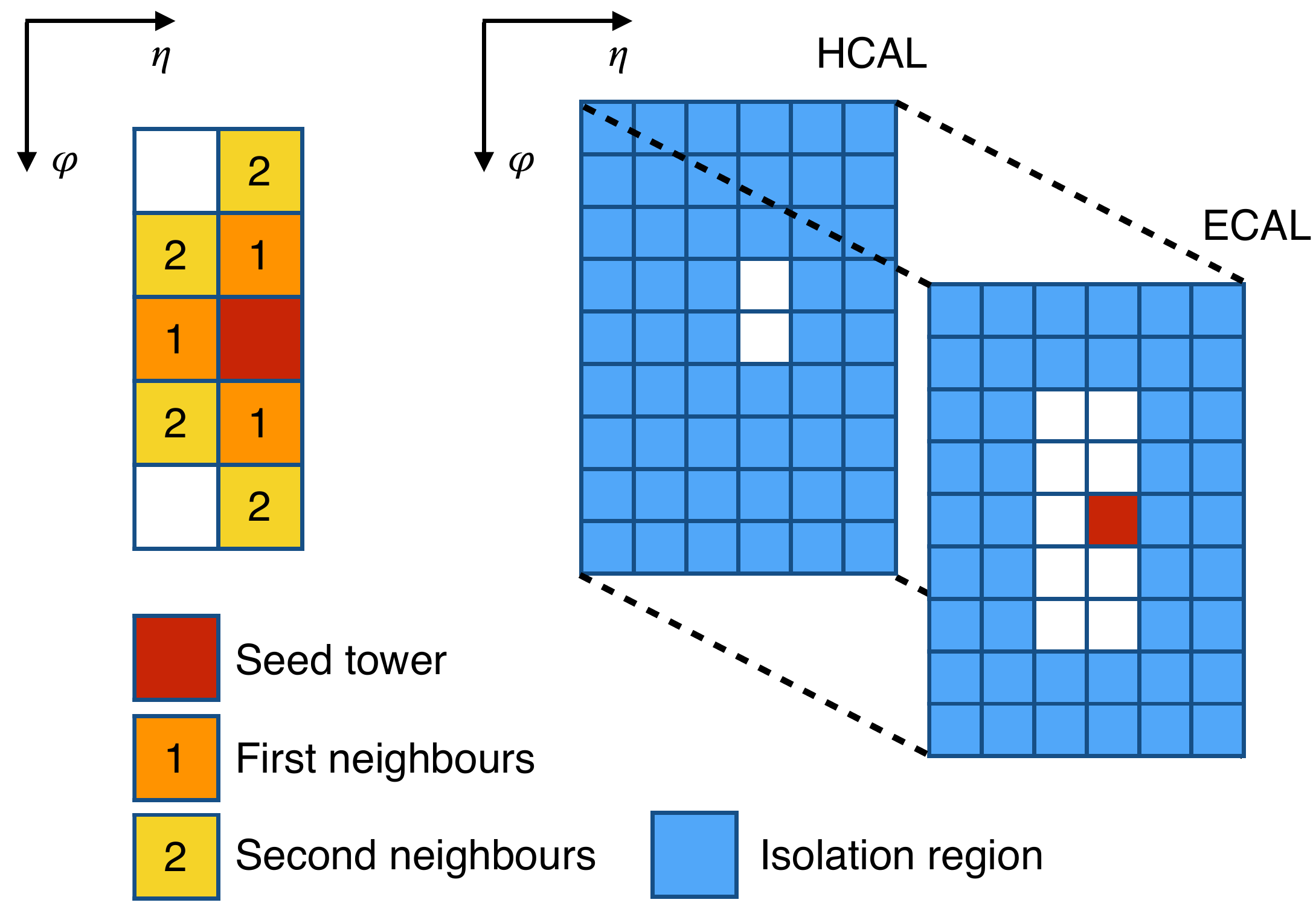}

		\caption{The Level-1 $\egamma$ clustering algorithm and
		isolation definition. A candidate is formed by clustering
		neighboring towers (orange and yellow) if they can be linked to
		the seed tower (red). Each square represents a trigger tower. A
		candidate is considered isolated if the \ET in the isolation
		region (blue) is smaller than a given value. Details are given
		in the text.  } \label{fig:eg-algo} \end{figure}

Electrons ($\Pe$) and photons ($\PGg$) are indistinguishable to the Level-1
trigger since tracking information is not available. The $\egamma$
reconstruction algorithm proceeds by clustering total (ECAL plus HCAL) energy deposits
around a ``seed'' trigger tower defined as a local energy maximum above $\ET =
2\GeV$.  Clusters are built dynamically, \ie, including surrounding towers over
1\GeV without any predetermined cluster shape requirement, and further trimmed
to include  only contiguous towers to match the electron footprint in the
calorimeter and optimize the trigger response. The trimming process results in
various candidate shapes being produced that can be categorized and used for
identification purposes. As illustrated in Fig.~\ref{fig:eg-algo}, the maximum
size of the clusters is limited to 8 TTs to minimize the impact of pileup
energy deposits, while including most of the electron or photon energy. An
extended region in the $\phi$ direction is used to obtain better coverage of
the shower since the electron energy deposit extends along the $\phi$-direction
because of the magnetic field and bremsstrahlung. 

\begin{figure} \centering
	\includegraphics[width=0.44\textwidth]{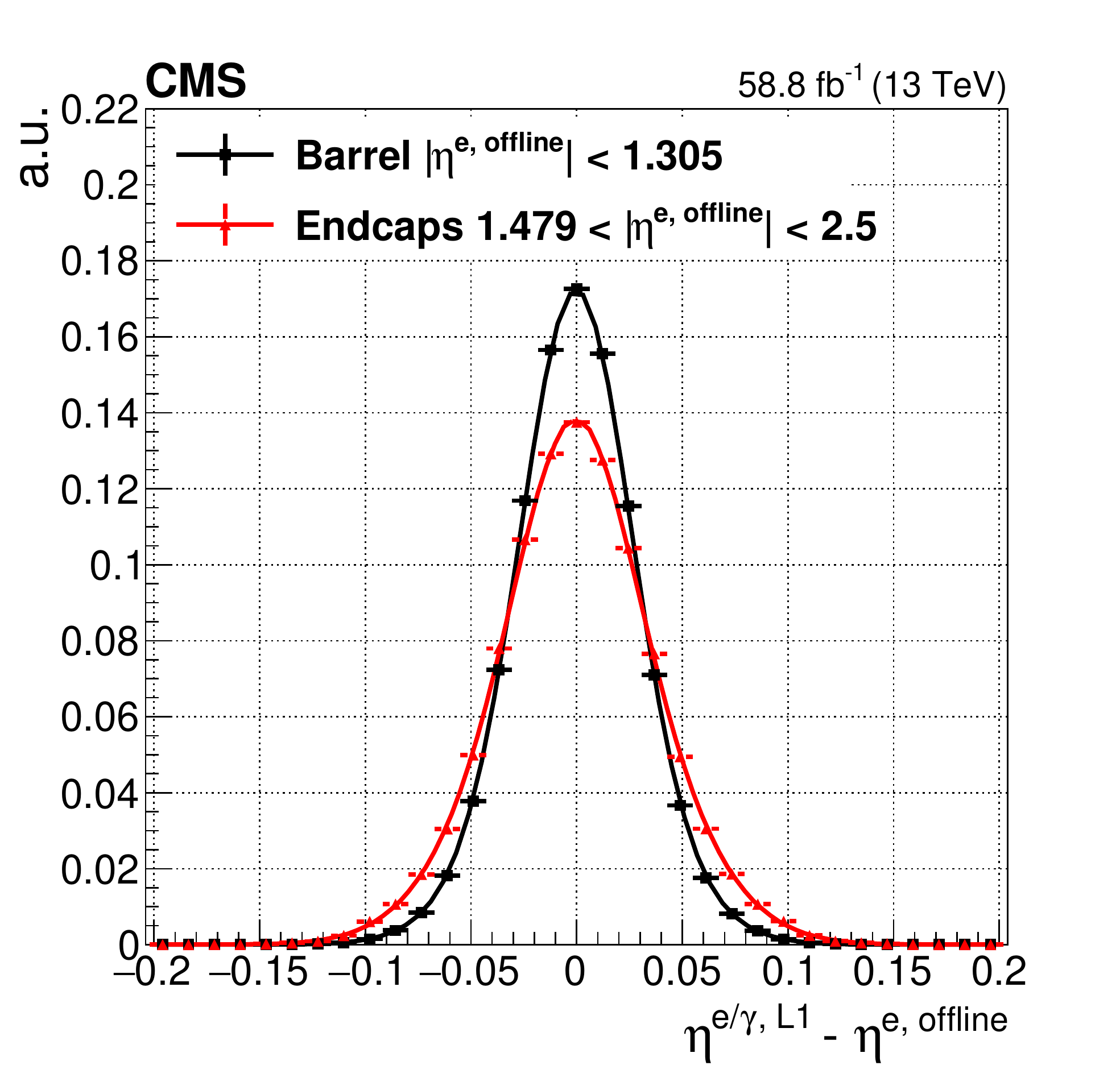}
	\includegraphics[width=0.44\textwidth]{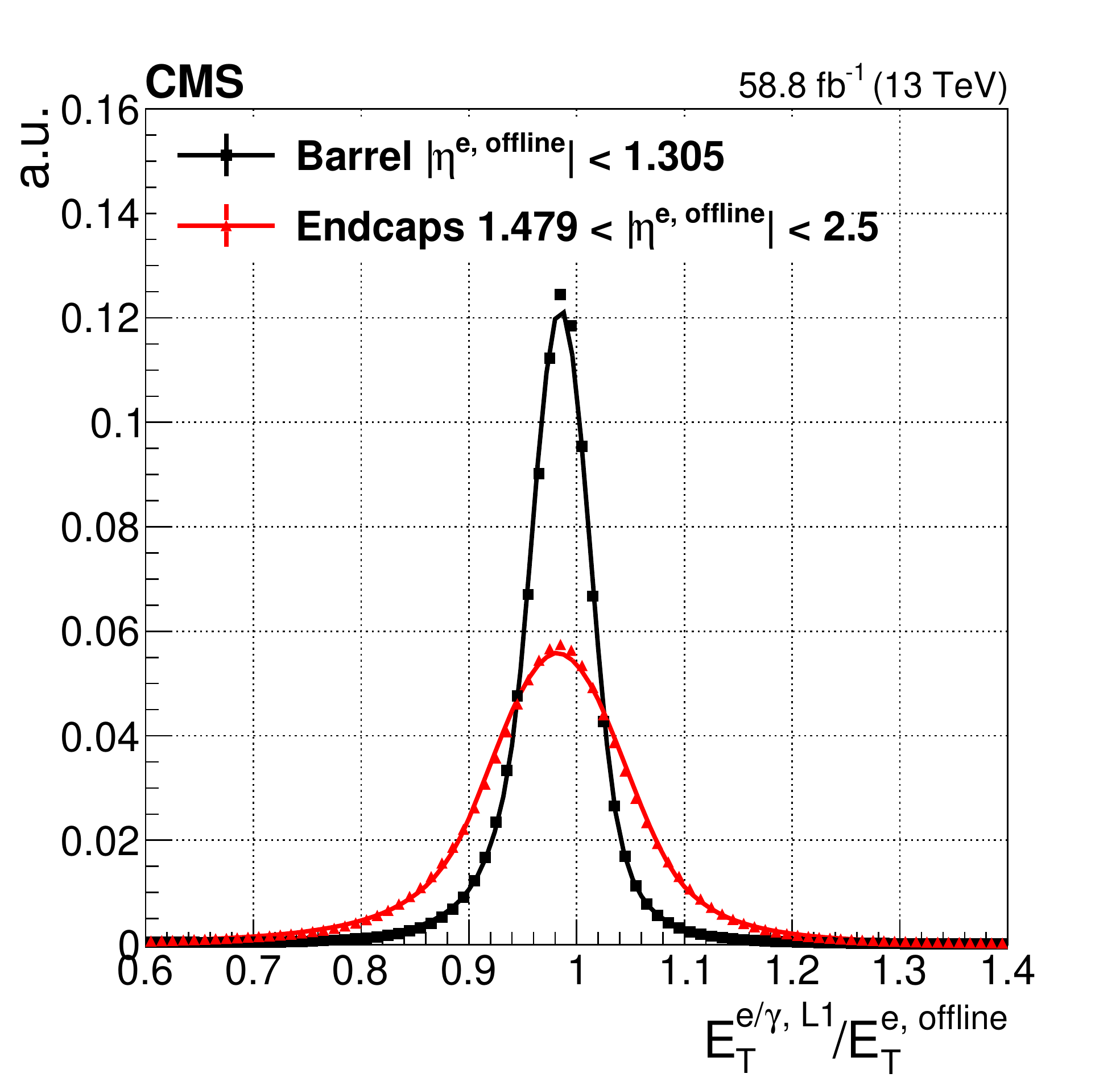} \caption{The
	pseudorapidity position of Level-1 $\egamma$ candidates with respect to
	the offline reconstructed electron position, separately for the barrel
	and endcap regions(\cmsLeft). The relative transverse energy of the
	Level-1 $\egamma$ candidates with respect to the offline reconstructed
	electron transverse energy, also separately for the barrel and endcap
	regions (\cmsRight). The functional form of the fits consists of a
	two-sided tail symmetric Crystal Ball function for the \cmsLeft plot and
	a combination of a Gaussian and an one-sided tail asymmetric Crystal
	Ball function for the \cmsRight plot.} \label{fig:eg-position-energy}
\end{figure}

The $\egamma$ candidate position is the energy-weighted position of the cluster
towers. Figure~\ref{fig:eg-position-energy} shows the  position and transverse
energy compared with those for objects reconstructed offline. Better position
resolution improves the computation of more sophisticated variables, such as
invariant masses at the \uGT level. The data used consist of events triggered
by a single electron trigger and tag-and-probe selections, which makes the
sample pure in $\PZ\to\Pe\Pe$ candidates, with the corresponding \pt
spectrum.  The resolution of the offline position is driven by the tracker
track uncertainty.

To reduce background rates, a shape veto is defined to reject the clusters least
compatible with a genuine $\egamma$ candidate such as pileup-induced energy
deposits. Additional identification criteria are also defined: \begin{itemize}
	\item The Fine Grain Veto Bit. This veto is used in the barrel to
		quantify the compactness of the electromagnetic shower within
		the seed tower and discriminates against hadron-induced showers.
\item The H/E veto. This veto requires a low ratio of HCAL to ECAL energy in the
seed tower. Different thresholds are used in the barrel and the endcap regions.
\end{itemize}

These identification variables are optimized to reduce the rate of misidentified
electrons while maintaining the maximum trigger efficiency for genuine
electrons, and are removed for candidates with $\ET > 128\GeV$.  

Isolation requirements are added to the identification criteria to produce a
collection of isolated Level-1 $\egamma$ candidates.  The isolation transverse
energy $\etiso$ corresponds to the \ET deposit in the $6{\times}9$ TT region in
$\eta{\times}\phi$ around the seed tower, from which the $\egamma$ \ET is
subtracted (illustrated in Fig.~\ref{fig:eg-algo}). To determine if an $\egamma$
candidate is isolated, a threshold stored in a LUT is applied to $\etiso$
depending on the $\ET^{\egamma}$, the $\eta$ position, and a pileup estimator
called $\ntt$. The latter is obtained by counting the number of TTs with
$\ET^\mathrm{TT}{\geq}0.5\GeV$ in the eight central $\eta$ rings of the
calorimeters ($\abs{\eta} \leq 0.34$).  The isolation threshold is optimized to
target a specific rate and efficiency for certain \ET ranges. Two working points
were derived using $\PZ\to\Pe\Pe$ collision events and a zero bias trigger
sample to estimate the rate. A loose set of isolation requirements is used for
candidates in trigger algorithms with intermediate \ET thresholds (between 20
and 30 \GeV), which are typically dielectron and cross-trigger algorithms.
For single electron algorithms, which apply energy thresholds on the electrons
above 30 \GeV, that are targeting events with a \PZ or a \PW boson, a
tighter set of isolation requirements is implemented.

The sum of the \ET of the seed and clustered towers is the raw \ET of the
$\egamma$ candidate. An additional energy calibration is performed in the
Layer-2 trigger with the scale factors derived from $\PZ\to\Pe\Pe$ collision
events. The raw energy is scaled with factors depending on the $\eta$ position
of the seed tower, the cluster shape, and the cluster \ET.  

\begin{figure} \centering
	\includegraphics[width=0.45\textwidth]{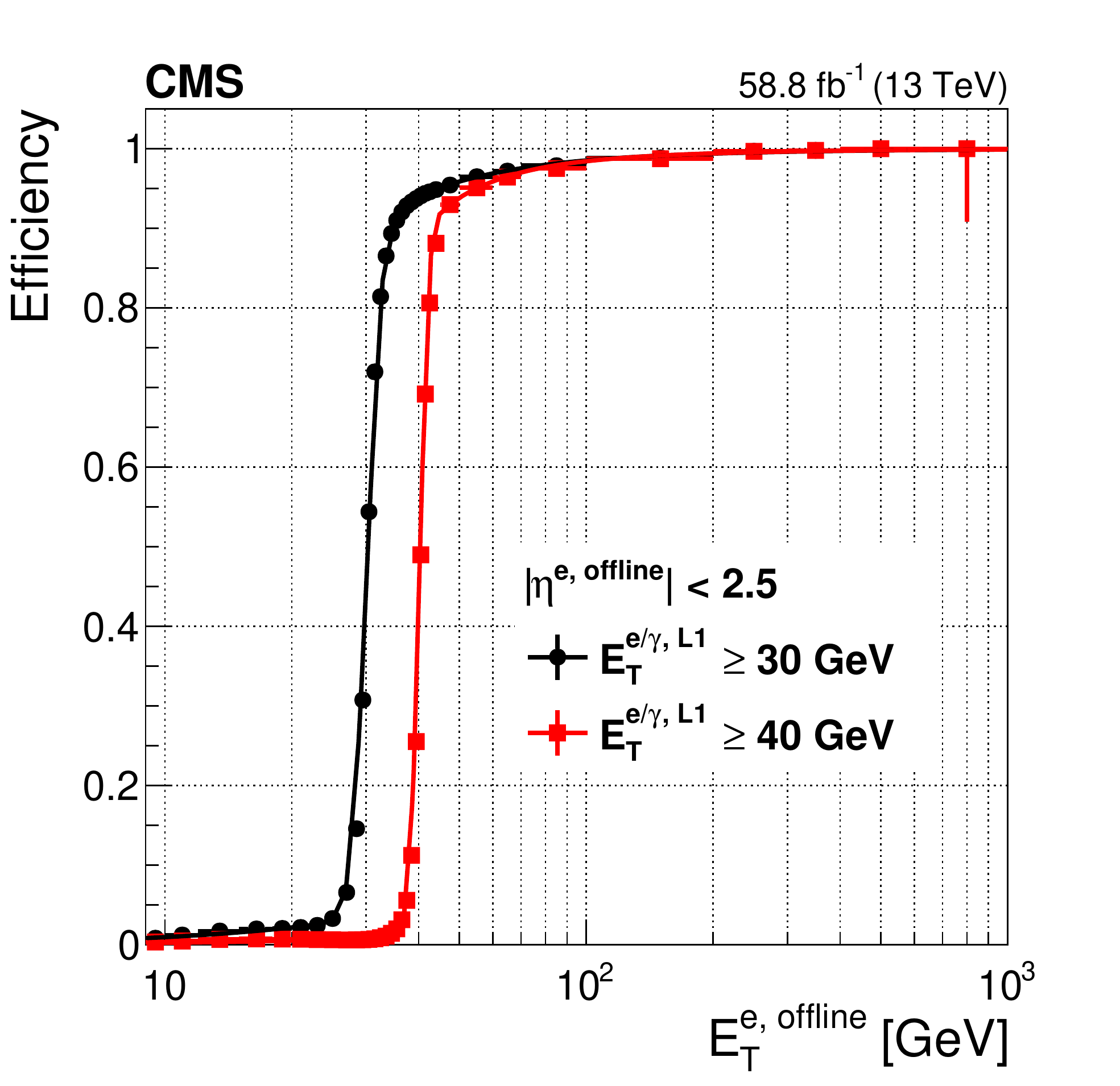}
	\includegraphics[width=0.45\textwidth]{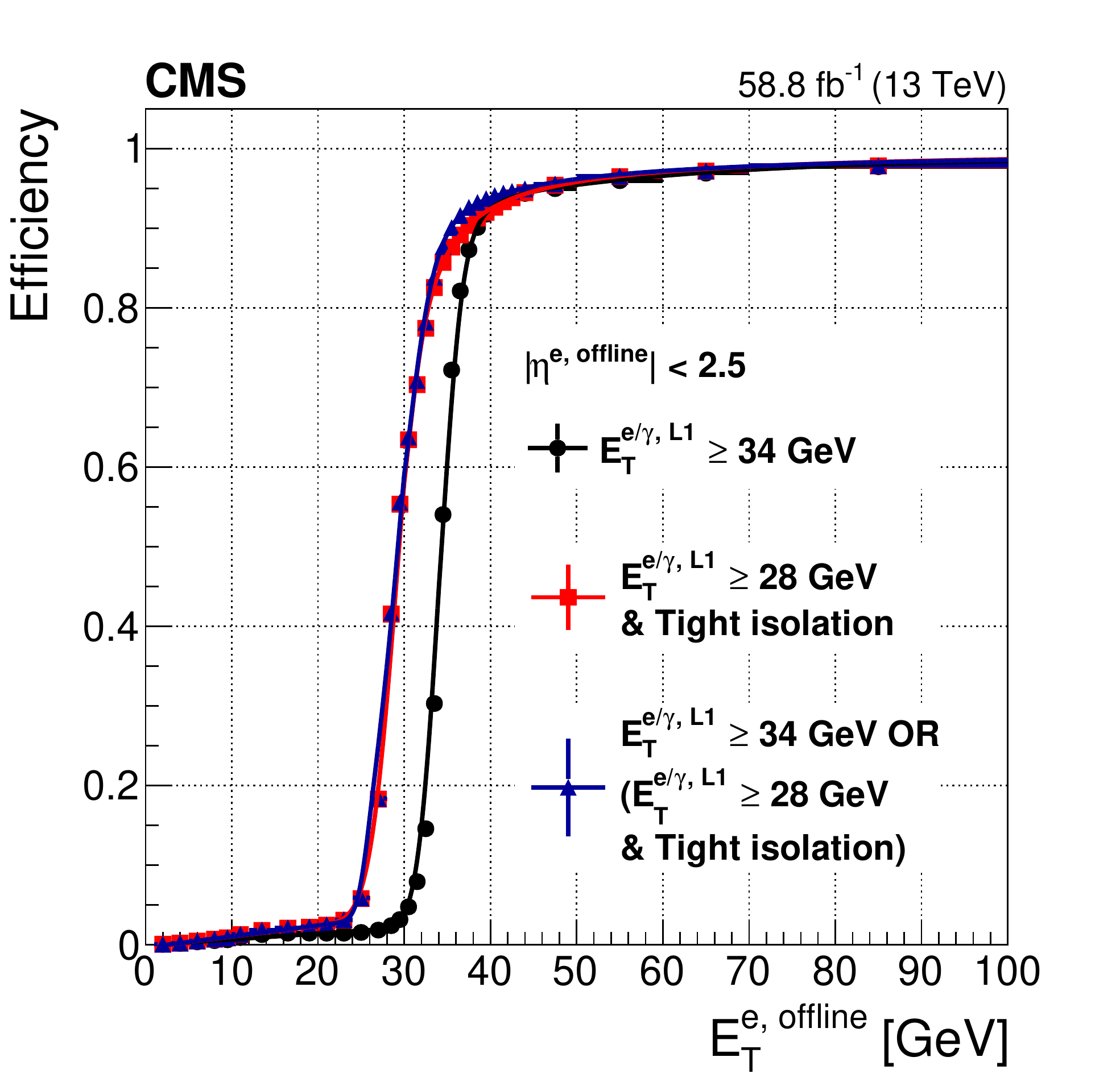} \caption{The
	Level-1 $\egamma$ trigger efficiency as a function of the offline
	reconstructed electron \ET for thresholds of 30 and 40\GeV (\cmsLeft).
	The Level-1 trigger efficiency as a function of the offline
	reconstructed electron \ET for two typical unprescaled algorithms used
	in 2018 (\cmsRight): an \ET threshold of 34\GeV in black, and of 28\GeV
	with the tight set of isolation requirements in red (as discussed in the
	text). The efficiency curve for the logical OR of the two algorithms is
	shown in blue. The functional form of the fits consists of a cumulative
	Crystal Ball function convolved with a polynomial or exponential
	function in the low \ET region.} \label{fig:eg-efficiency}
\end{figure}

\begin{figure} \centering
	\includegraphics[width=0.45\textwidth]{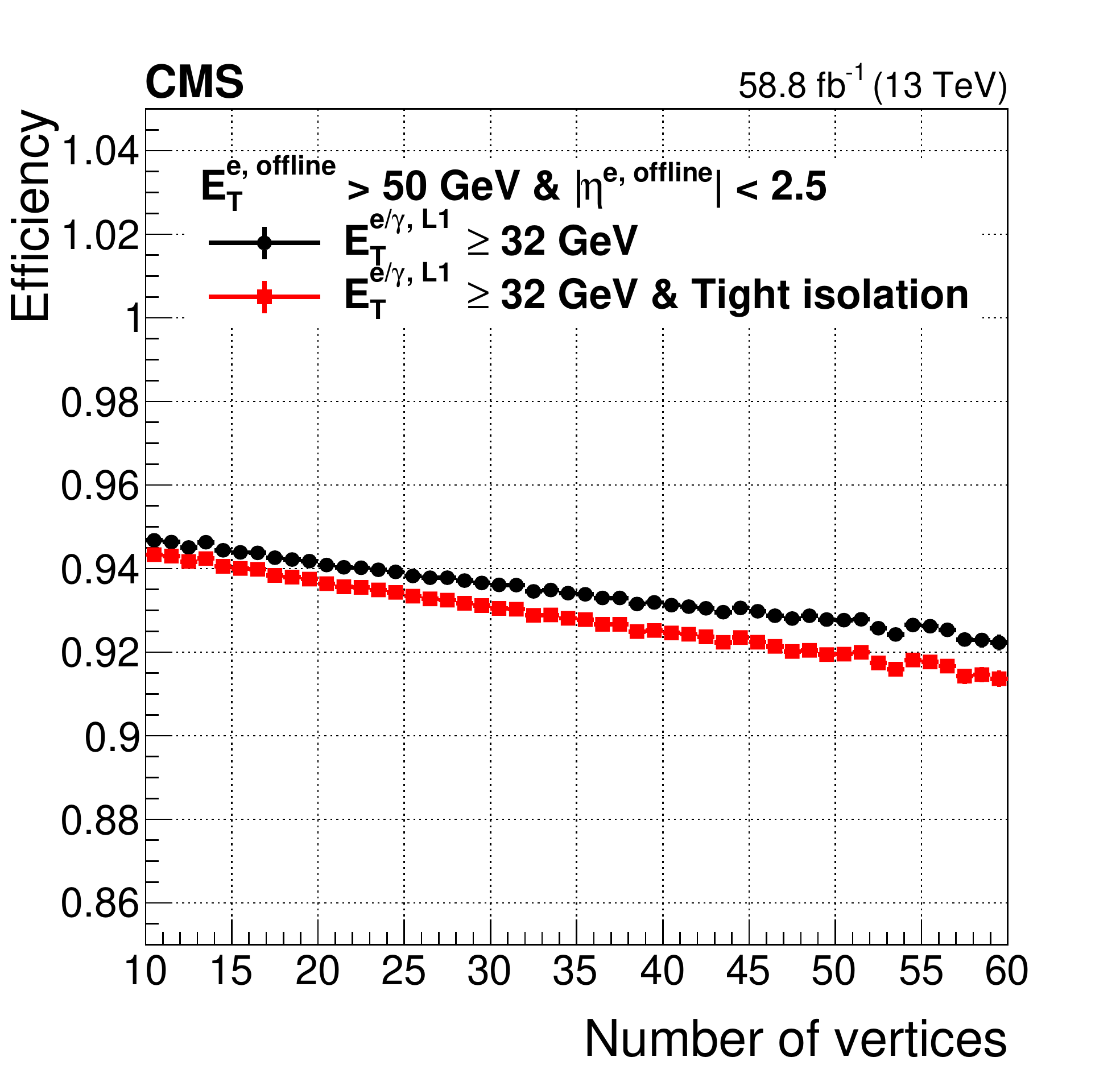}
	\includegraphics[width=0.45\textwidth]{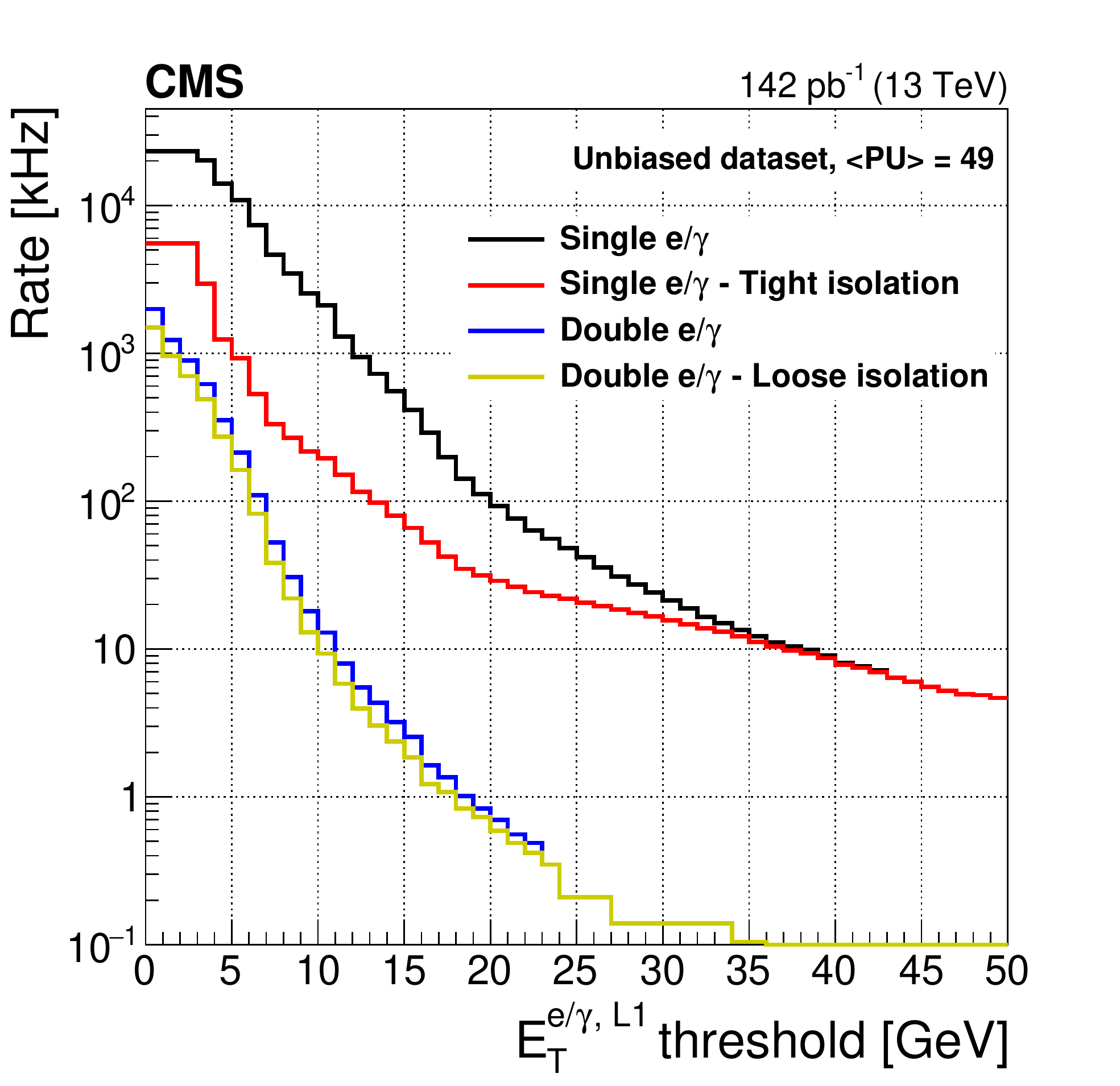} \caption{ The
	Level-1 $\egamma$ isolated trigger efficiency (\cmsLeft) as a function
	of the number of offline reconstructed vertices and the  Level-1 trigger rate
	(\cmsRight) as a function of the  \ET threshold applied on the
	candidate.  } \label{fig:eg-pileup} \end{figure}

The trigger efficiency of the upgraded $\egamma$ algorithm is shown in
Fig.~\ref{fig:eg-efficiency}. Performances for both the nonisolated and the
isolated Level-1 $\egamma$ triggers are provided. The studies are performed
using a tag-and-probe technique based on $\PZ\to\re\re$ events recorded in 2018
by an HLT trigger path requiring a tight electron with $\pt >32\GeV$. Both the
tag and the probe are offline electrons required to be within the ECAL fiducial
volume ($\abs{\eta}<1.4442$, or $\abs{\eta}>1.566$ and $\abs{\eta}<2.5$) and to pass the
loose electron identification criteria. In addition, the tag is required to have
a \pt above 30\GeV, and to be geometrically matched to the HLT electron
triggering the event within $\Delta R < 0.3$. All other reconstructed electrons
in the event passing the loose identification criteria are probe electrons. They
are geometrically matched to Level-1 $\egamma$ candidates with $\Delta R < 0.3$
and are used to evaluate the Level-1 $\egamma$ trigger efficiency. The
tag-and-probe electrons in the pair must not be within $\Delta R < 0.6$ of each
other. The invariant mass of the tag-and-probe electron system is required to be
between 60 and 120\GeV.  The trigger efficiency as a function of the number of
offline reconstructed vertices is shown in Fig.~\ref{fig:eg-pileup}. The
\cmsLeft plot shows the Level-1 $\egamma$ isolated trigger efficiency for a
32\GeV threshold as a function of the number of offline reconstructed vertices.
The trigger efficiency is also shown for the tight set of isolation
requirements. The \cmsRight plot shows in black (red) the  Level-1 trigger rate,
measured using an unbiased data set with an average pileup of 49, for a single
$\egamma$ algorithm as a function of the \ET threshold applied on the candidate
without (with) the tight set of isolation requirements. The same plot shows in
blue (yellow), the Level-1 trigger rate for a double $\egamma$ algorithm as a
function of the \ET threshold applied on the subleading $\egamma$
candidate without (with) the tight set of isolation requirements on the leading
$\egamma$ candidate (the \ET threshold on the leading candidates is always 10\GeV
higher). The rates of seeds with and without isolation converge
at high $\ET^{\egamma\mathrm{,\;L1}}$ because of the relaxation of the
isolation criteria with $\ET^{\egamma\mathrm{,\;L1}}$.

\subsection{The hadronic tau lepton trigger algorithm} \label{sec:tau} 

The hadronically decaying $\PGt$ lepton trigger algorithm efficiently
reconstructs $\PGt$ lepton decays to one, two, or three charged or neutral pions
($\PGt_\mathrm{h}$).  These pions may produce more than one cluster spatially
separated in $\phi$ because of the magnetic field.  Although the
$\PGt_\mathrm{h}$ energy deposit is typically more spread out than that of an
electron, the dynamic clustering developed for the $\egamma$ trigger is adapted
to reconstruct these individual clusters, which can subsequently be merged.

\begin{figure} \centering \includegraphics[width=0.85\textwidth]{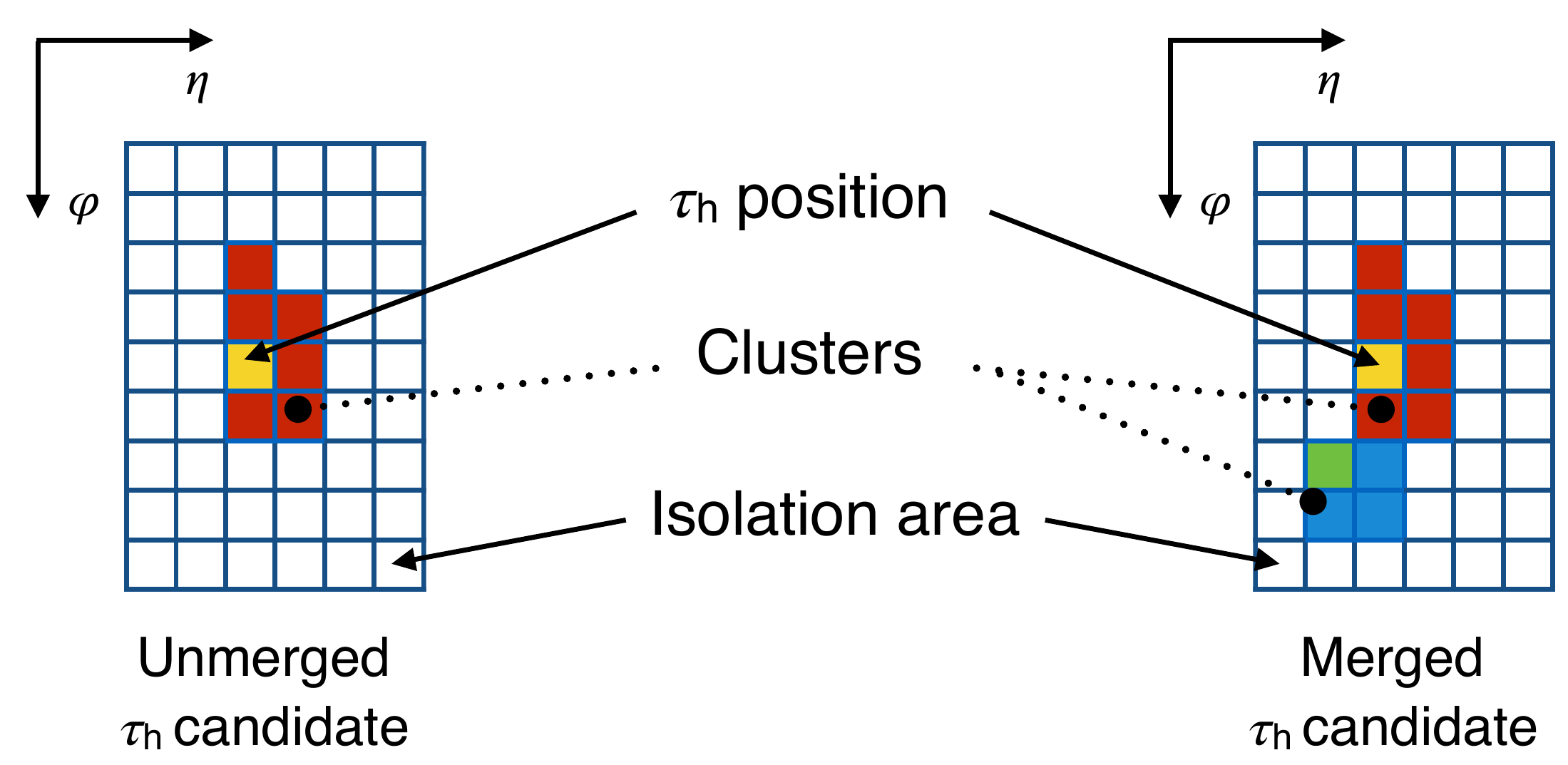}
	\caption{The Level-1 $\PGt$ clustering algorithm and isolation
	definition. The $\egamma$ dynamic clustering is used to reconstruct
	single clusters around local maxima or seeds (yellow and green), which
	can then be merged into a single $\PGt_\mathrm{h}$ candidate. Each
	square represents a trigger tower where the ECAL and HCAL energies are
	summed. A candidate is considered isolated if the \ET in the isolation
	region (white) is smaller than a chosen value.} \label{fig:tau-algo}
\end{figure} 

Figure~\ref{fig:tau-algo} illustrates the $\PGt$ lepton reconstruction
algorithm, which merges two neighboring clusters under some proximity
conditions. Hadronically decaying $\PGt$ leptons are typically low-multiplicity
jets, and have less surrounding hadronic activity than QCD-induced jets. The
candidate position is computed as an energy-weighted average centered around the
seed tower of the main cluster, giving four times better resolution than the
Run~1 $\PGt$ lepton trigger algorithm.  An isolation threshold, which depends on
the \ET and $\eta$ of the $\PGt$ lepton, and the $\ntt$ variable (as discussed
in Section~\ref{sec:egamma}), is applied to discriminate genuine $\PGt$ leptons
from QCD-induced jets. The isolation requirement is loosened for high $\ntt$ to
ensure constant $\PGt$ lepton identification efficiency as a function of pileup.
A relaxation of the isolation with \ET is also implemented to achieve the
maximum efficiency at high \ET. The isolation thresholds are stored in a LUT
that can be optimized to target a specific rate and efficiency for a given \pt 
range, \eg, for a $\PGt$ lepton pair from a Higgs boson decay. With the intense
LHC running conditions during Run 2, the working point for isolation is adjusted
to provide optimum efficiency even at the peak instantaneous luminosity of
$2.1\times10^{34}\percms$. The isolation optimization is performed on simulated
$\Z\to\PGt\PGt$ samples to evaluate the signal efficiency and on unbiased data
to estimate the rate.

\begin{figure} \centering
	\includegraphics[width=0.44\textwidth]{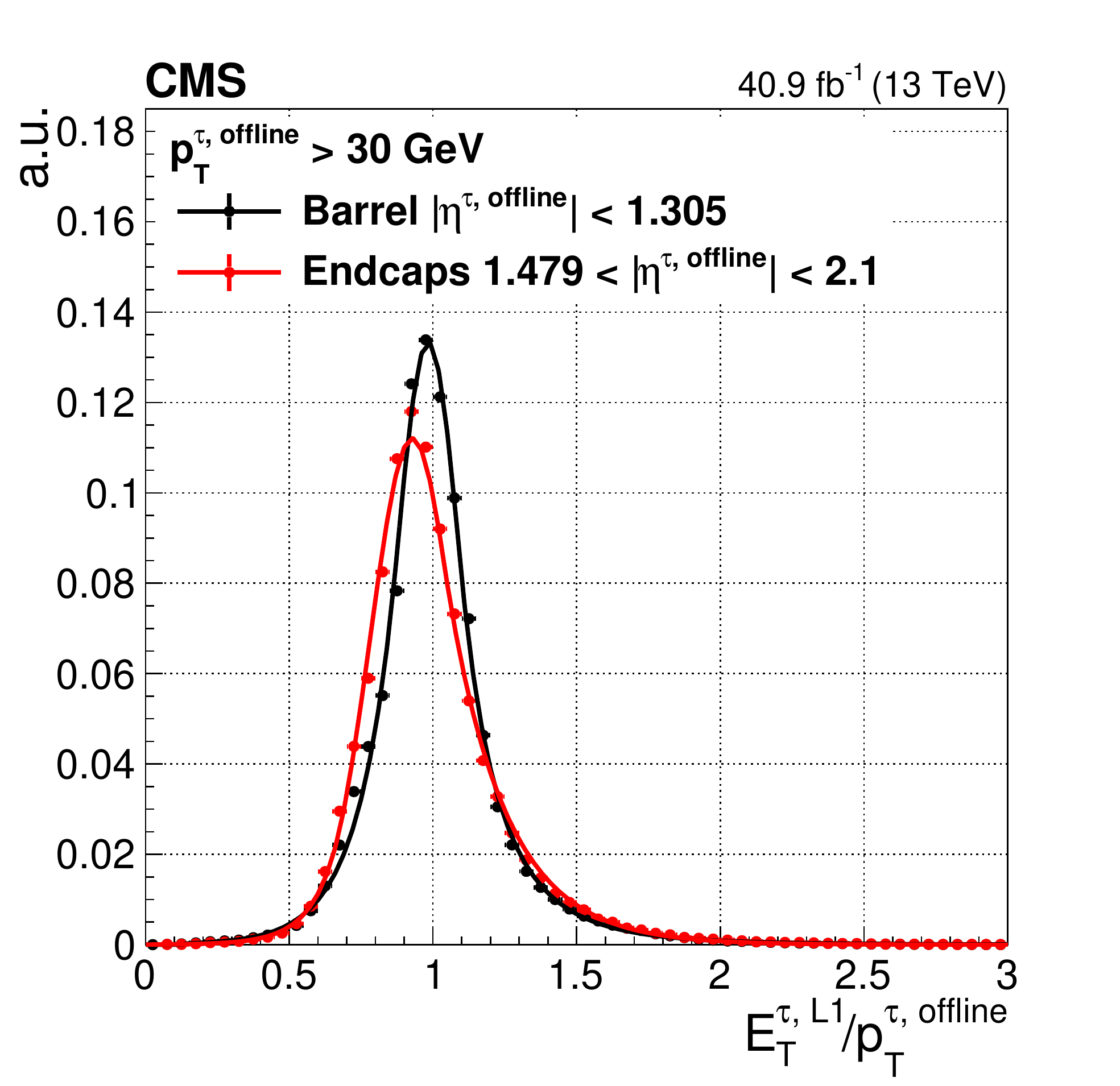}
	\includegraphics[width=0.44\textwidth]{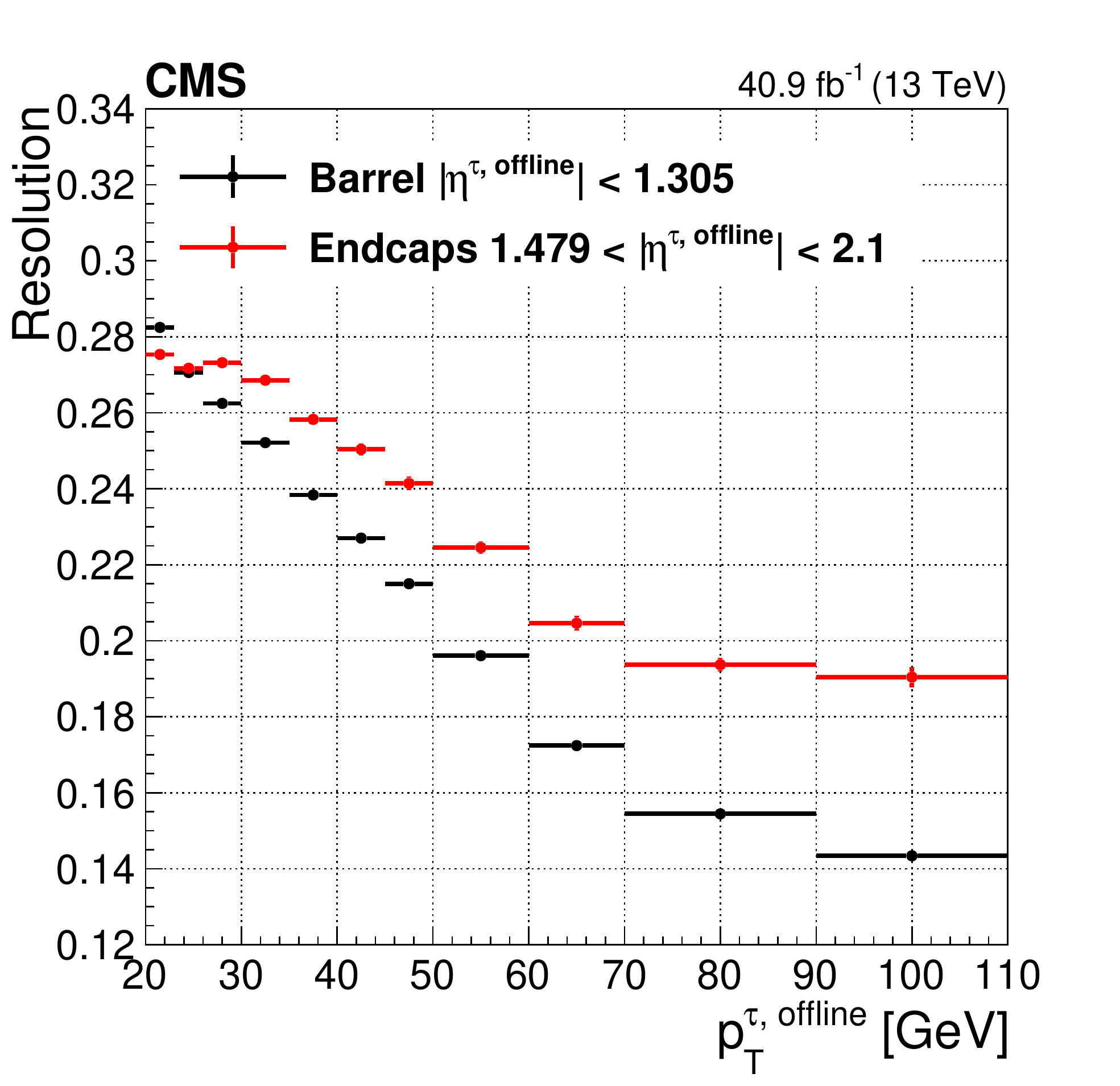} \caption{The
	Level-1 $\PGt$ trigger energy response with respect to the offline
	reconstructed $\PGt$ lepton \pt, as measured in 2017 data for the barrel
	and endcap regions (\cmsLeft). The fits consist of Crystal Ball
	functions. The resolution as a function of the offline $\PGt$ lepton \pt
	(\cmsRight), where the resolution is estimated by the root-mean-square
	of the $\ET^{\PGt,\text{ L1}}/\pt^{\PGt, \text{ offline}}$ distribution,
	divided by its mean, in bins of $\pt^{\PGt, \text{ offline}}$.}
\label{fig:tau-energy} \end{figure}

The $\PGt$ lepton \ET is calibrated using corrections that depend on the raw \ET
and $\eta$ of the candidate, the presence of a merged cluster, and an estimate
of the H/E fraction. The upgraded Level-1 $\PGt$ lepton trigger energy
resolution for barrel and endcap separately is shown in
Fig.~\ref{fig:tau-energy} (\cmsLeft).

By using a smaller number of TTs to reconstruct the energy deposit footprint of
the $\PGt$ lepton more precisely, the upgraded algorithm is more resilient
against pileup and allows more precisely adjustable thresholds for physics.
Figure~\ref{fig:tau-energy} (\cmsRight) shows the energy resolution of the
upgraded $\PGt$ trigger algorithm as a function of \pt.

The performance of the Level-1 $\PGt$ algorithm is measured in Run 2 data for
$\PGt$ leptons from $Z\to\PGt_{\PGm}\PGt_{\mathrm{h}}$ decays using a
tag-and-probe technique, where $\PGt_{\PGm}$ represents a decay to a muon and
neutrinos.  The measurement is performed in events that satisfy the single-muon
HLT path with a 27\GeV threshold on the muon \pt{}. The events contain a
well-identified and isolated $\PGm$-$\PGt_{\mathrm{h}}$ pair satisfying
transverse mass $\mT(\MET,\PGm) < 30\GeV$ and visible mass $40 <
m_\text{vis}(\PGt_{\mathrm{h}},\PGm) < 80\GeV$, where the computation of
$m_\text{vis}(\PGt_{\mathrm{h}},\PGm)$ only includes the visible decay products
of the $\PGt_{\mathrm{h}}$. The tag muon is required to have $\Delta R < 0.5$
to the HLT muon. The probe hadronically decaying $\PGt$ leptons are
reconstructed using the standard hadrons-plus-strip algorithm~\cite{CMS:taus},
and selected using a ``medium'' isolation criteria \cite{CMS:taus}, and are
required to satisfy $\pt > 20\GeV$ and $\abs{\eta} <  2.1$; discriminators are also
applied to reduce the contamination from muons and electrons. The details of
the offline $\PGt$ lepton reconstruction are described in
Ref.~\cite{CMS:taus}. The probes are matched to Level-1 hadronic $\PGt$
candidates within $\Delta R < 0.5$ and used for efficiency measurements.

\begin{figure} \centering
	\includegraphics[width=0.44\textwidth]{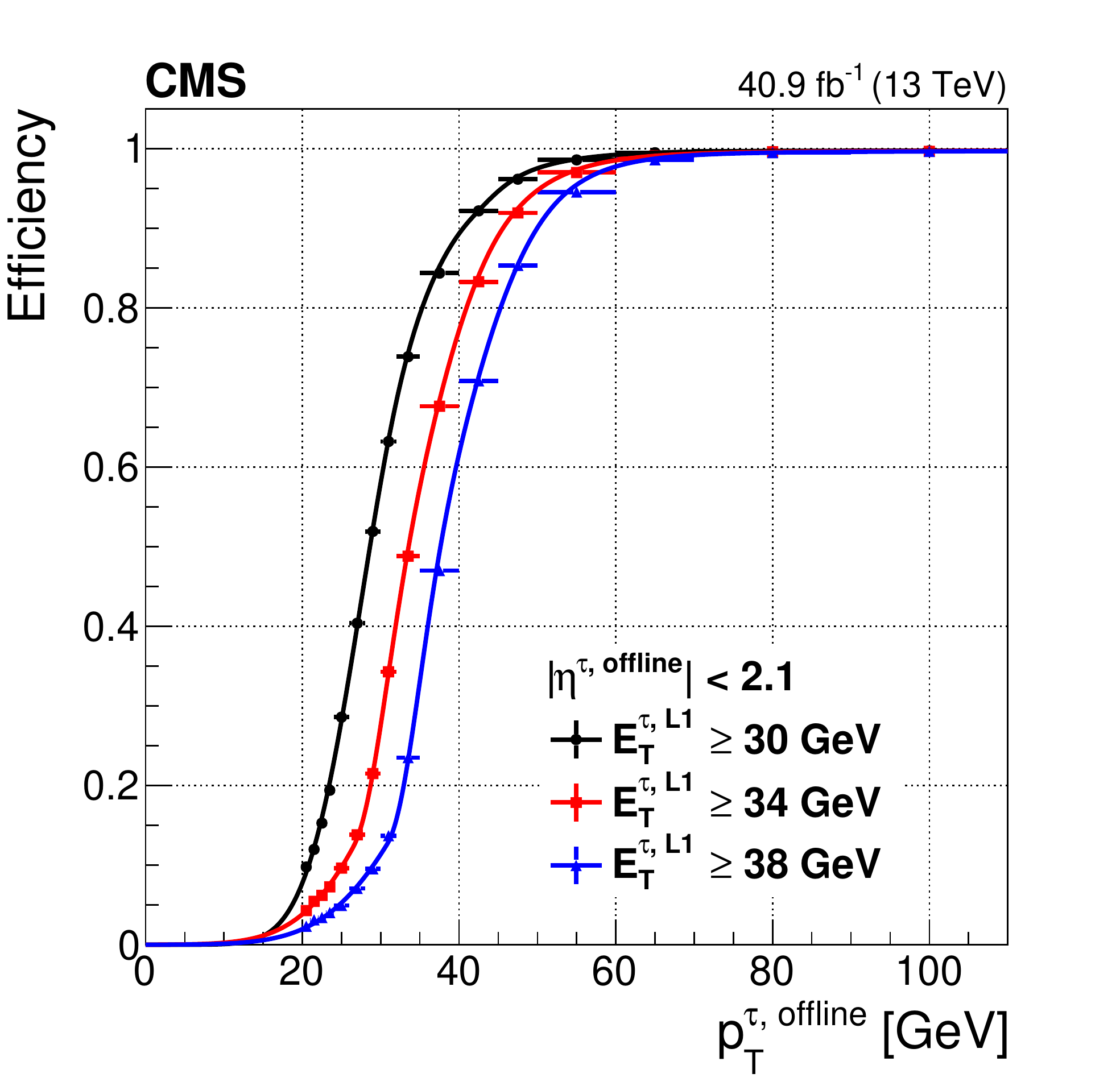}
	\includegraphics[width=0.44\textwidth]{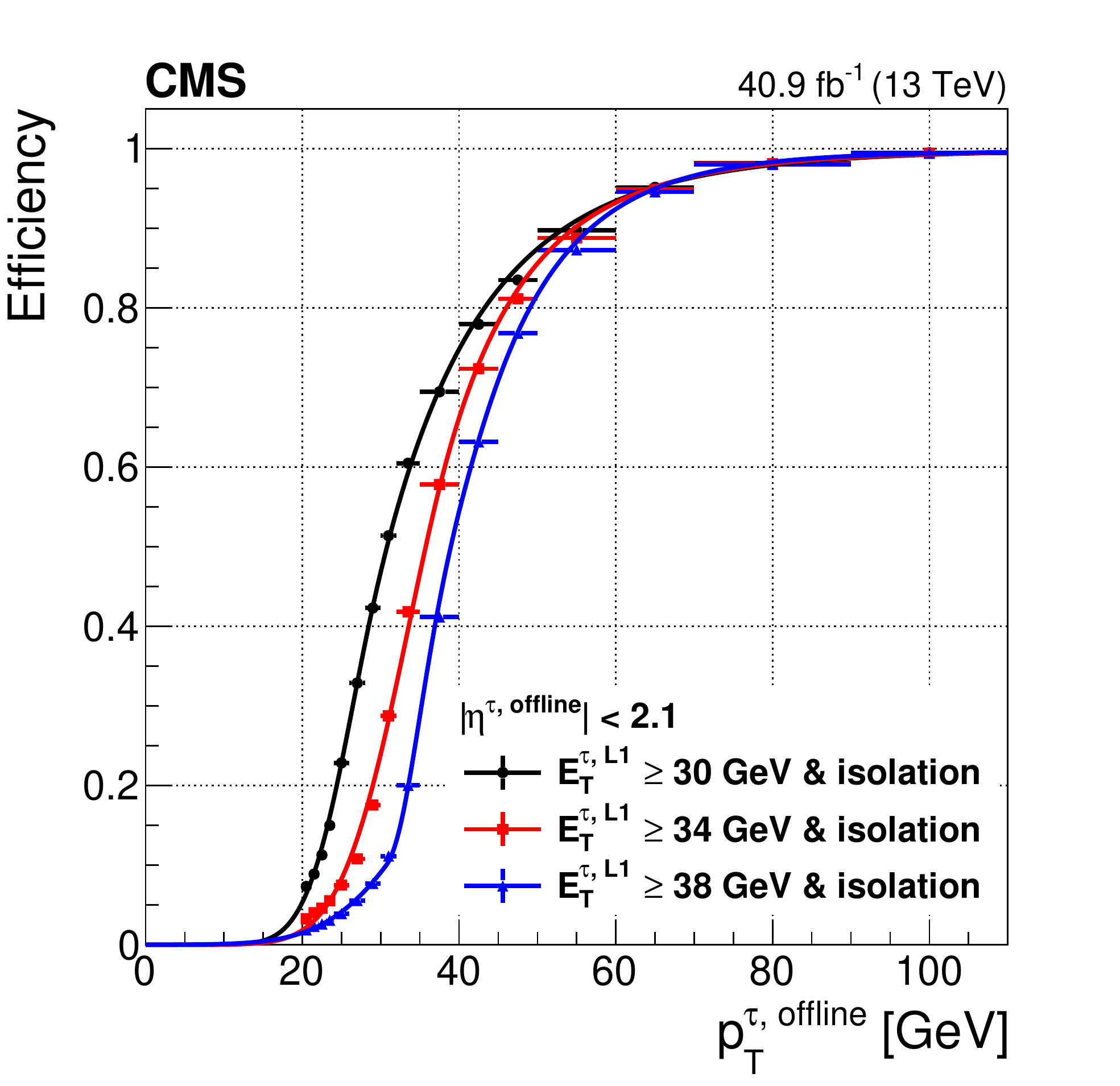} \caption{The
	Level-1 $\PGt$ trigger efficiency, as a function of the offline
	reconstructed $\PGt$ lepton \pt, for typical thresholds of 30, 34, and
	38\GeV (\cmsLeft). The Level-1 isolated $\PGt$ trigger efficiency, as a
	function of the offline reconstructed $\PGt$ \ET, for the same three
	thresholds (\cmsRight). The functional form of the fits consists of a
	cumulative Crystal Ball function convolved with an arc-tangent.}
\label{fig:tau-efficiency} \end{figure}

\begin{figure} \centering
	\includegraphics[width=0.44\textwidth]{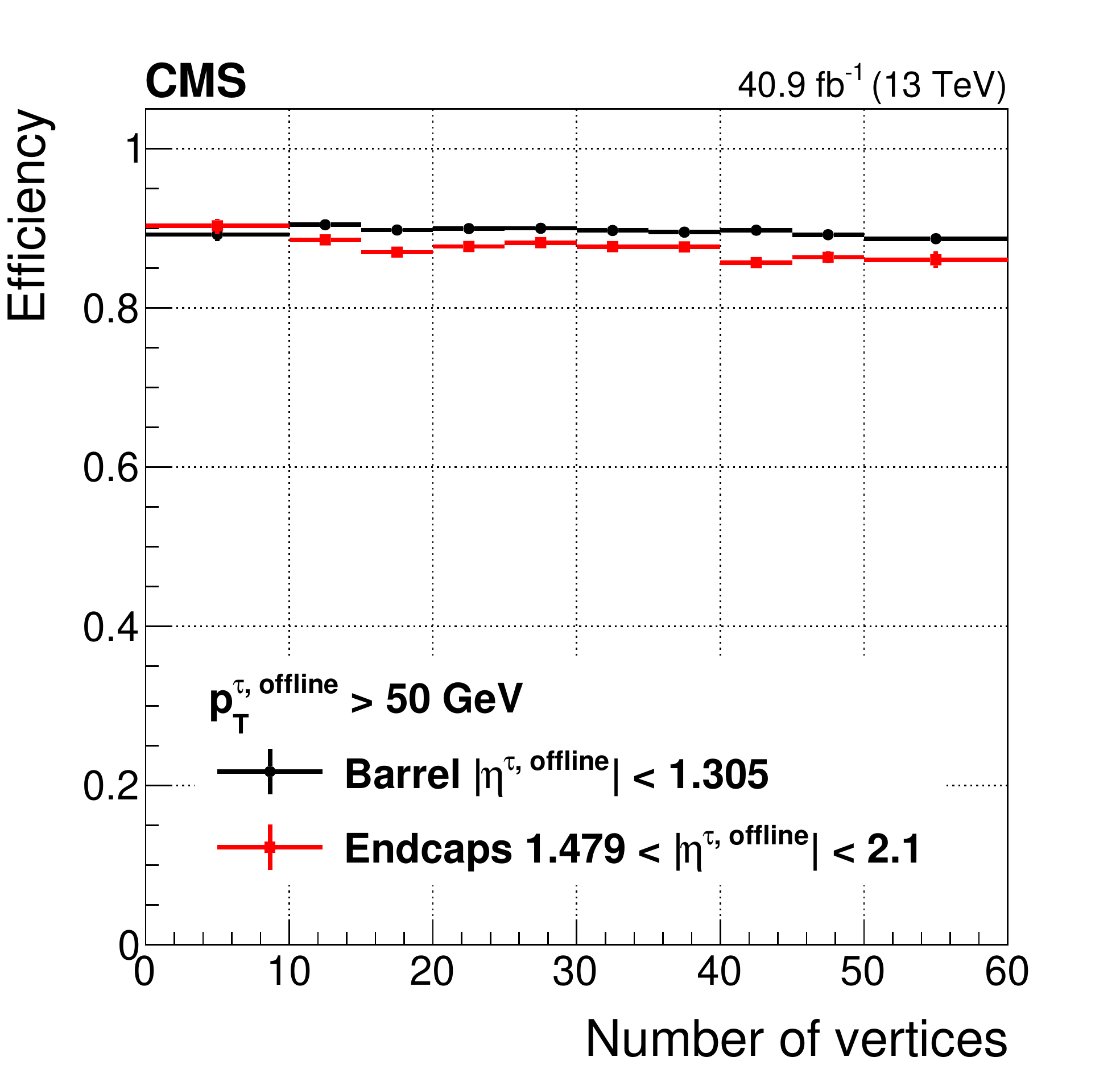}
	\includegraphics[width=0.44\textwidth]{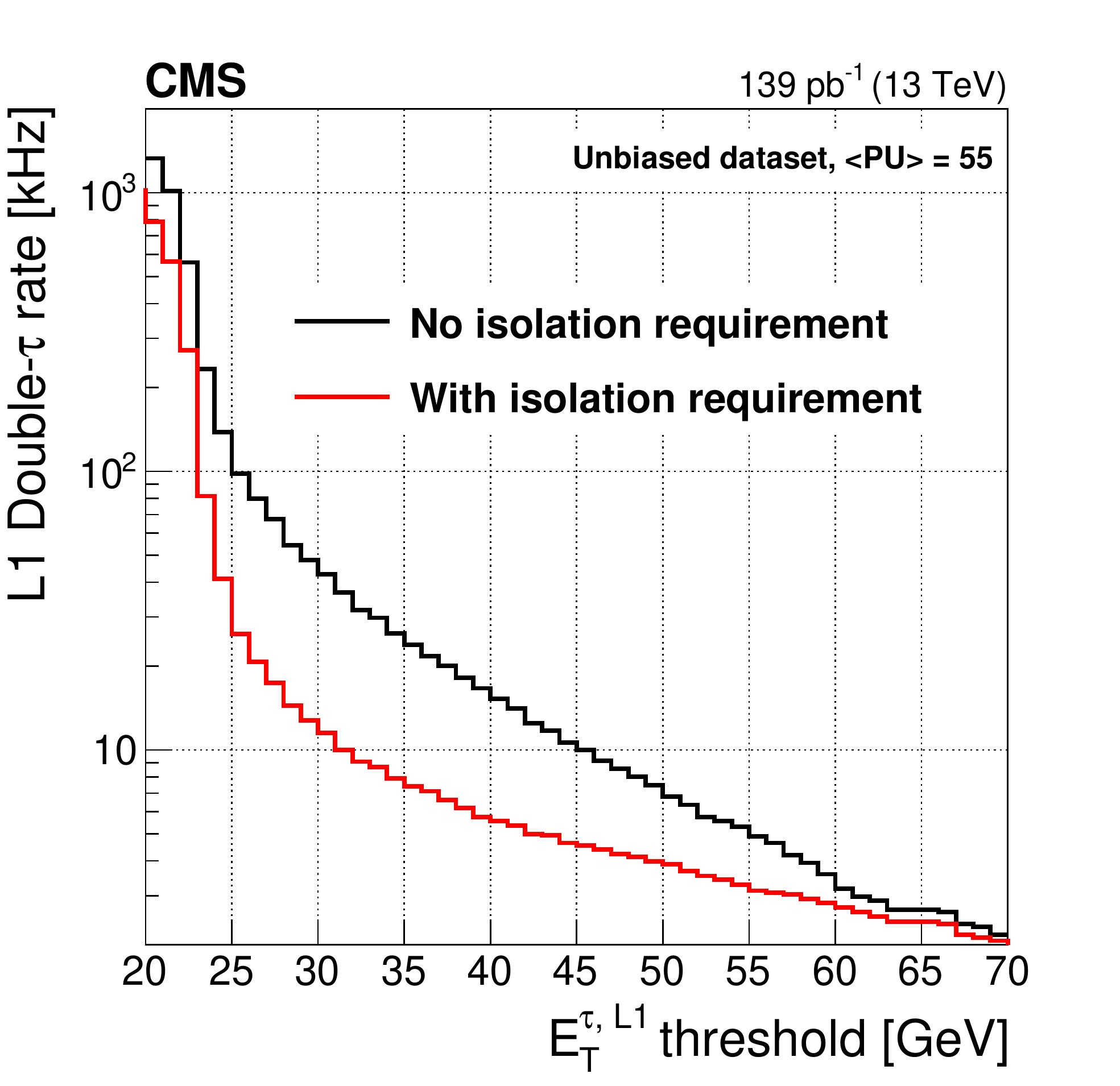} \caption{The
	integrated Level-1 selection efficiency for the isolated $\PGt$ trigger
	with $\ET {\geq} 30\GeV$, matched to an offline reconstructed and
	identified $\PGt$ lepton with $\pt > 50\GeV$, as a function of the
	number of offline reconstructed vertices~(\cmsLeft). The Level-1
	double-$\PGt$ trigger rate, as a function of the \ET threshold, for
	$\PGt$ candidates with and without an isolation requirement applied
~(\cmsRight). The rate is measured requiring  two $\PGt$ candidates with
	\ET larger than the bin value, in a unbiased data set with an average
	pileup of 55.} \label{fig:tau-pileup-rate} \end{figure}

The trigger efficiency, plotted as a function of the offline reconstructed \PGt
lepton \pt, is shown in Fig.~\ref{fig:tau-efficiency} for nonisolated and
isolated Level-1 $\PGt$ candidates. The relaxation of the isolation
identification criteria with \ET ensures that the efficiency reaches a plateau
value of 100\% at high \ET. The turn-on curves are obtained by matching
geometrically the $\PGt$ candidates reconstructed offline that pass all the
identification and isolation requirements of the $H\to\PGt\PGt$ analysis with
its Level-1 counterpart. The stability of the efficiency with respect to pileup
is illustrated in Fig.~\ref{fig:tau-pileup-rate} (\cmsLeft).
Figure~\ref{fig:tau-pileup-rate} (\cmsRight) shows the double-$\PGt$ rate as
function of the \ET threshold applied to both of the Level-1 $\PGt$ candidates.
The rate is measured in an unbiased data sample. For typical thresholds of
${\approx}30\GeV$, a significant rate reduction is achieved by using the
isolation requirement.

\subsection{The jet and energy sum trigger algorithms} 

The Level-1 jet reconstruction algorithm is based on square-jet approach
similar to that used in Run~1, but uses a $9{\times}9$ TT sliding window
centered on a local maximum, the jet seed, with $\ET > 4$\GeV. In the barrel,
the window size matches the anti-\kt \cite{antikt} clustering size of 0.4 used
in the offline jet reconstruction. A jet candidate must have a seed energy
greater than the TTs in the triangle above the diagonal of the $9{\times}9$
square window, and greater than or equal to the TTs in the triangle below the
same diagonal. This is to avoid double counting and to prevent TTs with the
same energy from vetoing one another when being considered as a jet seed. The
veto condition applied is antisymmetric along the diagonal of the $9{\times}9$
window to prevent TTs with the same energy from vetoing one another.  The jet
candidate energy is the sum of all TT energies in the $9{\times}9$ window.  In
addition to reconstructed jets, the total scalar sum of transverse energy over
all TTs, \ET{}, and the magnitude of the vector sum of transverse energy over
the same TTs, \MET{}, use trigger tower granularity. The total scalar
transverse energy of all jets, \HT{}, and the corresponding magnitude of the
vector sum \MHT{} are computed using Level-1 jets.

\begin{figure} \centering \includegraphics[width=0.5\textwidth]{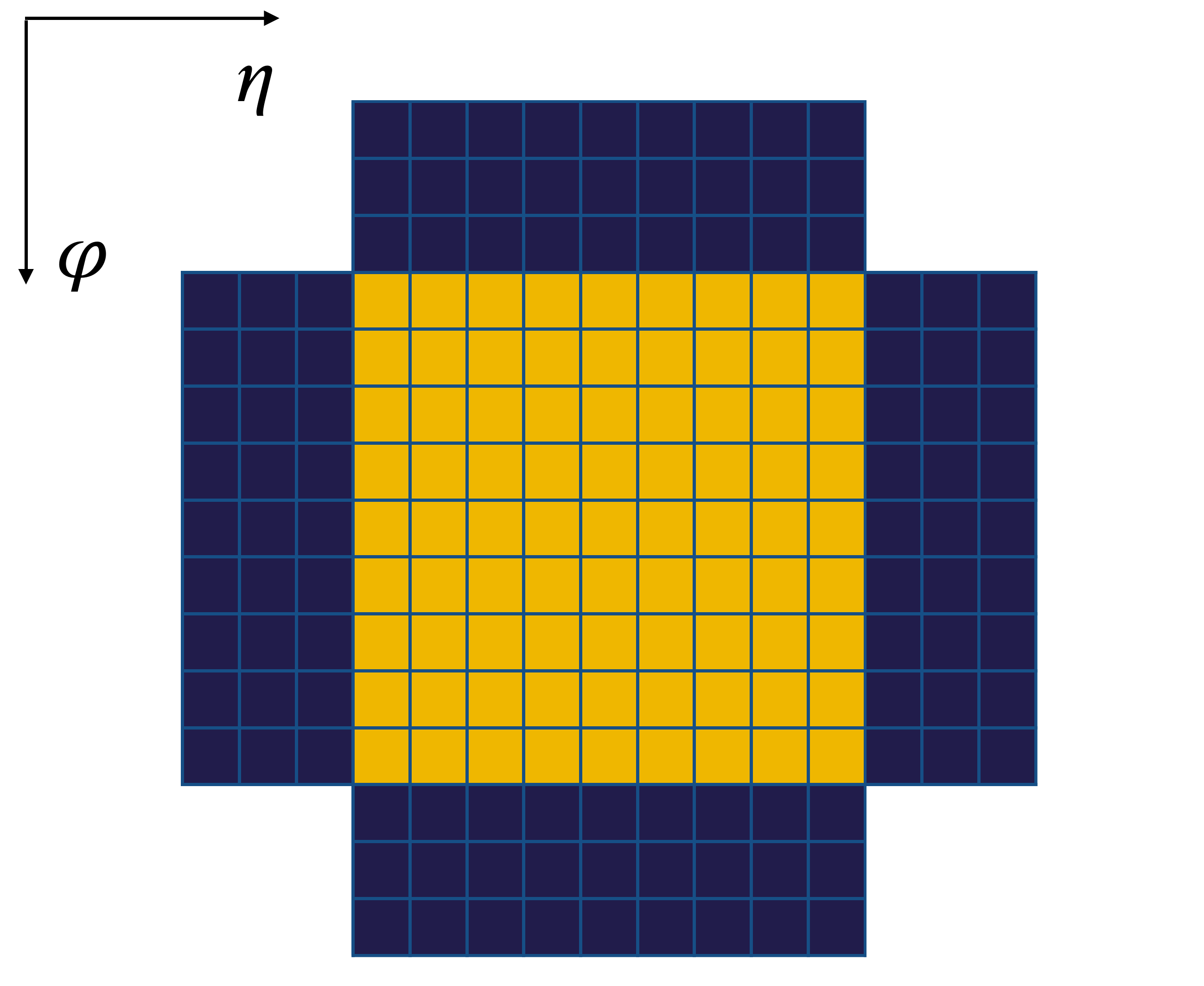}
	\caption{The area used by the jet pileup subtraction  algorithm to
	estimate the energy deposit from the local pileup, in blue, and the area
	used to measure the energy of the Level-1 jet, in orange.}
\label{fig:chunky_doughnut} \end{figure}

The estimated \ET from pileup, which is subtracted from each jet, is computed
locally on a jet-by-jet basis in each bunch crossing, to respond dynamically to
fluctuating pileup conditions. The chosen pileup subtraction algorithm provides
a significant rate reduction, while maintaining efficiency.
Figure~\ref{fig:chunky_doughnut} shows the regions that are used to estimate the
local pileup energy to be subtracted from the jet energy.  The pileup is
estimated using four $3{\times}9$ outer regions, one on each side of the
$9{\times}9$ jet square.  The pileup \ET is calculated as the energy sum of the
three lowest energy regions, so the \ET from an adjacent jet in the remaining
outer region is not subtracted from the \ET{}. Since this area for subtraction
(3 of 4 outer areas) equals the jet area, the implementation is simple.

To ensure consistent jet energy response, Level-1 jets are calibrated in bins of
jet \pt and $\eta$, since any loss or mismeasurement will depend on the energy
of the jet and the material it traverses. A dedicated LUT is derived from a QCD
multijet simulation that returns a \pt scale factor that is applied to each jet.
The LUT is derived by matching Level-1 jets to generator jets within $\Delta R <
0.25$, then fitting correction curves produced in bins of jet $\eta$ of
$1/\langle \ET^{\text{L1}}/\ET^{\text{gen}}\rangle$ as a function of $\langle
\ET^{\text{L1}}\rangle$. 

\begin{figure} \centering 
		\includegraphics[width=0.48\textwidth]{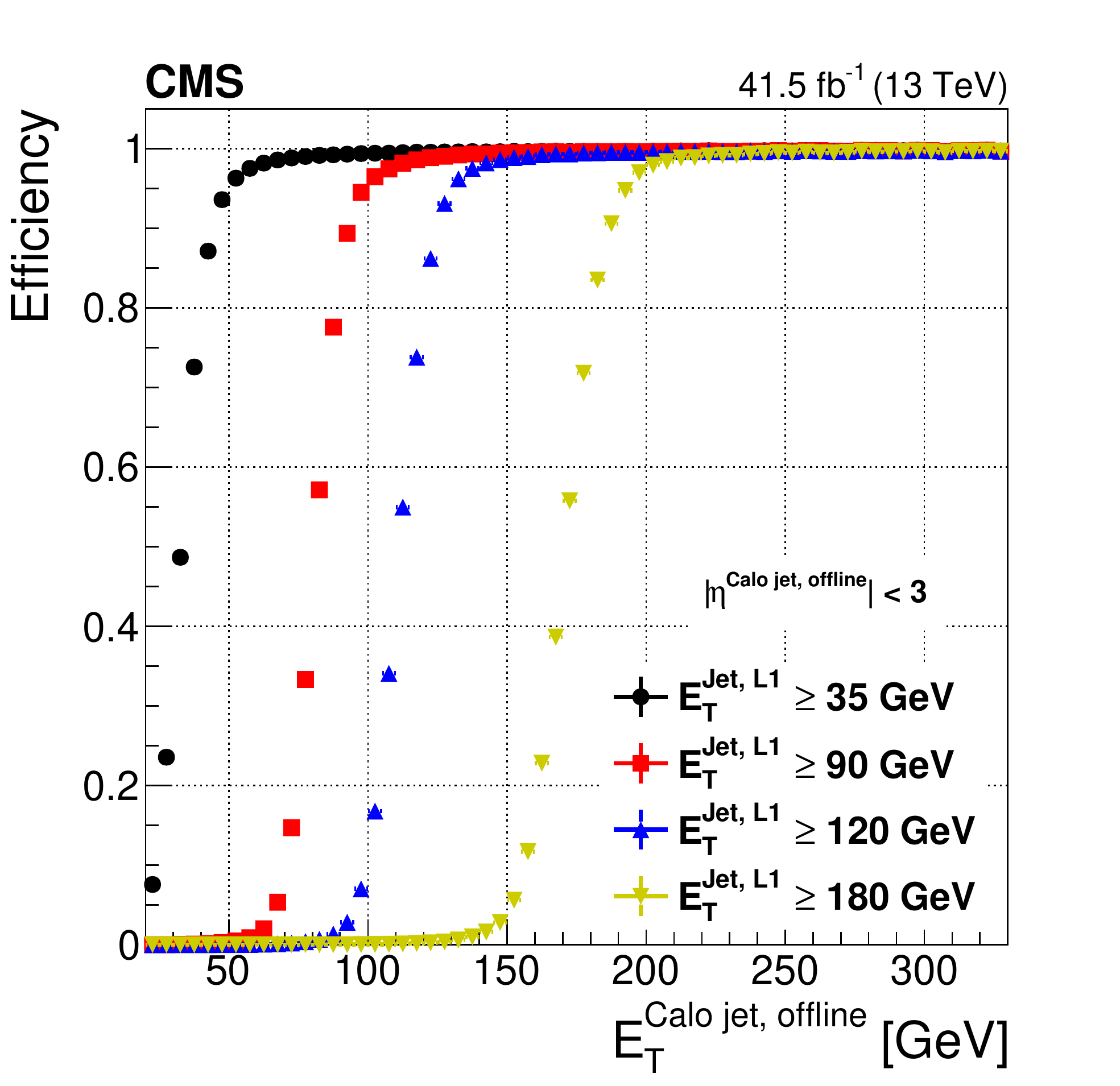}
		\includegraphics[width=0.48\textwidth]{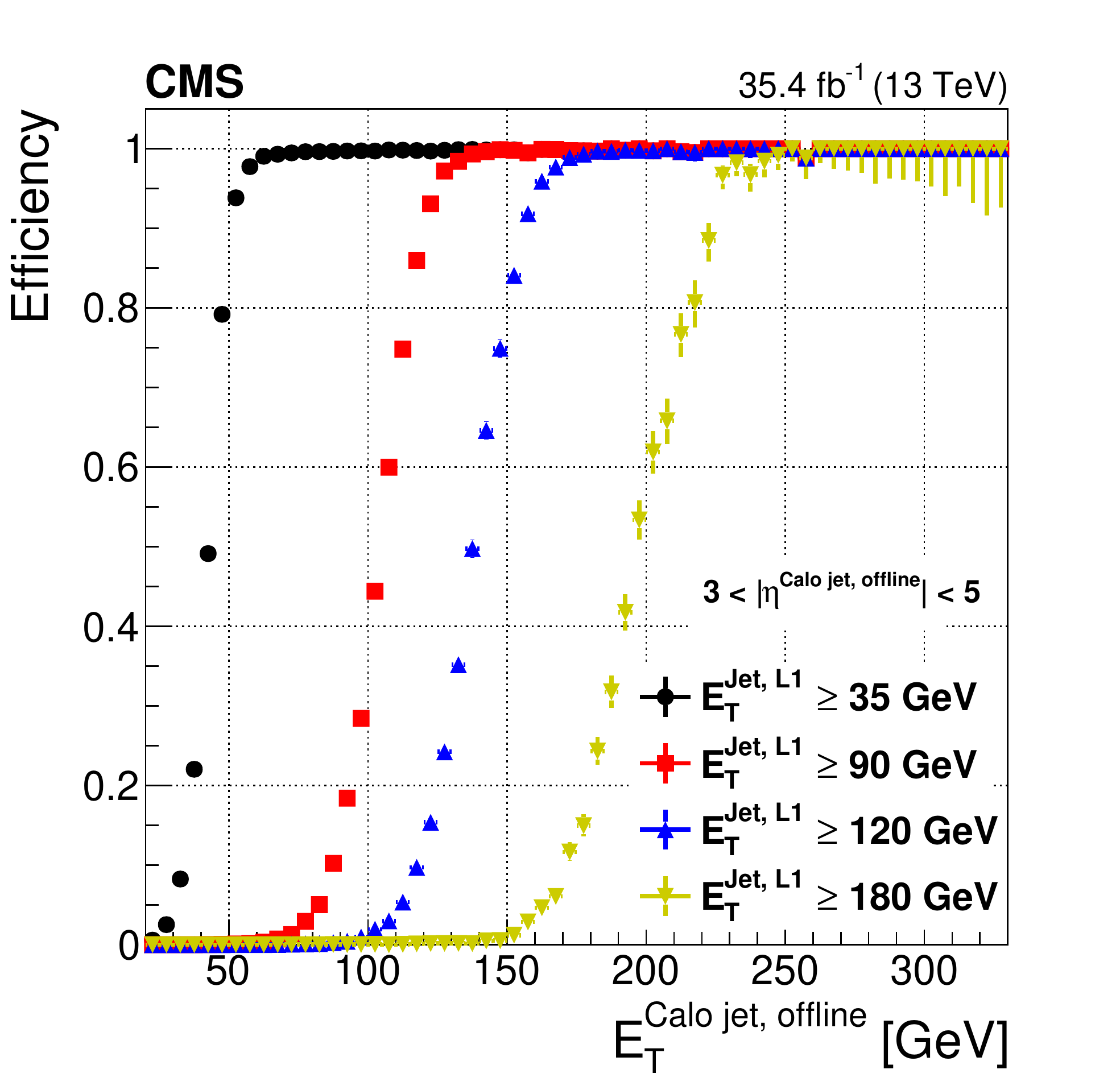}
	\caption{Efficiency curves for the Level-1 jet trigger for the barrel +
	endcap (left) and forward (right) pseudorapidity ranges.}
\label{fig:jetperf} \end{figure}

\begin{figure} \centering
		\includegraphics[width=0.48\textwidth]{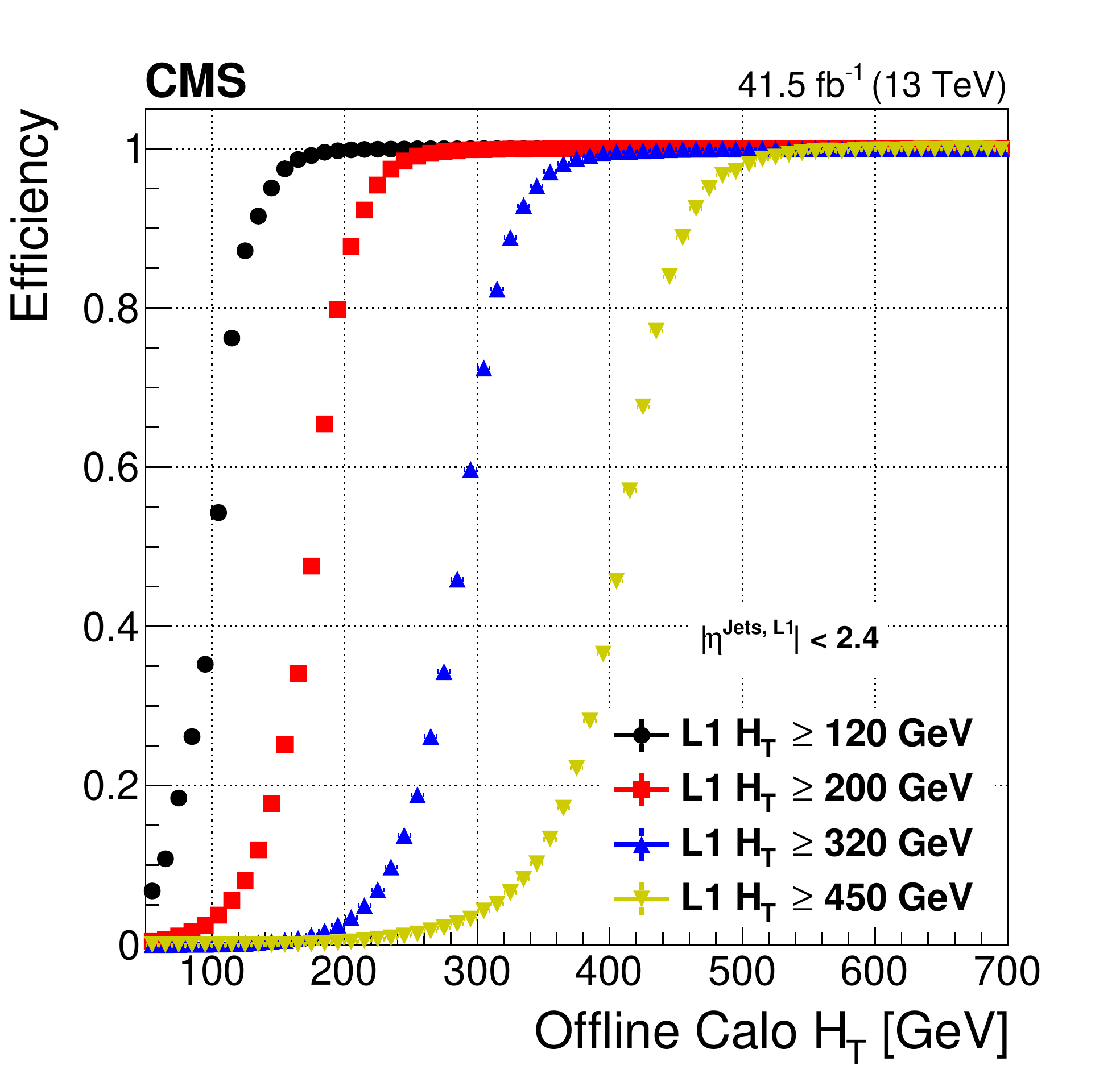}
		\includegraphics[width=0.48\textwidth]{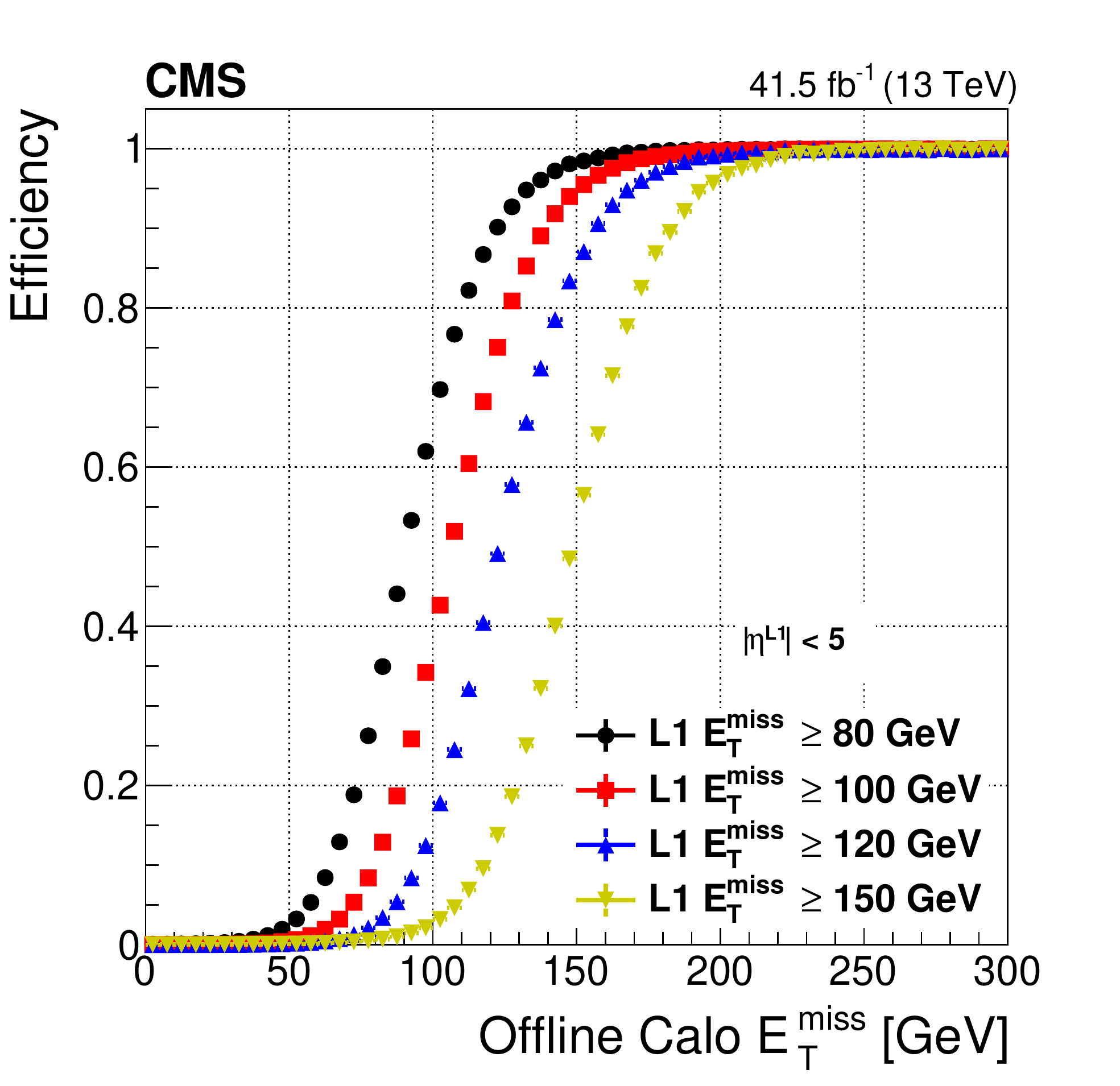}
	\caption{Efficiency curves for the scalar sum of jet energy with
	$\ET{\geq}30$\GeV (\cmsLeft) and missing transverse energy (\cmsRight)
	for various thresholds. The thresholds are  indicated as L1 \HT{} and L1
	\MET{} in the legends.} \label{fig:sumperf} \end{figure}

Figure~\ref{fig:jetperf} shows the performance of the Level-1 jet triggers in
the combined barrel and endcap region and in the forward region, measured using
an independent data sample collected with a single-muon trigger. The
efficiencies show a sharp turn-on and high efficiency for a number of
thresholds, representative of those used in Run~2 for various single-jet  and
multijet seeds. Figure~\ref{fig:sumperf} shows the efficiency curves for the
Level-1 $\HT$ and \MET{} triggers. The \MET trigger efficiency is measured using
events triggered by and reconstructed with a single muon, and is plotted as a
function of offline \MET{}, which is the magnitude of the negative vector sum of
the \pt of all calorimeter energy deposits, with $\abs{\eta} \leq 5.0$.

\begin{figure} \centering 
		\includegraphics[width=0.495\textwidth]{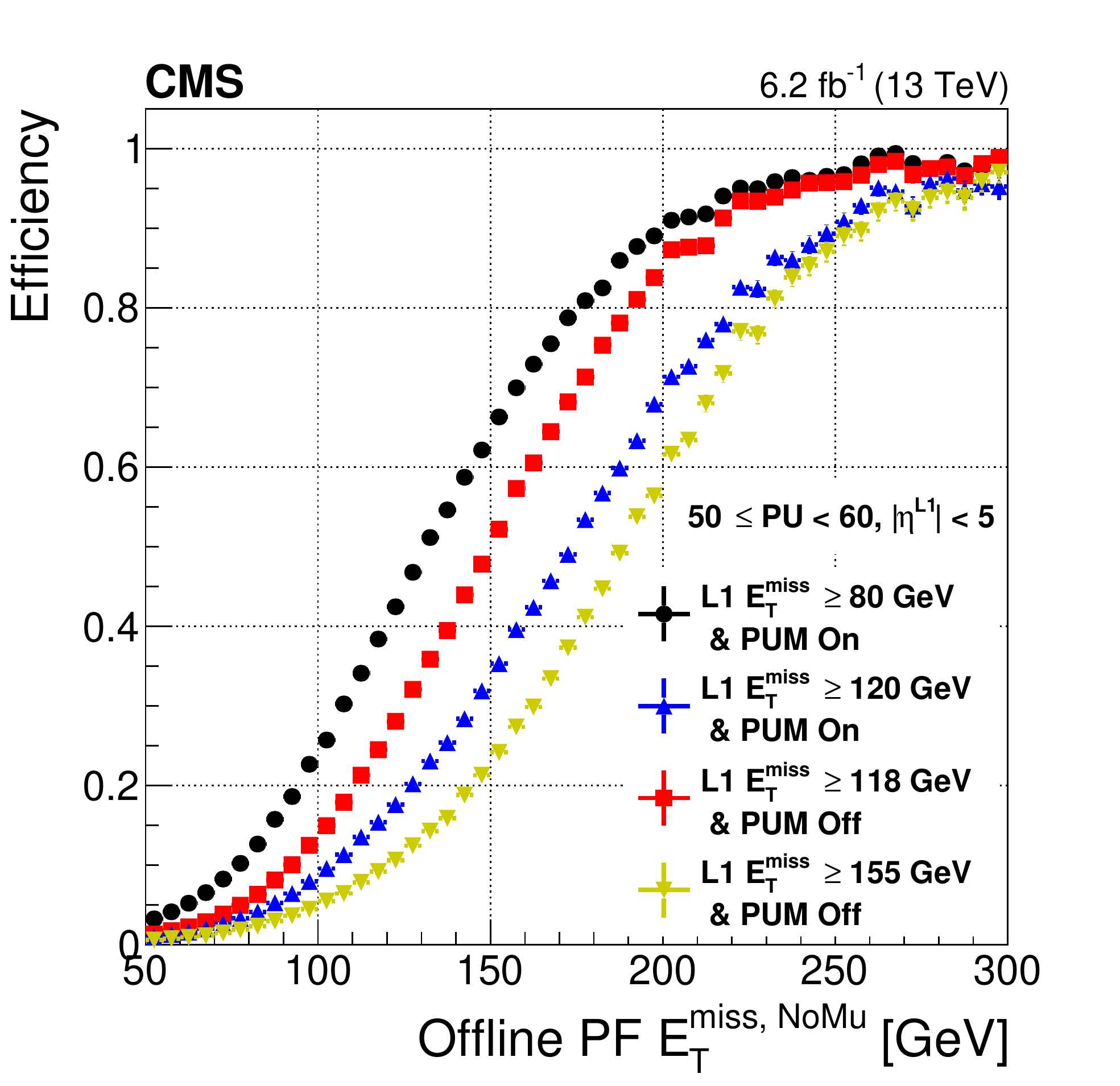}
		\includegraphics[width=0.495\textwidth]{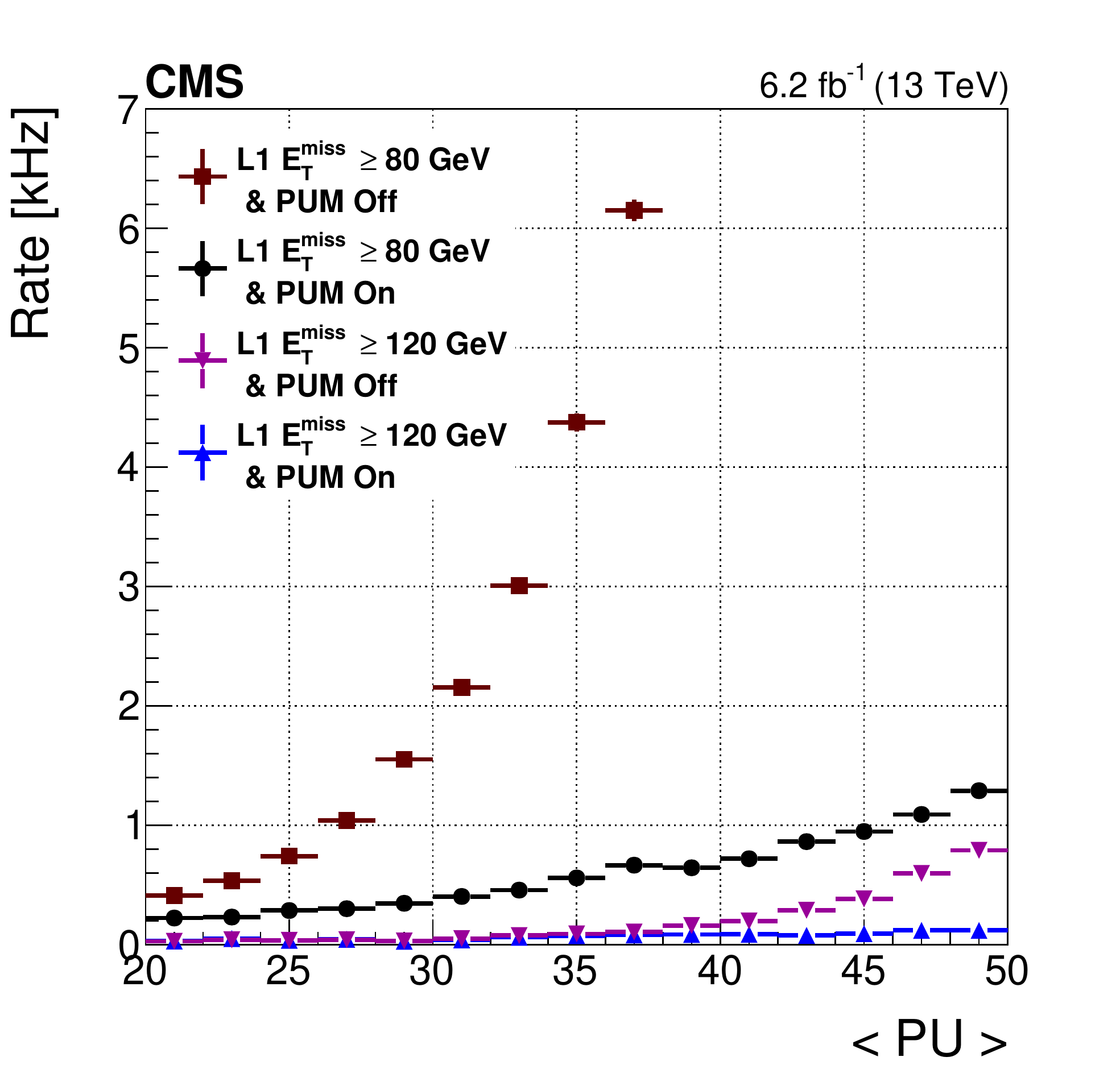}
	\caption{Efficiency curves with and without pileup mitigation (PUM)
	applied are compared (\cmsLeft) for the thresholds that give the same
	rate. These are shown as a function of the offline reconstructed
	particle flow missing energy excluding muons (PF
	$\MET{}^{\text{,NoMu}}$). Rate versus the average pileup per luminosity
	section is shown (\cmsRight) with and without pileup mitigation applied.  }
\label{fig:met} \end{figure} 

Toward the end of 2016 data taking, an increase in the instantaneous luminosity
revealed a significantly nonlinear dependence of the \MET rates on event pileup.
For 2017 and 2018 data taking, pileup mitigation was implemented and applied on an
event-by-event basis to the \MET algorithm.  The event pileup is estimated with
the variable  $\ntt$ (described in Section~\ref{sec:egamma}) and is used along
with the TT $\eta$ to retrieve from a LUT a pileup- and $\eta$-dependent \ET
threshold below which TTs do not enter the calculation of the \MET. The LUT was
derived using functions encoding the pileup estimate, the TT $\eta$, and the TT
width in $\eta$, since the pileup energy per TT increases with $\abs{\eta}$ and the
TT size. The functional form and corresponding constant factors were optimized
to give the best trigger efficiency, measured in single-muon triggered data, for
a fixed rate calculated from unbiased data. The LUT was also derived by
calculating the average TT \ET for each value of $\eta$ from unbiased data, and
this gave a similar performance to the function-based LUT. 

The improvement of the \MET trigger efficiency after using the pileup mitigation
algorithm is shown in Fig.~\ref{fig:met}, for events from 2018 single-muon
triggered data with pileup between 50 and 60. The rate of the Level-1 \MET{}
trigger with a threshold of 80 (120)\GeV with pileup mitigation enabled is the
same as the rate for a threshold of 118 (155)\GeV with pileup mitigation
switched off.  Also shown in Fig.~\ref{fig:met} is the pileup dependence for
fixed thresholds of the Level-1 \MET algorithm, with and without pileup
mitigation. Rate is calculated from unbiased data for 2855 filled bunches for
the Level-1 thresholds of 80 and 120\GeV, where the pileup shown is the average
pileup per luminosity section. Applying pileup mitigation, by excluding
low-energy TTs in events with significant pileup and reducing the contribution
from large TTs at large eta, provided a significant rate reduction while
maintaining trigger efficiency. This allowed the Level-1 \MET threshold to be
reduced, increasing sensitivity to a range of important physics channels. 

\subsection{Adjustments for heavy ion collisions} In heavy ion (HI) lead-lead
collisions, a large particle multiplicity variation is observed; although
peripheral collisions can result in only a few particles per interaction,
central events can produce large multiplicities equivalent to $\Pp\Pp$
collisions with pileup of 200--300.  While most of the algorithms developed for
$\Pp\Pp$ collisions were reused, the wide range of multiplicity required that
some of the Level-1 algorithms were optimized, and a few were developed
specifically for HI collisions.

To select low-\pt hadronic collisions efficiently, a minimum bias trigger was
developed based on a coincidence of energy deposits in the positive and negative
$\eta$ sides of the HF calorimeter. Using the same principle, an
ultra-peripheral collision (UPC) trigger was designed to be activated only in a
specific low-energy region. A high multiplicity UPC algorithm was also
developed, based on the imbalance between the positive and negative $\eta$ sides
of the sum of trigger tower \ET in the barrel calorimeter.

In addition, the parameters of the $\egamma$ algorithm were adapted by removing
the H/E constraint and adjusting the fine grain bit threshold. For optimal
performance in the HI environment, the jet pileup subtraction algorithm used for
proton collisions was replaced with an alternative, based on the average energy
in $\phi$-rings of the calorimeter.

\section{The global trigger} \label{sec:global} 

The \uGT combines information from both the \uGMT and the calorimeter
Layer-2, and it performs a trigger decision based on a menu of sophisticated
algorithms, as described in Section~\ref{sec:menu}.  The \uGT  is made compact
and reliable by merging the functionality formerly distributed across multiple
distinct boards into a single processor board type. The \uGT distributes its
processing across up to six of these common boards working independently of each
other. The outputs of the processing boards are merged before being sent to the
HLT. 

The \uGT began operation with one processing board in 2016 and was extended to
its final form of six processing boards by the beginning of 2017. The use of
multiple processing boards with larger FPGAs permitted the computation of more
high-level quantities, such as invariant or transverse masses, by using LUTs and
digital signal processors. In this way, it is possible to migrate
increasingly higher-level quantities from the HLT into the Level-1 trigger.

Occasionally, the LHC running parameters change on short notice, making it
operationally challenging to reoptimize the Level-1 trigger menu. The \uGT{}
calculates preview rates for each prescale column, so that the shift crew can
avoid premature enabling of prescale columns that would raise the Level-1 rate
above the limit.

A unique classification of certain physics objects input to the \uGT can be
difficult. For example, a hadronic jet could be separately reconstructed as both
a $\PGt$ lepton and a jet by the Layer-2 trigger. This poses a problem in
algorithms looking for both jets and $\PGt$ leptons.  The \uGT implements a
dedicated treatment to resolve ambiguities for all possible object combinations
between Level-1 objects, such as $\PGt$ leptons and jets. For example, in an
event with two jets, each having $\ET > 35\GeV$, and one $\PGt$ lepton with $\ET
> 45\GeV$, both jets must be separated by $\Delta R>0.2$ from the $\PGt$
candidate, which ensures that such an event contains at least three
nonoverlapping objects.

\subsection{Dedicated analysis triggers}

The large processing power available in the \uGT permits the implementation of
sophisticated analysis-targeted trigger algorithms. In this section, three types
of such algorithms are discussed.  The first type selects vector boson fusion
(VBF) events using the invariant mass of jet pairs. The second type targets the
production of low-mass dimuon resonances (\eg, \PgU{} decays), and the third
tags \PQb jet candidates using jet-muon coincidences.

\begin{figure} \centering
	\includegraphics[width=0.49\textwidth]{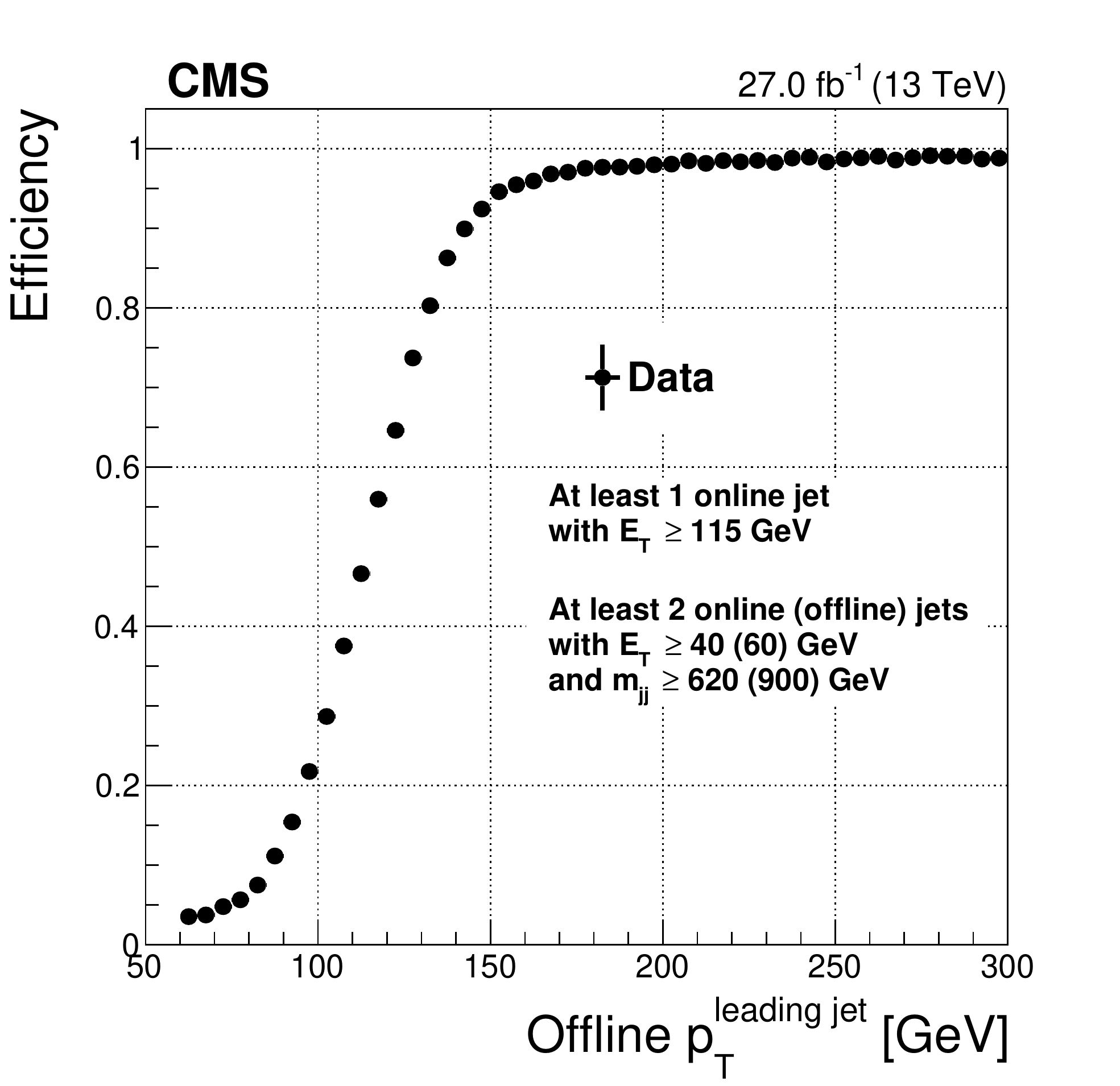}
	\includegraphics[width=0.49\textwidth]{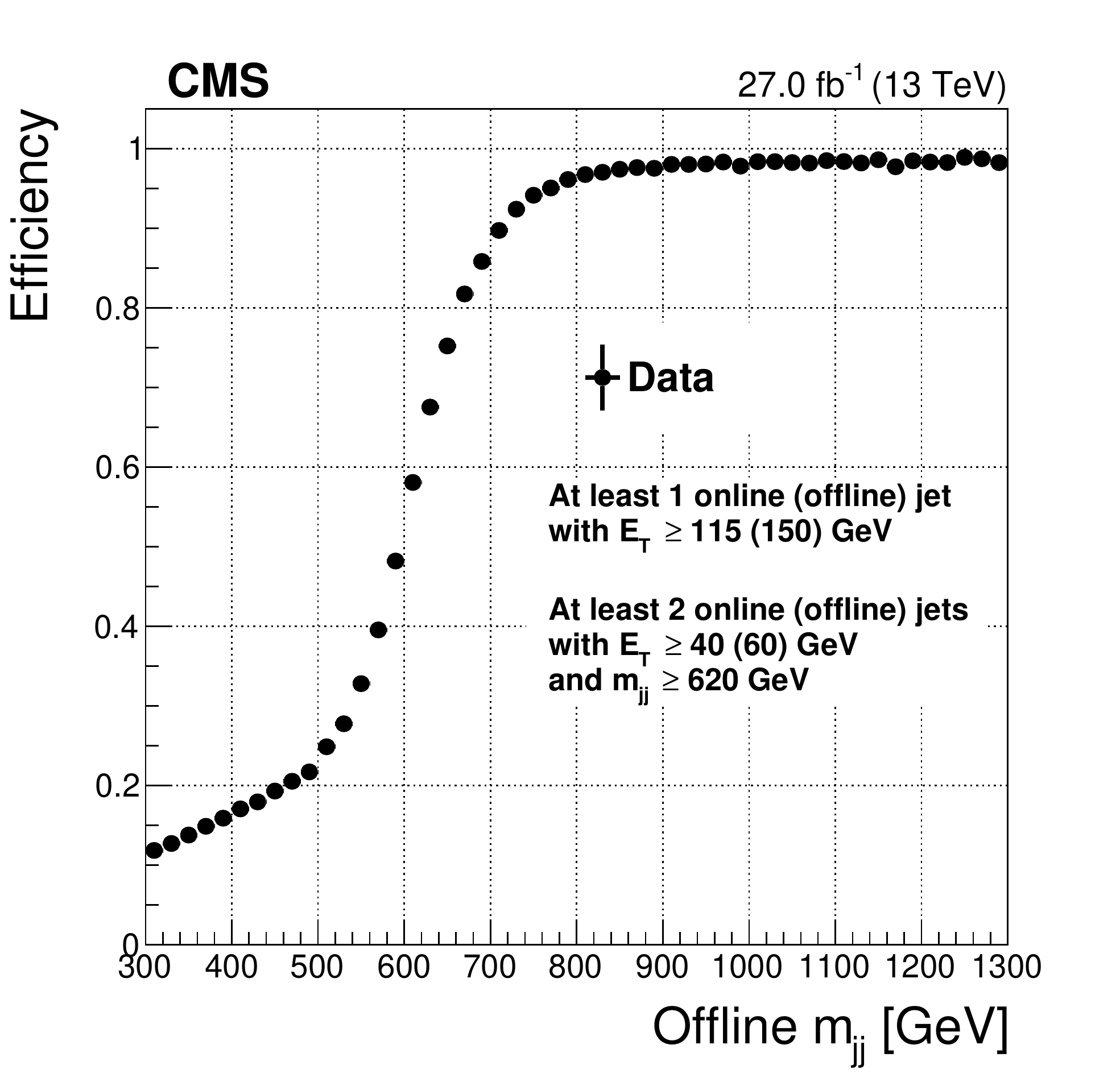}
	\caption{Efficiency of the Level-1 VBF trigger as a function of the
	offline leading jet \pt (\cmsLeft) and $\mjj$ (\cmsRight), estimated as
	the fraction of $\PH\to\PGt\PGt$ analysis-like offline events passing
	the Level-1 VBF trigger selection (the Level-1 and offline requirements
	applied are detailed in the plots). The efficiency is evaluated using
	2017 data.} \label{fig:VBFtrigger} \end{figure}

\paragraph*{Dedicated vector boson fusion trigger} Higgs boson production via
VBF occurs through the interaction of two \PW{} or \PZ{} bosons. The incoming
quarks only lose a small fraction of their energy in the interaction.  After
hadronizing, the outgoing quarks typically form jets in the forward direction,
with a large invariant mass and separation in $\eta$.  The VBF algorithm looks
for at least two jets with $\ET>115$ and $\ET> 35$\GeV and at least one pair of
jets with $\ET>35$\GeV each and an invariant mass greater than 620\GeV.  In the
\uGT{}, half of the squared mass is computed: \begin{linenomath}
	\begin{equation*} m_{j_1 j_2}^2/2 = \pt^{j_1} \pt^{j_2} [
	\cosh(\Delta\eta_{j_1 j_2}) - \cos(\Delta\phi_{j_1 j_2})],
	\end{equation*} where $\cosh(\Delta\eta_{j_1 j_2})$ and
	$\cos(\Delta\phi_{j_1 j_2})$ are obtained through dedicated LUTs using
	the $\eta$ and $\phi$ of the jets as inputs.  The algorithm can select
2- or 3-jet topologies, depending on whether the jet with $\ET{}>115\GeV$ enters
a pair with $m_{j_1 j_2} > 620\GeV$.  \end{linenomath} The performance of the
Level-1 VBF trigger algorithm was measured in 2017 data, using an unbiased
sample collected with a single-muon trigger. Figure \ref{fig:VBFtrigger} shows
that the efficiency, as functions of the offline leading jet \pt and the maximum
dijet invariant mass, reaches a high efficiency plateau for VBF-like events,
making it suitable as a lower rate and high efficiency trigger for VBF-like
topologies.  The Level-1 VBF trigger algorithms were used to seed HLT paths in
2017 and 2018, increasing the signal acceptance, especially for invisible Higgs
boson decays and $H\to\PGt\PGt$~\cite{CMS:higgs-invisible}.

\begin{figure} \centering \includegraphics[width=0.60\textwidth]{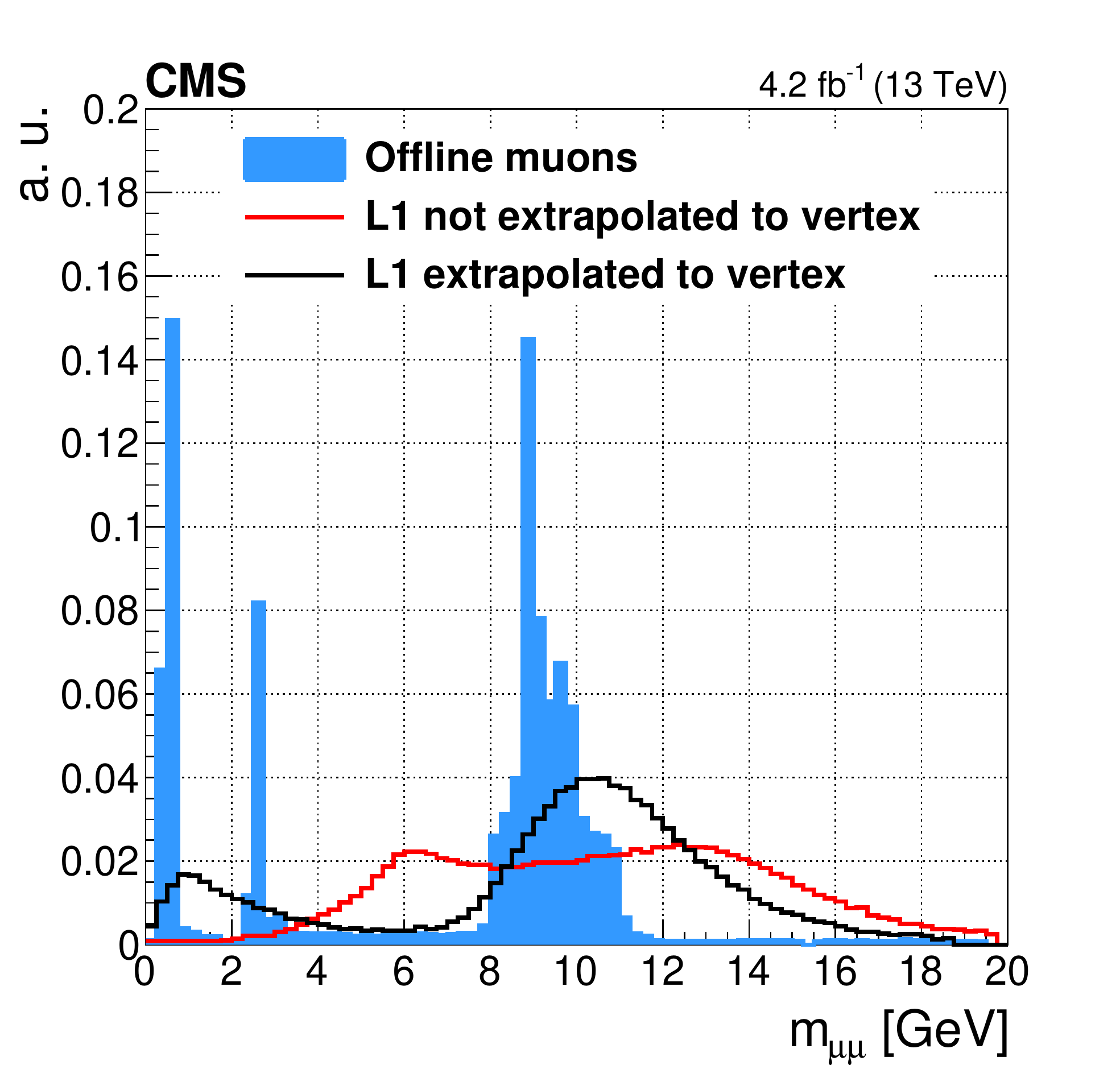}
	\caption{The offline and Level-1 $\mmumu$ spectra of oppositely charged
	muons, with and without extrapolation of the Level-1 track parameters to
	the nominal vertex, using a data set of low-mass dimuons.  The
	highest-mass resonance corresponds to the $\Upsilon$ mesons, and is
	clearly identifiable both offline and in Level-1, after extrapolation.
	The Level-1 $\mmumu$ spectrum is shifted higher compared with the
	offline spectrum because of \pt offsets designed to make the Level-1
	muon trigger 90\% efficient at any given \pt threshold.}
\label{fig:DiMuonMass} \end{figure}

\paragraph*{Low-mass dimuon triggers} The \pt thresholds for the usual dimuon
triggers are not well adapted to record dimuon resonances with masses less than
20\GeV. These thresholds are typically 15\GeV on the leading muon and 5\GeV on
the subleading muon, so they only select very boosted low-mass dimuon
resonances. To collect inclusive low-mass dimuon pairs at low enough rates, the
\uGT can compute the dimuon invariant mass $\mmumu$, using the same technique
described above in the case of the VBF trigger.  Seeds requiring $3 < \mmumu < 9
\GeV$ and $5 < \mmumu < 17 \GeV$ are included in the menu, as shown in
Table~\ref{tab:menu-slim}.  Figure~\ref{fig:DiMuonMass} shows the Level-1 and
the offline $\mmumu$ spectrum in Run 2 data collected with multi-muon triggers.
The 9.46\GeV $\Upsilon$ meson peak can be isolated quite distinctly after the
muon coordinates are extrapolated to the nominal vertex, as described in
Section~\ref{sec:uGMT}.  A recent example of a successful low-mass trigger is
the 5.6 sigma observation of $\PBzs \to \Pgmp \Pgmm$ with a branching fraction
of $2.9\pm0.7\pm0.2 \times 10^{-9}$ with a limit set on $\PBz \to \Pgmp \Pgmm <
3.6 \times 10^{-10}$ at $95\%$ confidence level~\cite{cms:btomumu}.

\paragraph*{\PQb jet tagging using muons}

A significant fraction of \PQb hadron decays produce muons. These are often in
the same direction as the rest of the products of the \PQb  hadron. The
Level-1 trigger includes a simple \PQb{}-tagging algorithm based on the
proximity of a muon to a jet.  For example, the \uGT implements seeds looking
for events with one $\pt>3\GeV$ muon and two $\ET>16\GeV$ jets, where the muon
is within $\Delta R<0.4$ of one of the jets. This new feature improves the
efficiency and reduces the rate of the already available \PQb jet tagging
seeds that were previously limited by the use of uncorrelated $\Delta \eta$ and
$\Delta \phi$ information between jets and muons. 

\section{Data certification and validation} \label{sec:certification} The
Level-1 trigger performance is monitored online by physicists working in shifts
for nonstop data-taking operational support, who are trained to recognize and
solve trigger problems. Trigger rates are continuously displayed for each
algorithm, as well as occupancy plots and energy distributions for each physics
object. Unexpected discrepancies compared with the reference distributions are
investigated promptly by Level-1 object experts who determine the appropriate
course of action. 

The Level-1 trigger system uses a two-step process to certify the collected
data. ``Express certification'' is typically performed within 24 hours, and
identifies any anomalous behavior of the trigger that may have passed unnoticed
during data taking. In the ``final certification'', high-quality data are
selected for physics analyses. The certification is performed for both collision
and cosmic ray data taking.

During express certification, the time evolution of the total output rate of the
Level-1 trigger is examined, taking into account information about the beam
conditions, prescale values applied, status of each subdetector, and dead time
(the recording time lost because the readout system is not ready to accept new
events). Individual rates of different trigger seeds targeting physics objects
are compared with reference rates as a function of pileup.

For each run, data quality monitoring (DQM) plots are produced, including
occupancy of muon and calorimeter trigger systems, physics object variables
(such as muon $\eta$ and $\phi$), and the timing of trigger seeds. The data are
also compared with an emulation of the Level-1 trigger reconstruction. The DQM
system performs statistical tests to identify distributions that differ from
expectations. Any abnormal rates or DQM distributions may indicate incorrect
functioning of some part of the Level-1 trigger system, which will be studied,
corrected (when possible), and taken into account in the final certification. 

The final Level-1 trigger certification is based on the comparison of the
efficiency and resolution measured for each type of Level-1 object to the
corresponding offline quantities, combined with the information from express
certification. The efficiencies are calculated for different types of trigger
seeds using a tag-and-probe method, and the resolutions are determined by
comparing the trigger-level kinematic variables with their offline reconstructed
counterparts, similarly to the performance studies presented in this paper. If
the efficiencies and resolutions show no significant deviation from the expected
performance, and the results of the express certification indicate that the
trigger operated successfully, the data is certified as valid for physics
analyses from the point of view of the Level-1 trigger. 

If a certain run does not pass the certification criteria, the source of the
performance loss is identified and analyzed. In general, trigger performance
losses are caused either by a malfunctioning Level-1 trigger subsystem itself,
or by missing or corrupted input from other detector subsystems. In case of a
severe performance loss, the data must be discarded independently of the origin
of the problem. To minimize the data loss, the certification is performed per
luminosity section.

In 2018, 1.36\% of the collision data collected by CMS was certified as ``bad''
by Level-1, but only 0.016\% was invalidated solely from Level-1 trigger issues.
The remainder included some other significant detector malfunction.

\section{Summary and conclusions} \label{sec:summary}

The CMS Level-1 trigger system was upgraded for Run~2 of the LHC. The system
improved in performance and flexibility using high-bandwidth serial I/O links
for data transfer and large, modern field-programmable gate arrays for
reconfigurable algorithms. Maintenance improved with increased standardization
through the use of the MicroTCA telecommunications standard and common hardware
designs for its components.

The new trigger hardware provides improved $\egamma$ isolation performance,
substantially more efficient \PGt lepton identification, improved muon
transverse momentum resolution, and the ability to reconstruct jets with finer
calorimeter granularity. New features, such as pileup subtraction and invariant
mass calculations,  expand the trigger design possibilities. These improvements
help to control trigger rates and keep thresholds at lower levels than would be
required with the previous system despite the significantly increased LHC
energy, luminosity, and pileup in Run~2. The adoption of more powerful trigger
processors led to the deployment of more advanced trigger algorithms, targeting
specific analyses, resulting in significant improvements in physics capability
compared to Run~1.

The upgraded Level-1 trigger system operated during Run 2 with high efficiency
for all physics objects, and adapted to the rapidly changing LHC running
conditions. As a result, the trigger efficiency was stable and independent of
the evolving LHC parameters. Special LHC running conditions and heavy-ion data
taking were accommodated effectively as well, exploiting the full capability and
flexibility of the trigger system.

The upgraded system improved the energy and momentum resolution, and the
identification efficiency and background rejection of the Level-1 physics
objects.  This significantly lowered the rate at a given threshold compared with
the Run 1 system, thereby allowing similar trigger requirements to fit within
the unchanged Level-1 rate limit.

An analysis of Run 2 data shows that the trigger rate reduction and efficiency
gain benefited  the physics program of the CMS Collaboration under conditions of
increased LHC energy, luminosity, and pileup. An example includes the $\PH\to
\PGt \PGt$ analysis \cite{cms:higgs-taus}, which shows a significant improvement
in trigger efficiency; other Higgs boson decay channel analyses maintained a
similar trigger efficiency despite the harsher beam conditions. Moreover, all
analyses looking for large transverse missing energy (\MET), including searches
for dark matter, supersymmetry~\cite{cms:ra1}, and invisible Higgs boson
decay~\cite{CMS:higgs-invisible}, were only possible in Run~2 because of the
improved resolution of the Level-1 \MET and the pileup mitigation algorithm.
Searches for low-mass dimuon resonances exploited the invariant mass requirement
for reducing the rate and lowering the muon momentum
requirement~\cite{cms:btomumu}.

\begin{acknowledgments} 
\hyphenation{Bundes-ministerium Forschungs-gemeinschaft Forschungs-zentren Rachada-pisek}
We congratulate our colleagues in the CERN accelerator departments for the excellent performance of the LHC and thank the technical and administrative staffs at CERN and at other CMS institutes for their contributions to the success of the CMS effort. In addition, we gratefully acknowledge the computing centers and personnel of the Worldwide LHC Computing Grid for delivering so effectively the computing infrastructure essential to our analyses. Finally, we acknowledge the enduring support for the construction and operation of the LHC and the CMS detector provided by the following funding agencies: the Austrian Federal Ministry of Education, Science and Research and the Austrian Science Fund; the Belgian Fonds de la Recherche Scientifique, and Fonds voor Wetenschappelijk Onderzoek; the Brazilian Funding Agencies (CNPq, CAPES, FAPERJ, FAPERGS, and FAPESP); the Bulgarian Ministry of Education and Science; CERN; the Chinese Academy of Sciences, Ministry of Science and Technology, and National Natural Science Foundation of China; the Colombian Funding Agency (COLCIENCIAS); the Croatian Ministry of Science, Education and Sport, and the Croatian Science Foundation; the Research and Innovation Foundation, Cyprus; the Secretariat for Higher Education, Science, Technology and Innovation, Ecuador; the Ministry of Education and Research, Estonian Research Council via PRG780, PRG803 and PRG445 and European Regional Development Fund, Estonia; the Academy of Finland, Finnish Ministry of Education and Culture, and Helsinki Institute of Physics; the Institut National de Physique Nucl\'eaire et de Physique des Particules~/~CNRS, and Commissariat \`a l'\'Energie Atomique et aux \'Energies Alternatives~/~CEA, France; the Bundesministerium f\"ur Bildung und Forschung, the Deutsche Forschungsgemeinschaft (DFG) under Germany's Excellence Strategy -- EXC 2121 ``Quantum Universe" -- 390833306, and Helmholtz-Gemeinschaft Deutscher Forschungszentren, Germany; the General Secretariat for Research and Technology, Greece; the National Research, Development and Innovation Fund, Hungary; the Department of Atomic Energy and the Department of Science and Technology, India; the Institute for Studies in Theoretical Physics and Mathematics, Iran; the Science Foundation, Ireland; the Istituto Nazionale di Fisica Nucleare, Italy; the Ministry of Science, ICT and Future Planning, and National Research Foundation (NRF), Republic of Korea; the Ministry of Education and Science of the Republic of Latvia; the Lithuanian Academy of Sciences; the Ministry of Education, and University of Malaya (Malaysia); the Ministry of Science of Montenegro; the Mexican Funding Agencies (BUAP, CINVESTAV, CONACYT, LNS, SEP, and UASLP-FAI); the Ministry of Business, Innovation and Employment, New Zealand; the Pakistan Atomic Energy Commission; the Ministry of Science and Higher Education and the National Science Center, Poland; the Funda\c{c}\~ao para a Ci\^encia e a Tecnologia, Portugal; JINR, Dubna; the Ministry of Education and Science of the Russian Federation, the Federal Agency of Atomic Energy of the Russian Federation, Russian Academy of Sciences, the Russian Foundation for Basic Research, and the National Research Center ``Kurchatov Institute"; the Ministry of Education, Science and Technological Development of Serbia; the Secretar\'{\i}a de Estado de Investigaci\'on, Desarrollo e Innovaci\'on, Programa Consolider-Ingenio 2010, Plan Estatal de Investigaci\'on Cient\'{\i}fica y T\'ecnica y de Innovaci\'on 2017--2020, research project IDI-2018-000174 del Principado de Asturias, and Fondo Europeo de Desarrollo Regional, Spain; the Ministry of Science, Technology and Research, Sri Lanka; the Swiss Funding Agencies (ETH Board, ETH Zurich, PSI, SNF, UniZH, Canton Zurich, and SER); the Ministry of Science and Technology, Taipei; the Thailand Center of Excellence in Physics, the Institute for the Promotion of Teaching Science and Technology of Thailand, Special Task Force for Activating Research and the National Science and Technology Development Agency of Thailand; the Scientific and Technical Research Council of Turkey, and Turkish Atomic Energy Authority; the National Academy of Sciences of Ukraine; the Science and Technology Facilities Council, UK; the US Department of Energy, and the US National Science Foundation. 
Individuals have received support from the Marie-Curie program and the European Research Council and Horizon 2020 Grant, contract Nos.\ 675440, 752730, and 765710 (European Union); the Leventis Foundation; the A.P.\ Sloan Foundation; the Alexander von Humboldt Foundation; the Belgian Federal Science Policy Office; the Fonds pour la Formation \`a la Recherche dans l'Industrie et dans l'Agriculture (FRIA-Belgium); the Agentschap voor Innovatie door Wetenschap en Technologie (IWT-Belgium); the F.R.S.-FNRS and FWO (Belgium) under the ``Excellence of Science -- EOS" -- be.h project n.\ 30820817; the Beijing Municipal Science \& Technology Commission, No. Z191100007219010; the Ministry of Education, Youth and Sports (MEYS) of the Czech Republic; the Lend\"ulet (``Momentum") Program and the J\'anos Bolyai Research Scholarship of the Hungarian Academy of Sciences, the New National Excellence Program \'UNKP, the NKFIA research grants 123842, 123959, 124845, 124850, 125105, 128713, 128786, and 129058 (Hungary); the Council of Scientific and Industrial Research, India; the HOMING PLUS program of the Foundation for Polish Science, cofinanced from European Union, Regional Development Fund, the Mobility Plus program of the Ministry of Science and Higher Education, the National Science Center (Poland), contracts Harmonia 2014/14/M/ST2/00428, Opus 2014/13/B/ST2/02543, 2014/15/B/ST2/03998, and 2015/19/B/ST2/02861, Sonata-bis 2012/07/E/ST2/01406; the National Priorities Research Program by Qatar National Research Fund; the Ministry of Science and Higher Education, project no. 02.a03.21.0005 (Russia); the Tomsk Polytechnic University Competitiveness Enhancement Program and ``Nauka" Project FSWW-2020-0008 (Russia); the Programa de Excelencia Mar\'{i}a de Maeztu, and the Programa Severo Ochoa del Principado de Asturias; the Thalis and Aristeia programs cofinanced by EU-ESF, and the Greek NSRF; the Rachadapisek Sompot Fund for Postdoctoral Fellowship, Chulalongkorn University, and the Chulalongkorn Academic into Its 2nd Century Project Advancement Project (Thailand); the Kavli Foundation; the Nvidia Corporation; the SuperMicro Corporation; the Welch Foundation, contract C-1845; and the Weston Havens Foundation (USA).   \end{acknowledgments}

\bibliography{auto_generated}

\clearpage \appendix

\section{Level-1 trigger prefiring} \label{sec:prefiring}

Since the beginning of Run 2, a slowly developing shift in the shape of the ECAL
pulses was observed. This effect, which manifests itself as an increasing offset
in the timing calibration of the pulses, is radiation-induced and is related to
the transparency loss of the ECAL crystals.  Because of this, the endcap
crystals at highest pseudorapidity are most affected. This timing calibration
offset is compensated offline via regular pulse shape and timing calibration
measurements, but was not corrected online in the formation of the ECAL TPs.
With time, the accumulated offset brought the endcap pulses to the limit of the
region where the trigger bunch-crossing assignment would be affected. Once this
was realized, in early 2018, the endcap timing delays in the ECAL front-end
electronics were corrected, and the   pulse synchronization was optimized.
However, in 2016--2017, a gradually increasing fraction of ECAL TPs at $\abs{\eta} >
2.5$ had wrongly associated an energy deposit to the previous bunch crossing
(BX~$-1$). When such a misassignment occurs it causes several effects on the
data. First, it may lead the Level-1 trigger system to ``prefire'', \ie, to
accept the earlier collision in BX~$-1$, whereas the collision in BX~0 is the
one of interest. Secondly, when the misassigned TP energy is not large enough to
pass the trigger condition, it induces a bias in the energy measurement of
calorimeter deposits in the trigger chain and offline.

Prefiring happens, \eg, when an ECAL TP, whose \ET exceeds the threshold of
the single electron trigger, is assigned to BX~$-1$; or when the misassignment
of an ECAL TP leads to a large \MET{} reconstructed at Level-1 in BX~$-1$.
Prefiring of Level-1 triggers represents a problem in their combined effect with
the CMS trigger rules. These are the conditions that prevent buffer overflows in
special cases. Triggers rules are enforced immediately after the final decision
of the global trigger (\uGT{}). The most commonly enforced trigger rules prevent
the issuance of more than one Level-1 trigger acceptance decision in three
consecutive bunch crossings, or more than two Level-1 trigger acceptances in 25
consecutive BXs.  Thus, when a trigger accepts the event in BX~$-1$, the
interesting event in BX~0 will not be accepted. The readout event in BX~$-1$
will likely be rejected by the HLT since it is unlikely to reconstruct any
interesting physics objects.  The main consequence of prefiring is therefore an
inefficiency in recording potentially interesting events. 

The measurement of the prefiring rate requires the use of a special set of
events called ``unprefirable'' events.  An event in BX~0 is unprefirable when
the event in BX~$-3$ is accepted by the Level-1 trigger: the trigger rules veto
events in BX~$-2$ and BX~$-1$. For every triggered event, all Level-1 objects
and \uGT decision bits are stored in a window of ${\pm}2 \text{BX}$. Therefore,
from a set of selected unprefirable events, the prefiring probability can be
computed for a specific analysis selection. The rate of unprefirable events is
very small compared with the total number of events in any given primary data
set, about 0.1\%. Ad hoc corrections at the analysis level are applied to
correct for this effect. One of the most affected analyses is the search for
invisible decays of a Higgs boson produced via VBF, with energetic forward jets.
Their measurements from an unbiased data sample result in a correction of about
1\% for $\mjj$ of 200\GeV and up to 20\% for $\mjj$ larger than 3.5\TeV
\cite{CMS:higgs-invisible}.

Secondary effects of the TP time shift are a potential bias in the energy
measurement of the calorimeter deposit in the trigger chain.  If the energy of
early TPs is large enough to create a Level-1 object that prefires a Level-1
trigger path, the event in BX~0 is lost. In contrast, if BX~$-1$ is not
accepted, a residual effect on BX~0 is still present because the information
about the TPs associated with BX~$-1$ is lost. This residual effect biases the
energy of several Level-1 objects and causes a degradation of the trigger
efficiency turn-on . Standard trigger efficiency measurements and scale factors
generally applied in physics analyses account for this effect. 

A second bias arises because of the impact on the ECAL selective readout logic.
The TP inputs are used by the ECAL selective readout units  to decide whether a
certain region of the detector needs to be read out or not (zero-suppressed).
Crystals associated with the early TP will be read out by the ECAL data
acquisition system in zero-suppression mode, injecting a bias into the
HLT/offline energy measurement. For high-\pt jets this effect is expected to be
small because the zero-suppression thresholds are low. This energy bias is
mostly recovered by the residual jet energy corrections applied at the analysis
level.

\cleardoublepage \section{The CMS Collaboration \label{app:collab}}\begin{sloppypar}\hyphenpenalty=5000\widowpenalty=500\clubpenalty=5000\input{TRG-17-001-authorlist.tex}\end{sloppypar}
\end{document}

%% file: TRG-17-001-authorlist.tex
\vskip\cmsinstskip
\textbf{Yerevan Physics Institute, Yerevan, Armenia}\\*[0pt]
A.M.~Sirunyan$^{\textrm{\dag}}$, A.~Tumasyan
\vskip\cmsinstskip
\textbf{Institut f\"{u}r Hochenergiephysik, Wien, Austria}\\*[0pt]
W.~Adam, F.~Ambrogi, B.~Arnold, H.~Bergauer, T.~Bergauer, M.~Dragicevic, J.~Er\"{o}, A.~Escalante~Del~Valle, M.~Flechl, R.~Fr\"{u}hwirth\cmsAuthorMark{1}, M.~Jeitler\cmsAuthorMark{1}, N.~Krammer, I.~Kr\"{a}tschmer, D.~Liko, T.~Madlener, I.~Mikulec, N.~Rad, J.~Schieck\cmsAuthorMark{1}, R.~Sch\"{o}fbeck, M.~Spanring, S.~Templ, W.~Waltenberger, C.-E.~Wulz\cmsAuthorMark{1}, M.~Zarucki
% ...
\vskip\cmsinstskip
\dag: Deceased\\
1:~~Also at Vienna University of Technology, Vienna, Austria\\
% ...
